\pdfoutput=1
\documentclass[twocolumn]{aastex701}
\usepackage[T1]{fontenc}
\usepackage[utf8]{inputenc}
\usepackage{amsmath}
\usepackage{comment,txfonts}
\usepackage{graphicx}
\usepackage{color}
\usepackage{multirow}

\graphicspath{{./}{figures/}}

\begin{document}

\author[orcid=0000-0001-6463-2929,sname='Kanon Nakazawa']{Kanon Nakazawa}
\altaffiliation{The University of Tokyo}
\affiliation{Department of General Systems Studies, The University of Tokyo, Meguro, Tokyo 153-8902, Japan}
\email[show]{kanon-nakazawa@g.ecc.u-tokyo.ac.jp}  

\author[orcid=0000-0003-3290-6758, sname='Kazumasa Ohno']{Kazumasa Ohno} 
\altaffiliation{National Astronomical Observatory of Japan}
\affiliation{Division of Science, National Astronomical Observatory of Japan, Mitaka, Tokyo 181-8558, Japan}
\email{ohno.k.ab.715@gmail.com}

\date{February 2026}

\title{Sulfur Enrichment in Close-in Exoplanet Atmospheres Induced by Pebble Drift across the Salt Line}

\begin{abstract}
Observations of JWST have revealed that several close-in exoplanets have sulfur-rich atmospheres through SO$_2$ detections.
Atmospheric sulfur is often thought to originate from solid accretion during planet formation, whereas recent simultaneous detections of SO$_2$ and NH$_3$ challenge this conventional scenario. 
In this study, we propose that ammonium salts, such as NH$_4$SH tentatively detected in comets and molecular clouds, play a significant role in producing sulfur-rich disk gases, which serve as the ingredient of giant planet atmospheres. 
We simulated the radial transport of dust containing volatile ices and ammonium salts, along with the dissociation, sublimation, and recondensation of these materials, thereby predicting the atmospheric chemical structures and transmission spectra of planets inheriting these compositions. Assuming that ammonium salts sequester 20\% of the elemental nitrogen and sulfur budgets, our results reveal that they enhance sulfur and nitrogen abundances in disk gases to $2$--$10$ times the solar values near the salt dissociation line.
Photochemical simulations demonstrate that SO$_2$, NS, H$_2$S, NO, and NH$_3$ become the dominant N and S chemical species in the atmospheres on planets that inherited the gas compositions inside H$_2$O snowline.  SO$_2$ features clearly appear in the infrared transmission spectra when the salt-bearing grains enhance the sulfur abundance of disk gas by pebble drift. 
Our model provides a novel scenario that explains the SO$_2$ detected in some exoplanet atmospheres solely from disk gas accretion. 
Volatile‑element ratios, particularly N/S and C/O, would provide a key to disentangle our scenario from the conventional solid‑accretion scenario.
\end{abstract}

\keywords{%
  \uat{Atmospheric composition}{2120} --- %
  \uat{Planetary atmospheres}{1244} --- %
  \uat{Protoplanetary disks}{1300}}

\section{Introduction}
\label{sec:intro}
Planetary atmospheric compositions record the history of planet formation processes and elemental transport in protoplanetary disks. The James Webb Space Telescope (JWST), operational since 2022, is capable of identifying previously undiscovered chemical species in exoplanetary atmospheres owing to its unprecedented wavelength coverage and spectral resolution \citep[e.g.][]{2023Natur.614..659R, 2023Natur.614..664A, 2024Natur.625...51D}, enabling us to precisely constrain atmospheric elemental ratios such as O/H and C/O that provide clues to disk compositions and planet formation \citep[e.g.,][]{2011ApJ...743L..16O,2014ApJ...794L..12M,2018A&A...613A..14E,2021A&A...654A..72S,2024MNRAS.535..171P,2025arXiv250616060O}.

Atmospheric sulfur has gained special attention in the current exoplanet community since the discovery of SO$_2$ in the JWST Transiting Exoplanet Community Early Release Science Program.
\citet{2023Natur.614..659R} and \citet{2023Natur.614..664A} reported significant SO$_2$ in the transmission spectrum of hot Saturn WASP-39b observed by JWST NIRSpec PRISM and G395H, which is the first discovery of sulfur-bearing species from exoplanetary atmospheres.
In hot Saturn/Neptune upper atmospheres, SO$_2$ is produced via photochemical reaction between H$_2$S and OH/H radicals produced by the photolysis of H$_2$O \citep{2016ApJ...824..137Z, 2021ApJ...923..264T}.
Thus, SO$_2$ provides clues to study atmospheric photochemistry in exoplanets.
Subsequent studies detected SO$_2$ in other close-in exoplanets, such as WASP-107b \citep{2024Natur.625...51D,2024Natur.630..836W,2024Natur.630..831S} and GJ3470b \citep{2024ApJ...970L..10B}.
Besides the discovery of SO$_2$, \citet{2024Natur.632..752F} detected H$_2$S in the atmosphere of HD189733b with moderate metallicity, thereby constraining the C/S ratio to be a sub-stellar value.

It has been believed that the sulfur seen in exoplanet atmospheres originates from solid accretion rather than from sulfur-rich nebular gas acquired during runaway gas accretion.
Classically, the dominant sulfur reservoirs in protoplanetary disks are suggested to be refractory compounds such as FeS (troilite) or S$\rm _n$, with volatile carriers contributing no more than $\sim$ 10 \% of the total sulfur budget \citep{2005Icar..175....1P, 2011A&A...536A..91J, 2019ApJ...885..114K, 2021ApJS..257...12L}. Hence, the SO$_2$ detected by JWST is generally interpreted as a signature of solid accretion \citep[e.g.,][]{2023ApJ...952L..18C}: accretion of planetesimals or pebbles containing refractory sulfur deposits sulfur to the atmosphere through dissolution \citep{2021ApJ...909...40T, 2022ApJ...937...36P}. 
Core erosion or dilution, inferred from Jupiter \citep{2018A&A...610L..14V, 2019ApJ...872..100D, 2022Icar..37814937H}, is another possible mechanism for mixing refractory elements into the envelope, though it still relies on a solid sulfur reservoir.

While solid accretion is believed as the main source of atmospheric sulfur, it remains unclear if it is a unique route to acquire abundant atmospheric sulfur during planet formation.
In particular, \citet{2024Natur.625...51D} and \citet{2024Natur.630..836W} reported abundant NH$_3$ in the atmosphere of WASP-107b in addition to SO$_2$, which poses a question on the conventional interpretation.
This is because nitrogen was expected to be largely depleted in disk solids due to the high volatility of N$_2$---likely the main nitrogen reservoir in protoplanetary disks \citep{2021PhR...893....1O}---in a wide orbital range \citep[e.g.,][]{2016ApJ...833..203P,2023ApJ...946...18O}.
Thus, except for the distant orbits, solid accretion acts to selectively enrich sulfur compared to nitrogen in the atmosphere, which is in tension with supersolar abundances of both sulfur and nitrogen inferred for WASP-107b \citep{2024Natur.630..836W}.

Here, we highlight the possible importance of semi-volatile species such as ammonium hydrosulfide (NH$_4$SH) as a source of atmospheric sulfur.
NH$_4$SH is one of several ammonium salts identified by the Rosetta/ROSINA in the coma of comet 67P.
The mass spectrometric analysis of cometary dust impacts revealed dissociation fragments of NH$_4$HCO$_2$, NH$_4$CN, and related compounds, with NH$_4$SH showing the strongest spectral signature \citep{2020NatAs...4..533A, 2022MNRAS.516.3900A}. Ammonium salts are now emerging as significant nitrogen carriers in disks, and they may represent tens of percent of the cosmic nitrogen budget \citep{2020Sci...367.7462P, 2020NatAs...4..533A}. In particular, NH$_4$SH alone could represent up to 20\% of the universal sulfur budget \citep{2025A&A...693A.146S}. Crucially, these salts are less volatile than NH$_3$ or H$_2$S but more volatile than FeS: NH$_4$SH sublimes at temperatures comparable to water ice \citep{2002Icar..155..393L}, and also other ammonium salts dissociate at 150--250 K \citep{2011A&A...535A..47D, 2016ApJ...829...85B, 2019ApJ...878L..20P}. Their enhanced thermal stability allows them to survive inward drift to planet-forming regions and potentially enrich the inner disk gas in sulfur and nitrogen, as similar to the water enrichment caused by pebble drift \citep{2017MNRAS.469.3994B,2021A&A...654A..71S,2025arXiv250616060O}.

In this study, we propose a novel hypothesis: ammonium salts provide an alternative pathway for close-in exoplanets to acquire sulfur-rich atmospheres.
Dust containing ammonium salts releases sulfur-containing vapors in addition to NH$_3$ when they cross the salt dissociation line (``salt line'' hereafter), subsequently enriching disk gases in sulfur and nitrogen.
To investigate this hypothesis, we model the transport of elements through the inward drift of dust containing ammonium salts and volatile ices, followed by vapor release through the sublimation and dissociation of these dust species. 
Furthermore, we calculate the atmospheric chemical compositions based on the computed disk compositions, thereby predicting the signatures of salt-related disk chemistry in transmission spectra of exoplanetary atmospheres.
Our study provides an end-to-end framework that bridges material transport within protoplanetary disks to observations of exoplanetary atmospheres.

This study is organized as follows. In Section \ref{sec:NSmodel}, we describe our model, which includes calculations for material transport in the disk, the chemical structures of planetary atmospheres formed there, and resulting transmission spectra. Section \ref{sec:NSresult} presents the results of our numerical calculations, while Section \ref{sec:NSdiscussion} discusses their implications. Section \ref{sec:NSconclusion} summarizes our findings.

\section{Model}
\label{sec:NSmodel}

\begin{figure*}[t]
\includegraphics[width=\hsize]{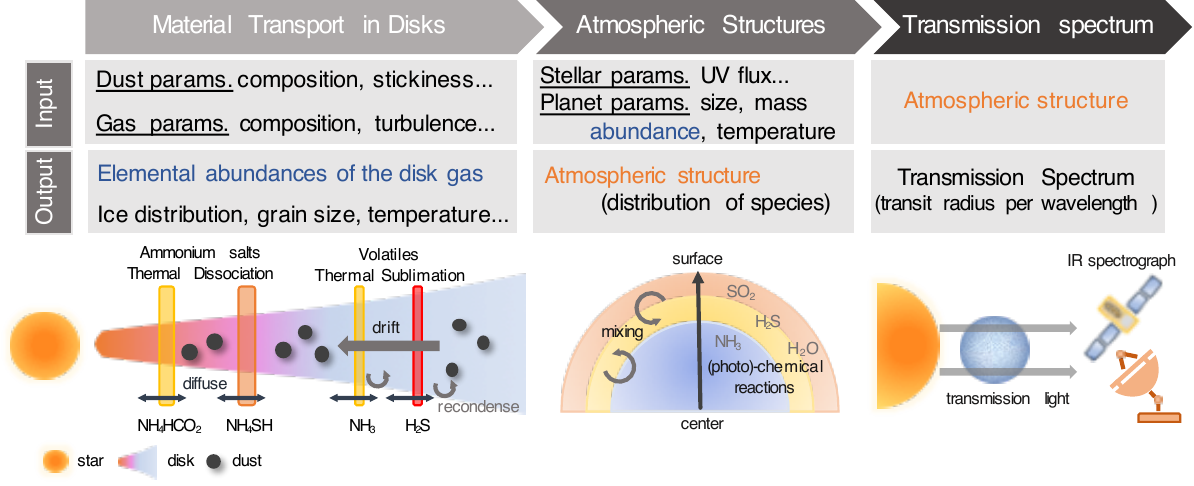}
\caption{The calculation flow in our model. We model the dynamics of gas and dust containing volatile ices and semi-volatile salts to calculate the C, O, N, and S abundances in the disk. Subsequently, we simulate the evolution of the planetary atmospheric structure that inherits the disk's composition and predict its transmission spectrum. The colored input parameters at the top of the figure correspond to the results obtained from the previous calculation step.}
\label{fig:calc_flow}
\end{figure*}

\subsection{model overview}
\label{{subsec:model_overview}}
In this section, we outline our model framework, which is illustrated in Figure \ref{fig:calc_flow}.
We first calculate disk compositions by taking into account material transport through gas diffusion and radial drift of dust that accommodate volatile ices and semi-volatile salts. 
We then conduct photochemical calculations based on the elemental abundances simulated by our disk model. 
The calculated atmospheric chemical structures are postprocessed to calculate the transmission spectrum. 
In what follows, we explain each modeling part in detail.

\subsection{Material transport in disks}
\label{sec:model_disk_calc}
\subsubsection{Overview}
The aim of our disk modeling is to explore the conditions under which nitrogen and sulfur are enriched in the disk gases by calculating material transport within the disk.
Our disk model simulates the evolution of gas surface density, dust surface density, grain size, and vapor surface densities released by the thermal sublimation of ices and dissociation of salts.  
Our disk model largely builds on the material transport model of \citet{2025PASJ..tmp...29N}, and thus we refer the readers to \citet{2025PASJ..tmp...29N} for details. 

\subsubsection{Gas Disk model}
\label{subsubsec:model_disk}
We adopt a 1D viscous disk model around a solar-mass star $M_{\odot}$ \citep{1974MNRAS.168..603L}. 
The disk gas surface density $\Sigma_{\rm g}(r,t)$ evolves through viscous diffusion, with gas viscosity parameterized by the $\alpha$ model \citep{1973A&A....24..337S}, which characterizes turbulent strength using the dimensionless parameter $\alpha$. In this study, we adopt a turbulent strength of $\alpha = 10^{-4}$. Such an $\alpha$ value is typical for protoplanetary disks, as inferred both from direct measurements using line-emission broadening and from estimates based on dust vertical settling \citep[e.g., ][]{2015ApJ...813...99F, 2017ApJ...843..150F, 2016ApJ...816...25P, 2021ApJ...912..164D, 2023MNRAS.524.3184P}.
For simplicity, we assume that $\alpha$ is constant in time and space.
The initial gas surface density is given by \citep{1974MNRAS.168..603L, 1998ApJ...495..385H}
\begin{equation}
    \Sigma_{\rm g, 0}(r) = \frac{M_{\rm disk}}{2\pi r_{\rm c}^2} \left(\frac{r}{r_{\rm c}}\right)^{-1} \exp{\left(-\frac{r}{r_{\rm c}}\right)}, 
\end{equation}
where $r$ is the distance from the star, $M_{\rm disk}$ is the initial total mass of the disk and $r_{{\rm c}}$ is the characteristic radius of the disk. We assume a compact disk with $M_{\rm disk} = 0.05 M_{\odot}$ and $r_{\rm c} = 50~\rm{au}$. 
We consider stellar irradiation for optically thick disks \citep[e.g.,][]{1970PThPh..44.1580K, 1997ApJ...490..368C}
and viscous accretion heating \citep[e.g.,][]{1994ApJ...421..640N, 2011ApJ...738..141O} to calculate the disk
mid-plane temperature $T(r,t)$.
For more explicit expressions, the reader is referred to Equations (7) and (14) of \citet{2021ApJ...916...72M}.
We adopt the time-dependent stellar luminosity prescribed by \citet{2023ApJ...949..119K}, an empirical fit to the stellar evolution model of \citet{2016A&A...593A..99F}.
For viscous accretion heating, we adopt a constant Rossland mean opacity $\kappa_{\rm R} = 4.5$ cm$^2$~g$^{-1}$ throughout the disk \citep{1985Icar...64..471P}. We assume that the background gas consists entirely of H$_2$.

\subsubsection{Dust evolution}
\label{subsubsec:model_dust}
The evolution of dust is calculated using the modified model of \citet{2025PASJ..tmp...29N}.
This model is based on the work of \citet{2016A&A...589A..15S} and deals with the evolution of the dust grain size according to the growth of particles that dominate the overall dust mass budget (single size approximation\footnote{In dust evolution simulations that employ the single-size approximation, we solve for the evolution of both the dust surface density and the dust mass (via the dust number surface density). The dust mass distribution at each radius is assumed to follow a narrow distribution peaked at a representative mass $m_{\rm d}$, and, assuming compact spheres, this representative mass is converted into a single characteristic grain size.}).
The dust evolves through collisional coalescence and fragmentation, inward drift due to the gas drag, and sublimation/dissociation/recondensation of volatiles and salts.

To investigate the evolution of disk elemental abundances, we introduce multi-component dust grains composed of rocks (including refractory minerals and ammonium salts) and volatile ices. The surface density of the dust is given by $\Sigma_{\rm d}(r,t) = \Sigma_{\rm d, rock}(r,t) + \Sigma_{\rm d, ice}(r,t)$, where $\Sigma_{\rm d, rock}$ and $\Sigma_{\rm d, ice}$ are the surface densities of the rock component and ice mantle, respectively, in a dust grain.
The evolution of $\Sigma_{\rm d, rock}$ is described by the following equation:
\begin{equation}
\label{Sigma_d_evolution}
     \frac{\partial \Sigma_{\rm d, rock}}{\partial t} +\nabla \cdot \mathcal{M_{\rm d}}= -\sum_i S_{\rm{salt}, i},
\end{equation}
where the dust mass flux, $\mathcal{M_{\rm d}}$, is given by
\begin{equation}
\label{Sigma_d_massflux}
   \mathcal{M_{\rm d}} = \Sigma_{\rm d}v_{\rm d} - D_{\rm d}\Sigma_{\rm g} \nabla \left(\frac{\Sigma_{\rm d}}{\Sigma _{\rm g}}\right),
\end{equation}
where $v_{\rm d}$ and $D_{\rm d}$ are the radial velocity and diffusion coefficient of the dust, respectively, and $S_{\rm{salt},i}$ represents the rate of decrease in the dust surface density due to ammonium salt dissociation (the subscript $i$ stands for the type of ammonium salt).
We apply the following equation to $S_{\rm{salt},i}$:
\begin{equation}
\label{S_salt}
    S_{\rm {salt}, i} = 
    \left\{
   \begin{array}{ll}
       {\displaystyle f_{i,\rm{dust}}}{\displaystyle\frac{\mathcal{M}_{\rm{d}}}{r\Delta r}}, & T \ge T_{\mathrm{salt},i}, \\
       0,        & \rm{otherwise},
   \end{array}
   \right.
\end{equation}
where $f_{i,\rm{dust}}$ is the mass fraction of salt $i$ in the rocky part of the grain, $\Delta r$ is the cell width in the simulation, and $T_{\mathrm{salt},i}$ is the dissociation temperature of salt $i$. Following \citet{2025PASJ..tmp...29N}, we assume that the salt is instantaneously and completely thermally dissociated from the rock and added to the vapor phase when the dust temperature exceeds the dissociation temperature of salt $i$.

We adopt an initial rock-to-gas ratio of 0.005 throughout the disk and set the initial ratio of rocks plus volatiles (ice and/or vapor) to gas to be 0.01 at each orbital radius. When all volatiles are in the ice phase, the resulting solid-to-gas ratio becomes 0.01, consistent with the standard ISM value \citep{1978ApJ...224..132B}. The initial spatial distribution of volatiles is described in Section \ref{subsubsec:model_volatiles}.

We simulate the number density of dust grains, $N_{\rm d}(r,t)$, using the model of \citet{2025PASJ..tmp...29N}. In this framework, $N_{\rm d}$ evolves through collisional growth and fragmentation. We characterize grain stickiness with fragmentation velocity $v_{\rm frag}$ and adopt a uniform value of  $v_{\rm frag} = 10$ m~s$^{-1}$ throughout the disk. The dust number density relates to grain size via $N_{\rm d} = \Sigma_{\rm d}/m_{\rm d}$, where $m_{\rm d} = 4/3 \pi a^{3}\rho_{\rm s}$ is the mass of a grain of a radius $a$ with the internal density of $\rho_{\rm s}$. For simplicity, we fix the internal density at $\rho_{\rm s} = 2.0$ g~cm$^{-3}$.

\subsubsection{Evolution of volatiles}
\label{subsubsec:model_volatiles}

\begin{figure*}[t]
    \begin{tabular}{cc}
    \hspace{-0.5cm}
      \begin{minipage}[t]{0.45\linewidth}
        \centering
        \includegraphics[width=\linewidth]{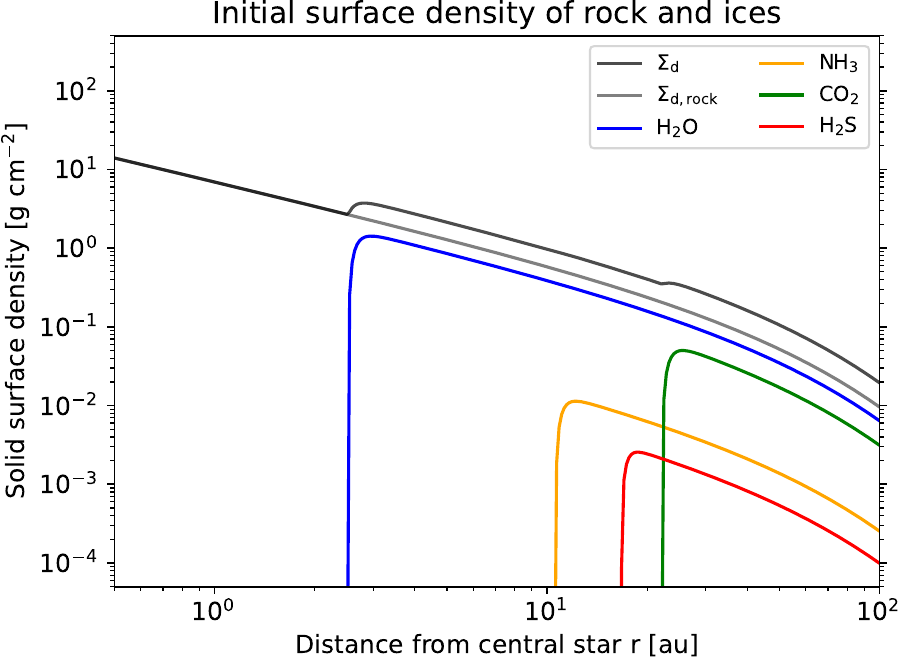}
      \end{minipage} &
        \hspace{0.5cm}
      \begin{minipage}[t]{0.45\linewidth}
        \centering
        \includegraphics[width=\linewidth]{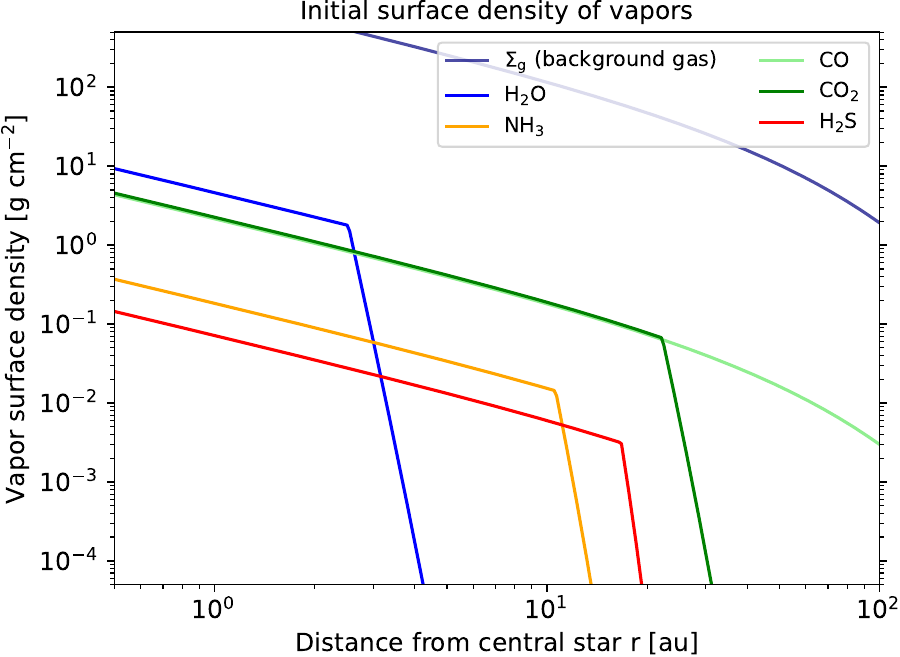}
      \end{minipage} \\
    \end{tabular}
    \caption{Initial dust surface density (left) and initial vapor surface density (right). Different colored lines show the surface density of different chemical species. In the left panel, the gray and black lines show the dust surface density of rocky component and the sum of rock and icy components, respectively.}
\label{fig:initial_ices_and_vapors}
\end{figure*}

We calculate the evolution of volatiles, including both ices and vapors produced by condensation/sublimation of ices and dissociation of salts. The surface density of volatile ices, $\Sigma_{\mathrm{ice,}i}(r,t)$, is calculated for each chemical species $i$, and $\Sigma_{\rm d, ice}$ is given by
\begin{equation}
    \Sigma_{\rm d, ice} = \sum_i \Sigma_{\mathrm{ice,}i}.
\end{equation}
The evolution of $\Sigma_{\rm ice,i}$ is described by the following equation, which has the same form as equation \eqref{Sigma_d_evolution}:
\begin{equation}
\label{Sigma_ice_evolution}
     \frac{\partial \Sigma_{{\rm ice},i}}{\partial t} +\nabla \cdot \mathcal{M}_{{\rm ice},i}= \dot{\Sigma}_{{\rm ice},i},
\end{equation}
where $\mathcal{M}_{{\rm ice},i}$ denotes the mass accretion rate of each ice component, obtained from equation \eqref{Sigma_d_massflux} by replacing the dust surface density with that of the corresponding ice species. The term $\dot{\Sigma}_{{\rm ice},i}$ represents the sublimation and recondensation rate for volatile ice, defined as \citep{2017A&A...602A..21S,2021A&A...646A..14H}
\begin{equation}
    \dot{\Sigma}_{{\rm ice},i} = R_{{\rm c},i} \Sigma_{{\rm ice},i} \Sigma_{{\rm vap},i} - R_{{\rm e},i} \Sigma_{{\rm ice},i},
\end{equation}
where $\Sigma_{\rm vap,i}$ is the vapor surface density of species $i$, and $R_{{\rm c},i}$ and $R_{{\rm e},i}$ are the condensation rate and sublimation rate, respectively, defined as
\begin{equation}
     R_{{\rm c},i} = 2 \sqrt{\frac{k_{\rm B}T}{\mu_{i}}}\frac{a^2}{m_{\rm d} H_{\rm g}(r,T)},
\end{equation}
\begin{equation}
     R_{{\rm e},i} = 2 \sqrt{2 \pi} \sqrt{\frac{\mu_{i}}{k_{\rm B}T}} P_{{\rm eq},i},
\end{equation}
where $k_{\rm B}$ is the Boltzmann constant, $\mu_{i}$ is the molecular weight of species $i$, $H_{\rm g}$ is the gas scale height, and $P_{{\rm eq},i}$ is the saturation vapor pressure of species $i$.
The saturation vapor pressure for each chemical species is determined experimentally and expressed as a polynomial of the form $\ln{P_{{\rm eq},i}} =A_{0,i}+\sum_{\ell=1}^{n_i} A_{\ell,i}/T_i^{\ell}$, where $\ell$ denotes an integer index and $A_{\ell, i}$ is the coefficient of the $\ell$-th order term for each species $i$. We adopt the polynomial coefficients from \citet{1991Icar...90..319L} for H$_2$O, \citet{2009P&SS...57.2053F} for H$_2$S, and \citet{2018A&A...614A...1W} for the remaining chemical species.
The surface density of the volatile vapor follows the similar equation of \eqref{Sigma_ice_evolution},
\begin{equation}
\label{Sigma_vap_evolution}
     \frac{\partial \Sigma_{{\rm vap},i}}{\partial t} +\nabla \cdot \mathcal{M_{\rm vap}}= \sum_i f_{i,j} S_{\rm{salt}, j} - \dot{\Sigma}_{{\rm ice},i},
\end{equation}
where $\mathcal{M_{\rm vap}}$ is
\begin{equation}
\label{Sigma_vap_massflux}
   \mathcal{M_{\rm vap}} = \Sigma_{{\rm vap},i}v_{\rm g} - D_{\rm g}\Sigma_{\rm g} \nabla \left(\frac{\Sigma_{{\rm vap},i }}{\Sigma _{\rm g}}\right).
\end{equation}
Equation \eqref{Sigma_vap_evolution} includes an additional source term for the dissociation of salts in the first term on the right-hand side. In this source term, $f_{i,j}$ represents the mass fraction of the component $i$ in salt $j$. For example, the dust releases NH$_3$ and H$_2$S vapors at the NH$_4$SH salt line where the 33 wt\% ($=\mu_{\rm NH_3}/\mu_{\rm NH4SH}=17/51$) of the dissociated NH$_4$SH salts end up in the NH$_3$ vapor (i.e., $f_{\rm NH_3,NH_4SH} = 0.33$), and the 67 wt\% ends up in the H$_2$S vapor (i.e., $f_{\rm H_2S,NH_4SH} = 0.67$).

Table \ref{tb:elements_carrier} summarizes the atomic fractions of each element carriers. We consider H$_2$O, CO, CO$_2$, NH$_3$, N$_2$, and H$_2$S as the volatile carriers of O, C, N, and S. Among these species, we assume CO and N$_2$ remain gaseous throughout the computational domain and are dynamically coupled to the background gas. We adopt the solar composition of \citet{2021AandA...653A.141A} as the reference elemental abundances.
We assume that silicates, FeS, and refractory carbon are entirely in the solid phase and do not contribute to the gas abundance.
We note that sublimation of the refractory material modifies the disk gas near the central star, as discussed in Section \ref{subsec:volatiles_enrichment}. However, this study focuses on the formation of gas giants at $r \gtrsim 1$ au, where these effects remain negligible.
The initial surface density distribution of the solids and vapors is shown in Figure \ref{fig:initial_ices_and_vapors}.
The initial vapor surface density at each orbit is given by
\begin{equation}
\label{eq:sigma_vap_init}
    \Sigma_{\mathrm{vap},0,i}(r) = \min\{\Sigma_{{\rm vap, sub}, i}(r), \Sigma_{{\rm vap, sat}, i}(r)\},
\end{equation}
where
$\Sigma_{{\rm vap, sub}, i} = x_i X_{\rm ref} \mu_i \Sigma_{\rm g}$ is the surface density when all of
species $i$ is in vapor form (here $x_i$ is the fraction of element X contained in species $i$ and
$X_{{\rm ref}}$ is the reference abundance of element X), and $\Sigma_{{\rm vap, sat}, i} = \sqrt{2\pi}H_{\rm g}P_{{\rm eq}, i}\mu_i/(k_{\rm B} T)$ is the surface density of that species under the saturated vapor pressure.
The initial surface density of volatile ices is given by
\begin{equation}
\label{eq:sigma_ice_init}
    \Sigma_{{\rm ice}, 0, i}(r) = \Sigma_{{\rm vap, sub}, i} - \Sigma_{\mathrm{vap},0,i}
\end{equation}
Using Equations \eqref{eq:sigma_vap_init} and \eqref{eq:sigma_ice_init}, a given carrier resides almost entirely in the vapor phase at radii sufficiently warmer than its sublimation temperature and in the ice phase at colder radii, with a smooth transition in surface density imposed by $\Sigma_{{\rm vap, sat}, i}$ in the vicinity of its snow line.

\begin{table*}[t]
    \centering
    \caption{Atomic Fractions of each Element Carriers Adopted in the Simulations}
    \footnotesize
    \begin{minipage}{\textwidth}
    \hspace*{-1.3cm}%
    \begin{tabular}{c|c|cc|c}
     Element  & Reference abundance$^{\rm a}$ & Species & Fraction & Note \\ \hline \hline
     \multirow{4}{*}{Nitrogen}
              & \multirow{4}{*}{$7.78\times10^{-5}$}
              & NH$_3$  & 0.1  &  \\
              &         & NH$_4$SH     & 0.04 & $T_{\rm salt, NH_4SH} = 150$ K \\
              &         & NH$_4$HCO$_2$& 0.16 & $T_{\rm salt, NH_4HCO_2} = 200$ K, Anion of this salt does not contribute to the disk gas composition \\
              &         & N$_2$        & 0.7  &  \\ \hline
     \multirow{3}{*}{Sulfur}
              & \multirow{3}{*}{$1.52\times10^{-5}$}
              & H$_2$S  & 0.1  &   \\ 
              &         & NH$_4$SH & 0.2  &   \\
              &         & FeS  & 0.7  & FeS does not contribute to the disk gas's sulfur abundance\\ \hline
     \multirow{4}{*}{Oxygen}
              & \multirow{4}{*}{$5.62\times10^{-4}$}
              & H$_2$O & 0.33 &   \\  
              &        & CO   & 0.1  & Similar to cometary ice composition, $0.3\times$~H$_2$O \citep{2011ARAandA..49..471M}  \\
              &        & CO$_2$ & 0.07 & Similar to cometary ice composition, $0.2\times$~H$_2$O \citep{2011ARAandA..49..471M}  \\
              &        & Others & 0.5 & These oxygen carriers do not contribute to the disk gas's oxygen abundance\\ \hline
     \multirow{3}{*}{Carbon}
              & \multirow{3}{*}{$3.31\times10^{-4}$}
              & CO   & 0.17 & 0.1 $\times$ O/C (O/C = 1.7, the protosolar value from \citet{2021AandA...653A.141A})\\
              &       & CO$_2$ & 0.12 & 0.07 $\times$ O/C (O/C = 1.7, the protosolar value from \citet{2021AandA...653A.141A})\\
              &       & Others & 0.71 & These carbon carriers do not contribute to the disk gas's carbon abundance \\
    \end{tabular}
    \vspace{0.1cm}
    
    \hspace*{-1.3cm}%
    
    {\footnotesize $^{\rm a}$ The solar composition from \citet{2021AandA...653A.141A}.}
    \end{minipage}
    \label{tb:elements_carrier}
\end{table*}

We consider NH$_4$SH and NH$_4$HCO$_2$ as representative ammonium salts contained in dust particles.
We assume that NH$_4$SH, which is suggested to be the most abundant salt in comets \citep{2022MNRAS.516.3900A}, accounts for 20\% of the sulfur budget, and that the sum of NH$_4$SH and NH$_4$HCO$_2$ account for 20\% of the nitrogen budget. 
If ammonium salts act as carriers that compensate for the nitrogen depletion of comets relative to the solar composition, then 10--30\% of the nitrogen budget could be stored in ammonium salts \citep{2020Sci...367.7462P, 2021PhR...893....1O}, and our assumption that 20\% of the nitrogen resides in salts is chosen within this range. The fraction of this salt reservoir that is in the form of NH$_4$SH is even more uncertain. \citet{2025A&A...693A.146S} experimentally measured spectra of H$_2$O--NH$_3$--H$_2$S mixtures under various conditions and compared them with observations of protostars and cold dense clouds. They suggested that NH$_4$SH could account for 10--20\% of the sulfur budget, provided that most of the salts responsible for the 6.85~$\mu$m NH$_4^+$ feature are in the form of NH$_4$SH. Our assumption for the NH$_4$SH budget corresponds to this upper limit.
The dissociation temperatures of NH$_4$SH and NH$_4$HCO$_2$ are assumed to be 150 K and 200 K \citep{2002Icar..155..393L, 2016ApJ...829...85B}, respectively.
We note that ammonium ion can bind to various anions --such as CN$^-$, F$^-$, and Cl$^{-}$ \citep{2020Sci...367.7462P, 2020NatAs...4..533A, 2022MNRAS.516.3900A}, and NH$_4$HCO$_2$ serves only as a proxy for this broader group. Therefore, we only consider the emission of NH$_3$ vapor in the salt line at $T =200$ K and ignore any contribution of the accompanying anions to the elemental budget of the disk gas.

\subsection{Calculations of atmospheric structure and transmission spectra}
\label{subsec:model_atmos}
Disk gas accreted onto a planet undergoes chemical evolution in the atmosphere, including photochemistry, which dictates the observable spectra of exoplanetary atmospheres. 
In this study, we calculate the atmospheric chemical structure of a planet that inherits the elemental abundances of the disk gas obtained from our disk model (Section \ref{sec:model_disk_calc}). We then calculate the transmission spectrum and assess the impact of ammonium salts on atmospheric observations.

\subsubsection{Atmospheric Structure}
\label{subsubsec:model_as}

We calculate the atmospheric structure using the open-source photochemical kinetics code VULCAN, which incorporates a H-C-N-O-S chemical network, including photochemical reactions \citep[][SNCHO\_full\_photo\_network implemented in VULCAN]{2017ApJS..228...20T, 2021ApJ...923..264T}. 
To obtain the atmospheric structure in the 1D vertical direction, VULCAN solves the following continuity equation: 
\begin{equation}
\label{eq:VULCAN-cont}
    \frac{\partial n_i(z,t)}{\partial t} = \mathcal{P}_i - \mathcal{L}_i - \frac{\partial \phi _i}{\partial z},
\end{equation}
where $n_i$ is the number density of chemical species $i$, $z$ is a coordinate in the vertical direction of the planet, $\mathcal{P}_i$ and $\mathcal{L}_i$ are the production and loss rates of species $i$, and $\phi_i$ represents the transport flux of $i$ induced by eddy diffusion and molecular and thermal diffusion. 
For the eddy diffusion coefficient $K_{\rm zz}$ , we adopt the formula assumed in \citet{2022ExA....53..279M}:
\begin{equation}
\label{eq:Kzz}
    K_{\rm zz}(z) = 5 \times 10^8 [P({\rm bar})]^{-0.5}\left(\frac{H_{\rm 1mbar}}{620~{\rm km}}\right) \left(\frac{T_{\rm eff}}{1450~{\rm K}}\right),
\end{equation}
where $K_{\rm zz}$ is in units of cm$^2$~s$^{-1}$, $P$ is atmospheric pressure (in bar), $H_{\rm 1mbar}$ is atmospheric pressure scale height at 1mbar (in km), and $T_{\rm eff} = T_{\rm star}\times (R_{star}/r)^{1/2}$ is the effective temperature of the planet. Here, $T_{\rm star}$ and $R_{\rm star}$ are the effective temperature and radius of the star, respectively.
We employ the stellar spectrum of HD 85512, a K-type main sequence star, compiled by the MUSCLES Treasury survey \citep[][version 2.2]{2016ApJ...820...89F, 2016ApJ...824..101Y, 2016ApJ...824..102L}.

To solve equations \eqref{eq:VULCAN-cont}--\eqref{eq:Kzz}, we need to determine the planetary gravity, the atmospheric Pressure--Temperature (P--T) profile, and the initial elemental abundances. Here, we assume the formation of a warm, Saturn-sized gas giant. 
For our fiducial case, we adopt a P--T profile of a gas giant corresponding to an equilibrium temperature $T_{\rm eq}=800$ K, an intrinsic temperature $T_{\rm in}=100 $ K, and a surface gravity of 10 m s$^{-2}$, as calculated by \citet{2023ApJ...946...18O} for 10$\times$ solar metallicity.
We adopt this high-metallicity atmospheric P--T profile because our framework envisions the planet accreting metal-rich disk gas produced by salt dissociation and ice sublimation.
To investigate the effect of $T_{\rm eq}$ on the atmospheric chemical structure, we also perform photochemical simulations for $T_{\rm eq} = 400$ K and 1200 K in Section \ref{subsec:AC_temp}.

We set atmospheric elemental abundances by extracting those of disk gas assuming that the planetary atmosphere inherits elemental abundances of disk gas. 
To investigate the impact of nitrogen and sulfur enrichment by ammonium salts on the atmospheric structure, we examine two cases: with and without ammonium salts in the disk. Because close-in gas giants typically show sub-stellar C/O ratios \citep{2024RvMG...90..411K, 2025arXiv250601800W}, our fiducial model assumes gas accretion inside the water snow line, with accretion at $r=1.5$ au and $t=0.5$ Myr, where water-vapor enrichment leads to a sub-stellar C/O ratio in the gas. 
As demonstrated later, the dissociation of ammonium salts greatly affects the chemical structure of planets formed near the water snowline, since the salt lines lie near the snowline.
We explore the atmospheric chemical structure of planets formed outside the water snow line in Appendix \ref{append:C/O}. 
Note that the planetary equilibrium temperature at the current orbit (400, 800, 1200 K) does not necessarily match the past disk temperature, as the planet can migrate during or after the disk dissipation. 

\begin{figure}[t]
\includegraphics[width=\hsize]{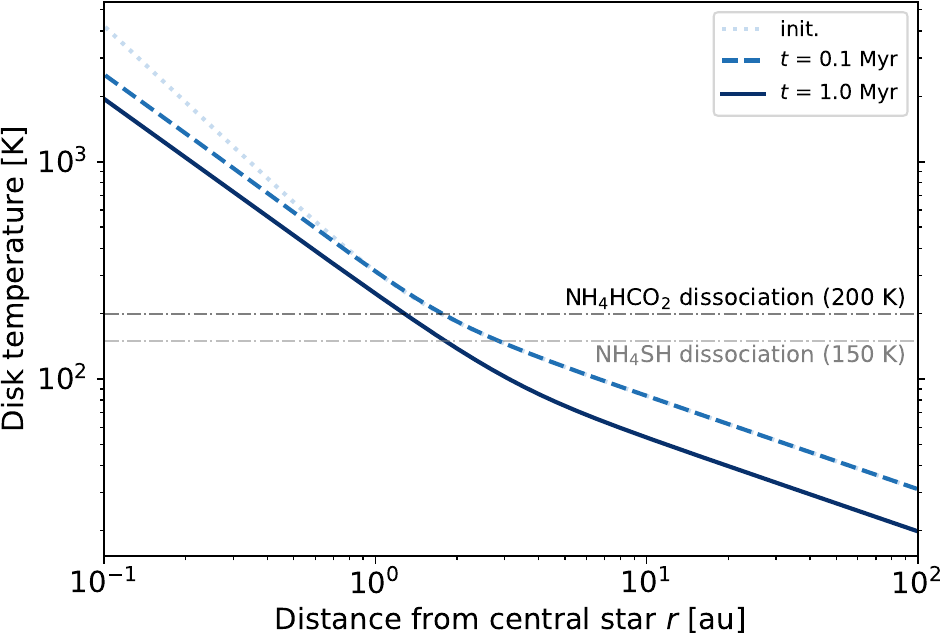}
\caption{Time evolution of the disk temperature $T$ from the fiducial model as a function of the distance from the central star $r$. The blue dotted, dashed, and solid lines are the snapshots at times $t = 0, 0.1$ and 1.0 Myr, respectively.}
\label{fig:time_evo_T}
\end{figure}
\begin{figure*}[t]
    \begin{tabular}{cc}
    \hspace{-0.5cm}
      \begin{minipage}[t]{0.45\linewidth}
        \centering
        \includegraphics[width=\linewidth]{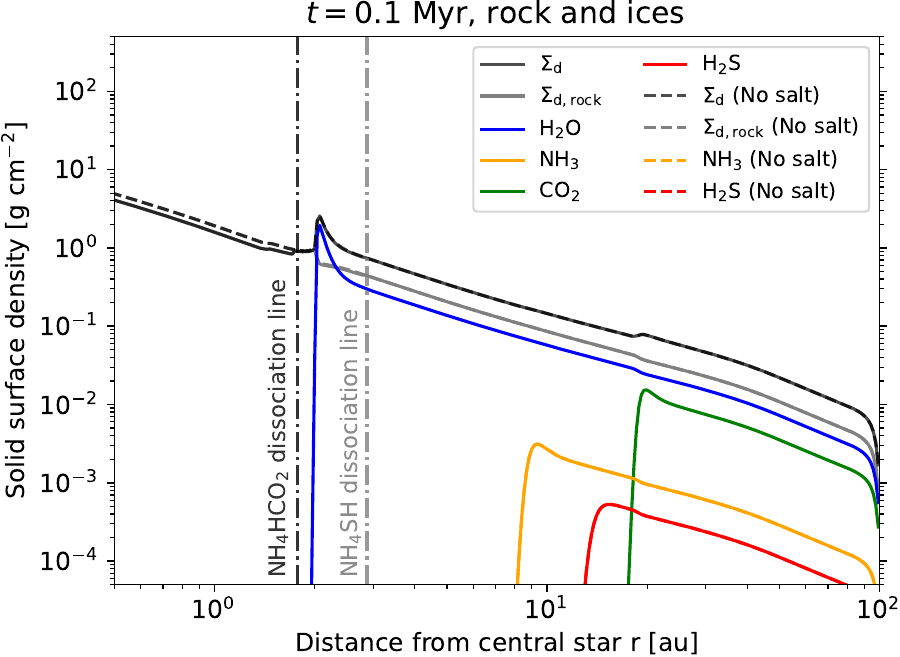}
      \end{minipage} &
        \hspace{0.5cm}
      \begin{minipage}[t]{0.45\linewidth}
        \centering
        \includegraphics[width=\linewidth]{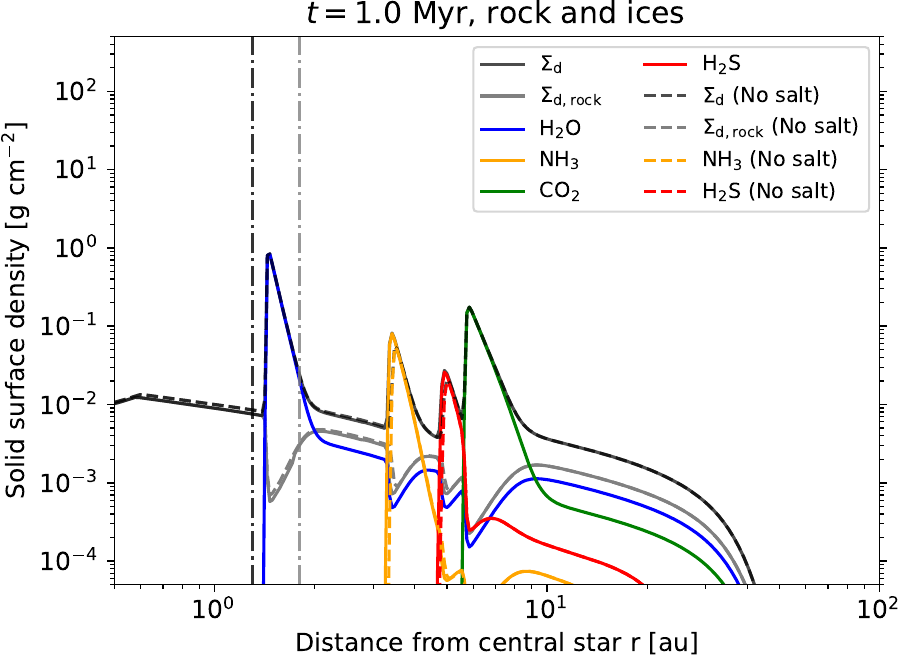}
      \end{minipage} \\
    \hspace{-0.5cm}
      \begin{minipage}[t]{0.45\linewidth}
        \centering
        \includegraphics[width=\linewidth]{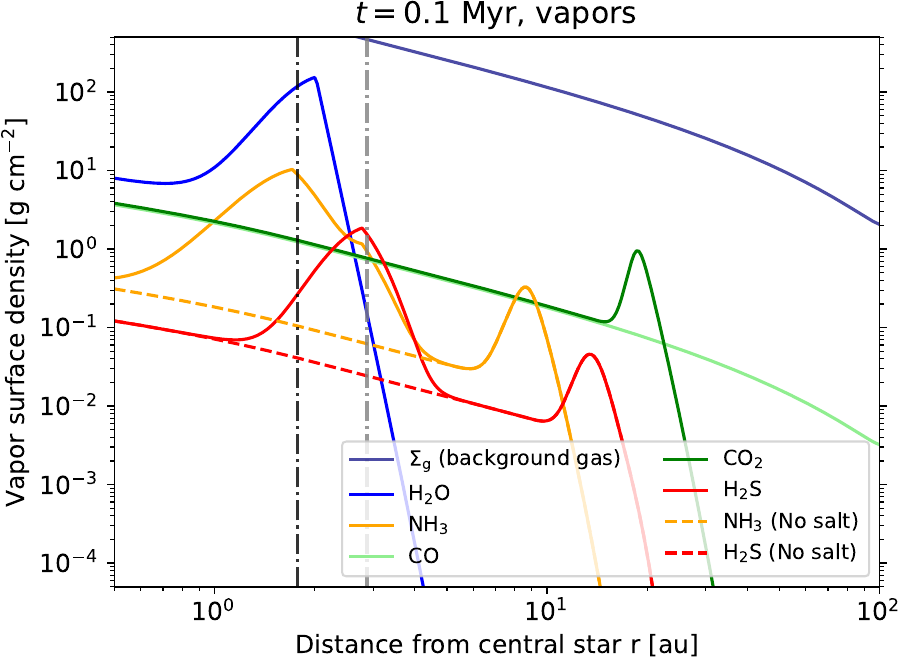}
      \end{minipage} &
        \hspace{0.5cm}
      \begin{minipage}[t]{0.45\linewidth}
        \centering
        \includegraphics[width=\linewidth]{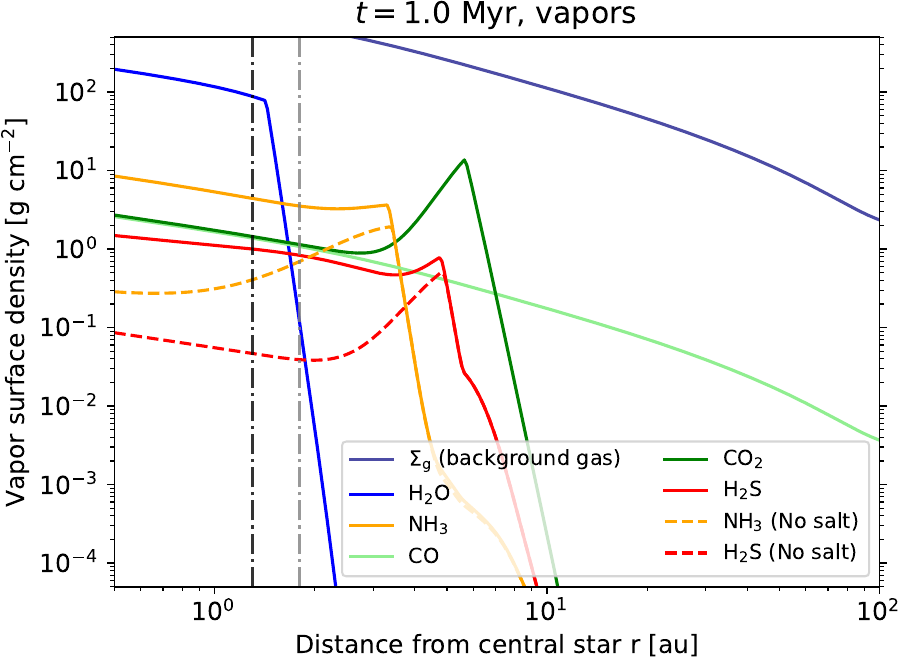}
      \end{minipage} \\
    \end{tabular}
    \caption{Evolution of dust surface density (upper) and vapor surface density (lower). The left and right panels are the snapshots at times $t = 0.1$ and 1.0 Myr. Solid and dashed lines correspond to models with and without salt, respectively. Vertical dash-dotted lines in the Figure represent the salt dissociation lines.}
\label{fig:evo_ices_and_vapors}
\end{figure*}

\begin{figure}[t]
\includegraphics[width=\hsize]{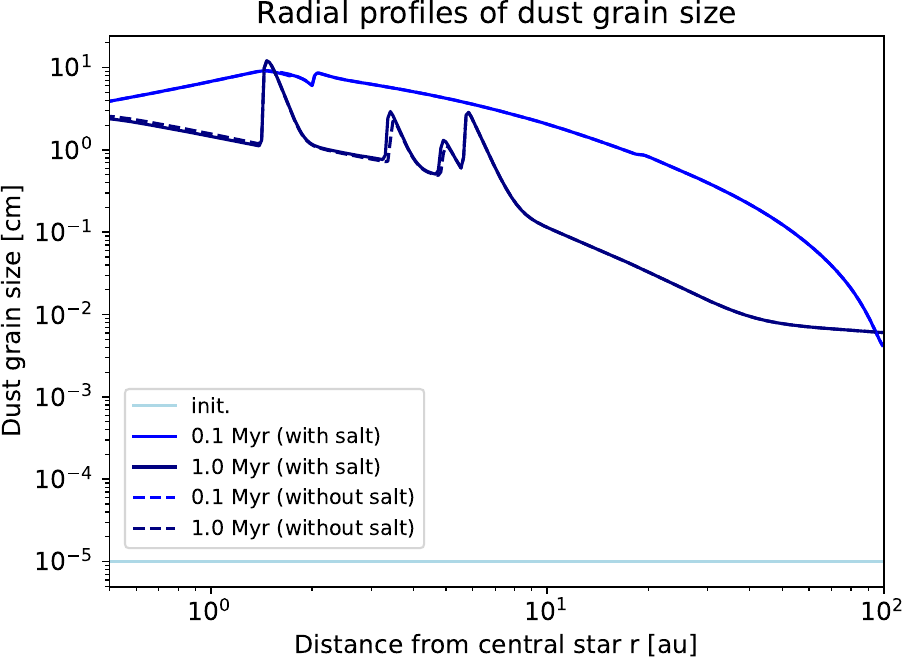}
\caption{Time evolution of the grain size $a$ as a function of the distance from the central star $r$. The light blue, blue and navy lines are the snapshots at times $t = 0, 0.1$ and 1.0 Myr, respectively. The solid and
dashed lines correspond to models with and without salts, respectively.}
\label{fig:time_evo_a}
\end{figure}

\begin{figure*}[t]
    \begin{tabular}{cc}
    \hspace{-0.5cm}
      \begin{minipage}[t]{0.45\linewidth}
        \centering
        \includegraphics[width=\linewidth]{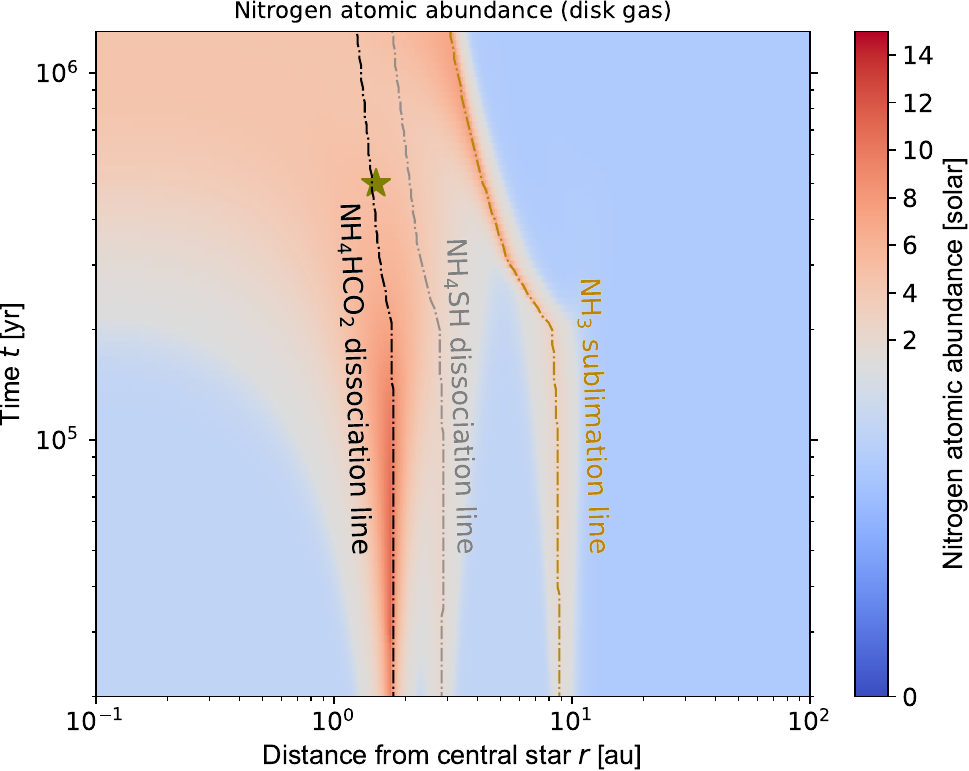}
      \end{minipage} &
        \hspace{0.5cm}
      \begin{minipage}[t]{0.45\linewidth}
        \centering
        \includegraphics[width=\linewidth]{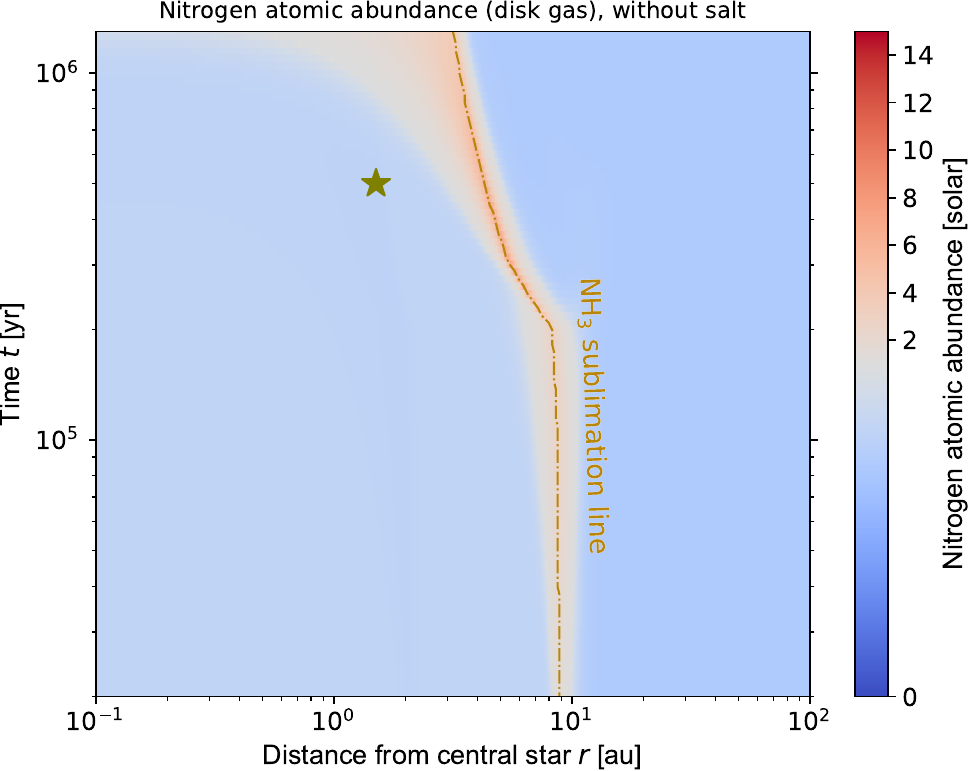}
      \end{minipage} \\
    \hspace{-0.5cm}
      \begin{minipage}[t]{0.45\linewidth}
        \centering
        \includegraphics[width=\linewidth]{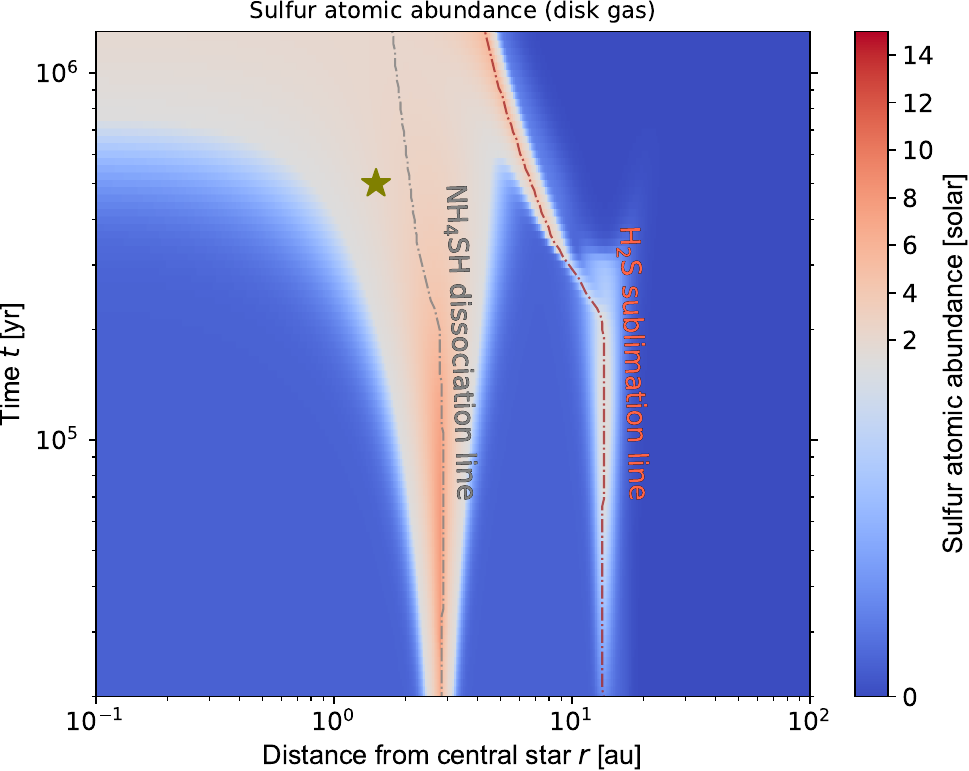}
      \end{minipage} &
        \hspace{0.5cm}
      \begin{minipage}[t]{0.45\linewidth}
        \centering
        \includegraphics[width=\linewidth]{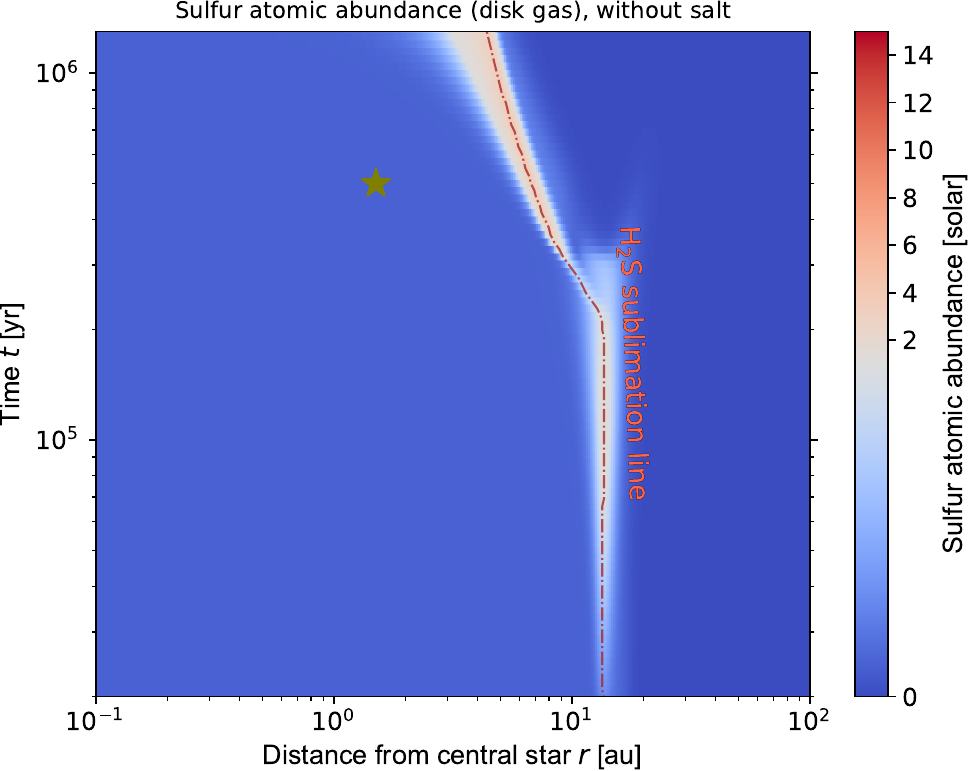}
      \end{minipage} \\
    \end{tabular}
    \caption{Space-time diagrams of the nitrogen atomic abundance (upper) and sulfur atomic abundance (lower) in the disk gas. The left and right panels correspond to models with and without salts, respectively. Dash-dotted lines in the figure represent salt dissociation lines and sublimation lines for NH$_3$ and H$_2$S. The abundance of each element is normalized to the solar value \citep{2021AandA...653A.141A}. The color scale is centered at unity (solar value), with blue indicating sub-solar abundances and red indicating super-solar abundances. The star symbol marks the planetary formation orbit and timing adopted in the fiducial model.}
\label{fig:space-time_abundance}
\end{figure*}
\subsubsection{Transmission Spectrum}
\label{subsubsec:model_ts}
We calculate the transmission spectra of the planets formed in disks with and without ammonium salts using the open-source radiative transfer code petitRADTRANS \citep{2019A&A...627A..67M}.
In this calculation, petitRADTRANS assumes a spherically symmetric atmosphere and calculates the transmission spectrum based on the 1-D vertical radius-temperature-composition profile, which we obtain from VULCAN, as outlined in Section \ref{subsubsec:model_as}.
The $P$-$T$ structure used in VULCAN is converted to an $r_{\rm p}$-$T$ profile under the assumption of hydrostatic equilibrium. The computational grid consists of 100 logarithmically spaced divisions for $P = 10^{-10}$ -- $10^{2}$ bar. Due to different grid structures between VULCAN and petitRADTRANS, the radius, temperature, and composition on each grid are interpolated linearly using the SciPy Python package \citep{2020SciPy-NMeth}. We consider H$_2$O, CO, CH$_4$, CO$_2$, H$_2$S, SO$_2$, COS, NS, NH$_3$, NO, and HCN as the molecular species that contribute to the transmission spectrum. The line list of each molecule are sourced from the ExoMol database \citep{2016JMoSp.327...73T}.
We assume a cloud-free atmosphere throughout this paper for simplicity.

\section{Result}
\label{sec:NSresult}

\subsection{Disk evolution}
\label{subsubsec:disk_evo}

First, we trace the evolution of the disk's temperature profile and the surface densities of individual species, identifying the orbital radii where enrichment occurs and quantifying how strongly each species concentrates at those orbits with versus without ammonium salts.
Figure \ref{fig:time_evo_T} shows the radial distribution of the disk mid-plane temperature at different times. Viscous heating predominates within 3 au, while irradiation heating becomes the dominant mechanism at greater distances. The disk cools over time as the stellar luminosity diminishes.
The salt lines of NH$_4$HCO$_2$ and NH$_4$SH are initially located at 2 au and 3 au, respectively.

The dissociation of salts greatly enhances the vapor surface density of NH$_3$ and H$_2$S at the inner orbits.
Figure \ref{fig:evo_ices_and_vapors} shows the evolution of solids and vapors of each species.
At 0.1 Myr, the sublimation and recondensation cycle generates peaks in both the solid and vapor surface densities of volatiles near their respective snow lines, with the exception of CO. Additionally, the dissociation of salts generates peaks in the vapor surface densities of NH$_3$ and H$_2$S at 2 au and 3 au, respectively.
At 1.0 Myr, most solid particles have drifted toward the central star. However, they are partially retained near the snow lines of each chemical species by repeated cycles of sublimation and recondensation. At the snow line of a given volatile species, the ice of that species becomes locally enhanced, whereas the surface densities of other solid components present on the same orbits exhibit a dip. This behavior arises because ice accumulation near the snow line promotes dust growth, increasing the Stokes number and hence the radial drift velocity, so that dust is efficiently removed from that region (Figure \ref{fig:time_evo_a}). As a result, only the volatile whose snow line lies at that location is efficiently retained, while other solids are transported further inward. Similar dip-like structures can also be seen in disk-composition evolution models that include dust transport \citep{2017MNRAS.469.3994B, 2019MNRAS.487.3998B}.
Moreover, because salts constitute only a small mass fraction of the rocks, the decrease in the solid surface density due to salt dissociation is minor.
Furthermore, the peaks in vapor surface density have been smoothed out by diffusion, leading to a uniform increase in vapor surface densities inside the snow lines.
The salt dissociation enhances the surface density of NH$_3$ and H$_2$S vapors by a factor of $\sim10$--$30$ inside the salt lines at $r\lesssim3~{\rm au}$ compared to the model without salts (dashed lines in Figure \ref{fig:evo_ices_and_vapors}). 

CO remains in the gas phase throughout the disk. The CO vapor surface density decreases due to diffusion within $r = r_c=50$ au and increases beyond 50 au, mirroring the evolution of the background gas surface density, although the overall change is not significant.

\subsection{Nitrogen and Sulfur abundances}
\label{subsubsec:disk_NS}

Here we analyze in detail how the salt dissociation impacts the evolution of nitrogen and sulfur abundances in disk gas.
Figure \ref{fig:space-time_abundance} presents a space-time diagram of nitrogen and sulfur atomic abundances in the disk gas. For nitrogen, NH$_3$ vapor accumulates at the salt lines of NH$_4$HCO$_2$ and NH$_4$SH, as well as at the NH$_3$ snow line, resulting in super-solar nitrogen abundances. The highest nitrogen abundance, $~\sim$ 10 times the solar value, occurs at the NH$_4$HCO$_2$ salt line. This NH$_3$ vapor then diffuses inward and outward across the disk over a timescale of $\sim 1$ Myr, uniformly increasing the nitrogen abundance inside the NH$_3$ snow line to $\sim$ 4 times the solar value. 
In the absence of ammonium salts, nitrogen enrichment in the disk gas remains confined to the orbit right above the NH$_3$ snow line during the early stage ($t \sim 0.1$ Myr) and to the region inside the NH$_3$ snow line at later stage. The maximum nitrogen abundance reaches only about half the level achieved by the model with salt dissociation.

The salt dissociation also greatly enriches the disk gas in sulfur.
The sulfur abundance peaks at the NH$_4$SH salt line at $\sim$ 0.1 Myr, reaching $\sim$ 6 times the solar value.
The H$_2$S vapor produced by NH$_4$SH dissociation subsequently diffuses to achieve $\sim 3\times$ solar sulfur abundance at $\lesssim3~{\rm au}$.
Without NH$_4$SH, sulfur enrichment in the disk gas remains confined to a narrow zone around the H$_2$S snow line throughout the first 1 Myr.

It should be stressed that salt-driven sulfur enrichment in disk gas operates in a broad orbital range of $\lesssim3~{\rm au}$.
This result contrasts with previous studies that predicted sulfur-rich disk gases only near the central star of $\lesssim0.3~{\rm au}$, inside the sublimation line of FeS \citep{2021A&A...654A..72S,2025arXiv250616060O}.
Our results indicate that the dissociation of ammonium salts generates sulfur-rich disk gases in a much broader orbital range than previously thought, even reaching the orbit near the water snowline.

Another notable feature of material transport by salts is that the sulfur abundance in the disk gas increases simultaneously with the nitrogen abundance near the water snow line.
Along the NH$_4$SH salt line, nitrogen and sulfur abundances exceed twice the solar value from $t \lesssim 0.1$ Myr.
The diffusion of NH$_3$ and H$_2$S vapors subsequently enhances the nitrogen and sulfur abundances in the inner disk, causing both to exceed twice the solar value inside the NH$_3$ snow line.

Ammonium salts dissociate into NH$_3$ and H$_2$S, which are more volatile than the salts themselves\footnote{We note that the present model ignores the reverse reaction of NH$_3$+H$_2$S$\longrightarrow$NH$_4$SH that forms ammonium salts---suggested to operate in the Jovian atmosphere for forming NH$_4$SH clouds \citep{1969Icar...10..365L}. This process may hinder the outward diffusion, but it unlikely affects NH$_3$ and H$_2$S vapor transport inside the salt lines.}; the resulting vapors diffuse inward and outward and create nitrogen- and sulfur-rich zones that span the water snow line. 
Meanwhile, water sublimation strongly modifies the C/O ratio, which has traditionally been a focus in the context of disk gas composition. In the following sections, our model for planetary atmospheric structure adopts the disk-gas composition inside of the water snow line, where the C/O ratio is sub-solar. In the fiducial model, where the planet forms at $t=0.5$ Myr and $r=1.5$ au, the disk gas composition inherited by the planet is (O/H, C/H, N/H, S/H)=(15, 0.27, 5.0, 2.0) $\times$ solar values and (O/H, C/H, N/H, S/H)=(15, 0.27, 0.8, 0.1) $\times$ solar values. in the with- and without-salts cases, respectively. We analyzes how changes in the C/O ratio influence planetary composition in appendix \ref{append:C/O}.

\subsection{Atmospheric Structure}
\label{subsubsec:atomos}

\begin{figure}[t]
\includegraphics[width=\hsize]{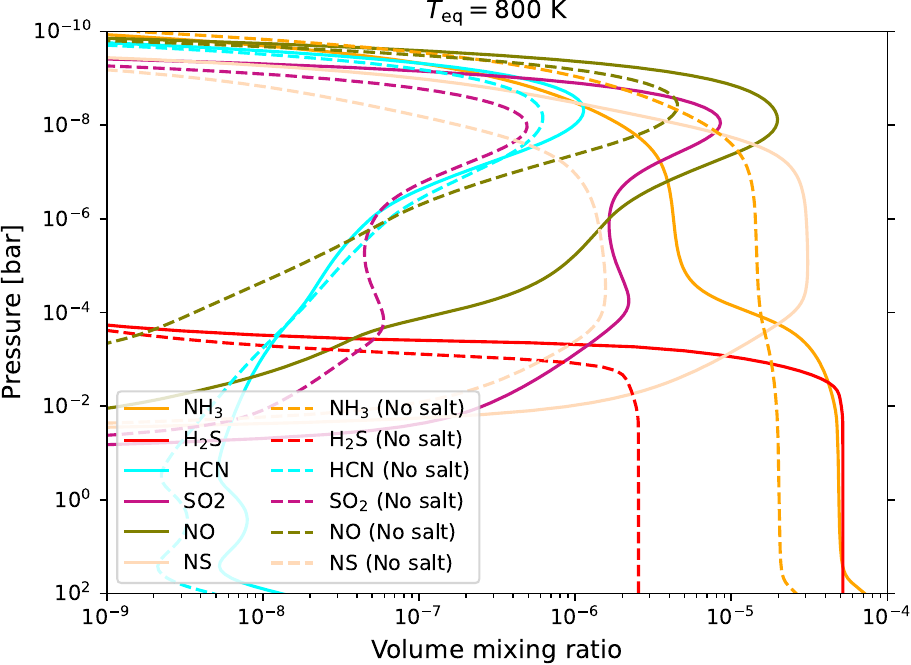}
\caption{Atmospheric structures calculated by VULCAN using the O, C, N, and S atomic abundances in the disk gas at $t=0.5$ Myr, $r = 1.5$ au. The planet's equilibrium temperature is 800 K. Solid lines depict atmospheric structures for the planet that formed in the salt-bearing disk with elemental abundances (O/H, C/H, N/H, S/H)=(15, 0.27, 5.0, 2.0) $\times$ solar values, whereas dashed lines correspond to planets formed in the disk without salts with (O/H, C/H, N/H, S/H)=(15, 0.27, 0.8, 0.1) $\times$ solar values.}
\label{fig:AS_fid}
\end{figure}

Salt dissociation enhances the abundance of nitrogen and sulfur of the disk gas near the salt lines, thereby modifying the formation rates of N- and S-bearing molecules in the atmospheres of planets that acquire the disk gas there.
Figure \ref{fig:AS_fid} presents the volume mixing ratios of N- and S-bearing molecules calculated by VULCAN for the atmosphere of the planet with $T_{\rm eq}=800~{\rm K}$ using the elemental abundances derived from the disk gas composition with and without salt dissociation shown in Section \ref{subsubsec:disk_evo}.
Accretion of gas enriched by salt dissociation raises the atmospheric volume mixing ratios of most nitrogen- and sulfur-bearing species by an order of magnitude or more across a broad pressure range ($10^{-8}$--$10^{-2}$ bar) relative to gas accreted without salt dissociation. For sulfur compounds, H$_2$S dominates at pressures greater than $10^{-4}$ bar, NS dominates between $10^{-4}$ and $10^{-8}$ bar, and SO$_2$ dominates at still lower pressures.
The mixing ratios of all three species increase by at least an order of magnitude by salts; for example, the salt model yields a maximum volume mixing ratio of $\sim 10^{-5}$ for SO$_2$ at $P \sim 10^{-8}$ bar, whereas the model without salts yields the maximum mixing ratio of only $\sim 5\times10^{-7}$ for SO$_2$.

Salt-driven nitrogen enrichment generally raises the mixing ratios of N-bearing molecules such as HCN and NO; however, NH$_3$ behaves differently. Salt dissociation enhances NH$_3$ mixing ratio at $P > 10^{-4}$ bar but reduces it at $P < 10^{-4}$ bar. This decline occurs because simultaneous sulfur enrichment consumes nitrogen radicals into NS production. Similarly to SO$_2$ and H$_2$S, the volume mixing ratio of NS increases by more than an order of magnitude when salts are present. In the uppermost atmosphere, nitrogen radicals are also consumed in NO formation, which suppresses NH$_3$ production. 

In this section, we examined planets that form inside the water snow line. H$_2$O sublimation enriches the disk gas in oxygen, lowering its C/O ratio to 0.02. This excess of oxygen  abundance suppresses the synthesis of carbon-bearing sulfur and nitrogen species such as CS, favoring oxygen-bearing molecules such as SO$_2$ instead. Appendix \ref{append:C/O} investigates the cases with $\rm{C/O} \sim 1$, corresponding to the planets formed outside the water snow line.

\subsection{Transmission Spectra}
\label{subsubsec:spectra}

\begin{figure*}[t]
      \hspace{0.925cm}
      \begin{minipage}[c]{0.85\linewidth}
        \centering
        \includegraphics[width=0.75\linewidth]{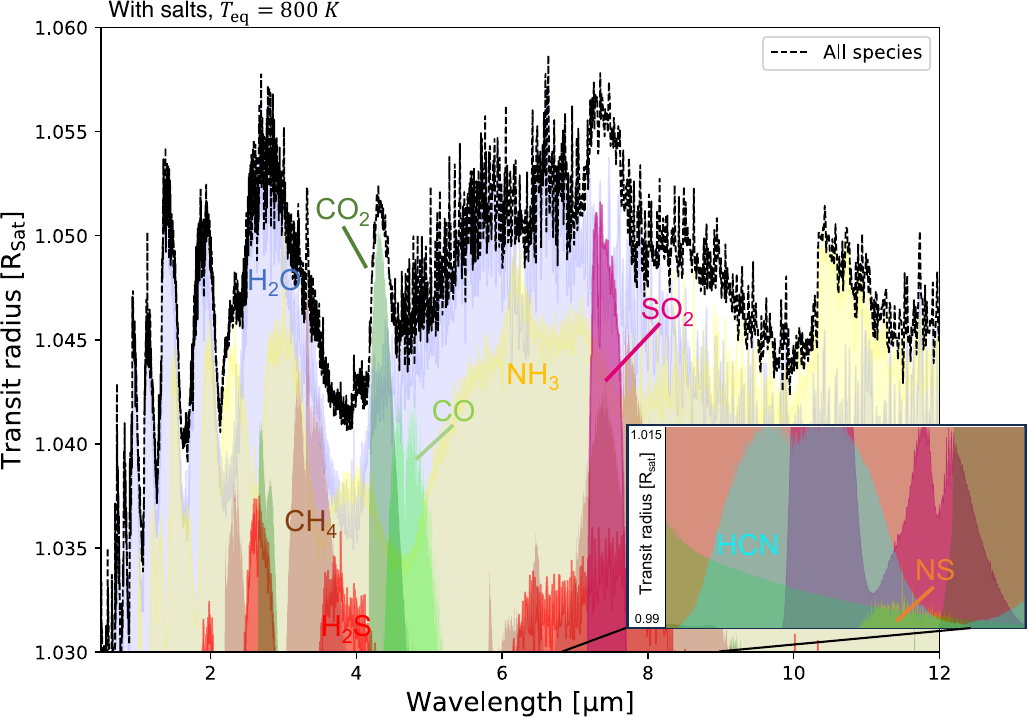}
      \end{minipage} \\
      \hspace{0.5cm}
      \begin{minipage}[c]{0.85\linewidth}
        \centering
        \includegraphics[width=0.7\linewidth]{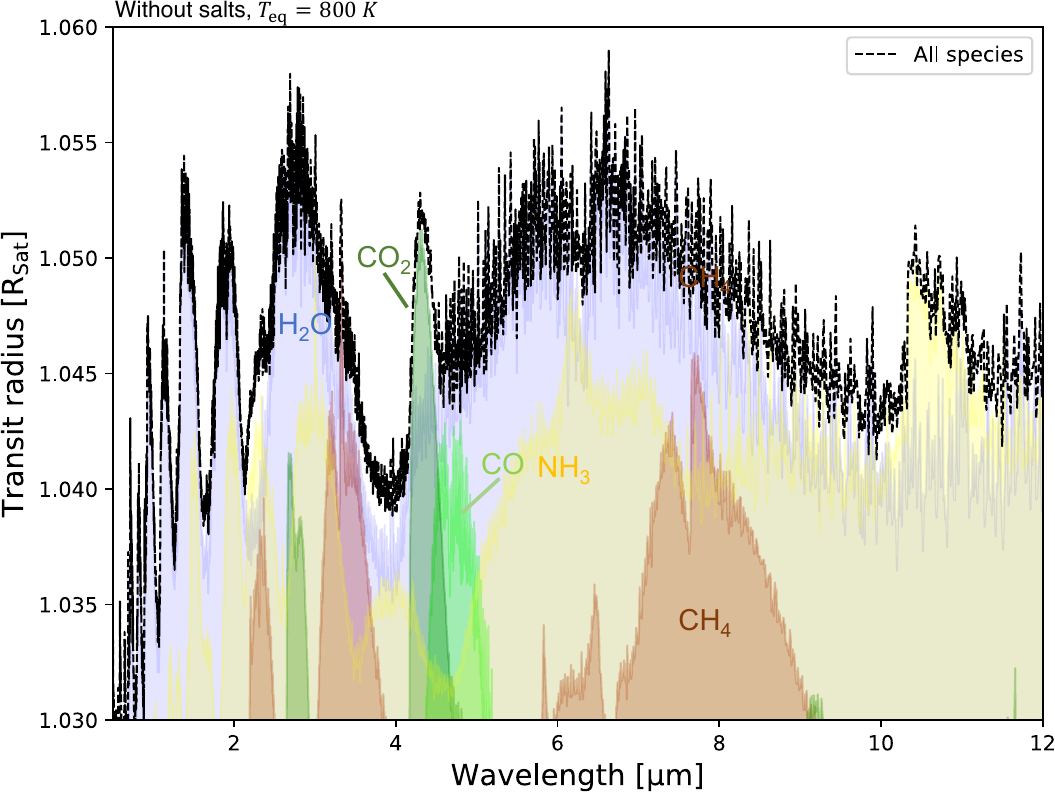}
      \end{minipage} \\
    \caption{Transmission spectra of a planet with an atmospheric structure inheriting the elemental abundances of the disk at 1.5 au and 0.5 Myr. The black line represents the combined transmission spectrum of all molecules (H$_2$, He, H$_2$O, CO$_2$, CO, CH$_4$, NH$_3$, H$_2$S, SO$_2$, CS, COS, NO, HCN, NS), while the shaded areas show the contributions of individual molecules. Upper panel: Spectrum for a planet formed in a salt-bearing disk. The corresponding atmospheric structure is shown in the upper left panel of Figure \ref{fig:AS_fid}. The inset in the lower right corner provides an enlarged view of the 7--9 $\rm \mu m$ bands in the range of 0.99--1.015 $R\rm _{sat}$ transit radius, highlighting the small contributions of HCN and NS. Lower panel: Spectrum for a planet formed in a disk without salt. The corresponding atmospheric structure is shown in the lower left panel of Figure \ref{fig:AS_fid}.}
\label{fig:TS_fid_innerH2O}
\end{figure*}

Here, we examine how the presence or absence of ammonium salts in protoplanetary disks affects the spectrum. Figure \ref{fig:TS_fid_innerH2O} displays the transmission spectrum for a planet whose atmosphere inherits the disk gas composition at $t=0.5$ Myr and $r=1.5$ au. The upper panel shows the case where dust contains ammonium salts, while the lower panel shows the case without salt. The corresponding atmospheric structures are depicted by the solid and dashed lines in Figure \ref{fig:AS_fid}, respectively.
In both cases with and without salts, the spectra are mainly sculpted by spectral features of H$_2$O, CO$_2$, and NH$_3$.
CH$_4$ has minor contributions to the overall spectral shapes owing to a very low carbon-to-oxygen ratio of C/O$=0.02$ in our hypothetical planets.

The most notable difference introduced by ammonium salts in protoplanetary disks is the emergence of prominent SO$_2$ feature.
In the model with ammonium salts, SO$_2$ absorption feature clearly appears at 7.0--8.0 $\rm{\mu m}$. 
In contrast, SO$_2$ does not affect the spectrum shape of planets formed in salt-free disks.
This result demonstrates that salt-dissociation chemistry in protoplanetary disks does affect whether planets formed inside the water snowline exhibit SO$_2$ feature in their atmospheric spectra.
In addition to SO$_2$, H$_2$S adds trace contributions through bands at 3.6--4.0 $\rm{\mu m}$ and 6.0--9.0 $\rm{\mu m}$. 
NH$_3$ produces the dominant spectral feature among nitrogen compounds.
However, the presence or absence of salts has little effect on NH$_3$ feature because salts raise NH$_3$ mixing ratio in the lower atmosphere by supplying additional nitrogen, whereas at higher altitudes the nitrogen forms NS with the co-delivered sulfur, reducing NH$_3$ (see Figure \ref{fig:AS_fid}).

Although the atmospheric structure calculations predict the production of abundant NS, NS barely contributes to the transmission spectrum in the range of 0.5--10 $\rm{\mu m}$. The inset in Figure \ref{fig:TS_fid_innerH2O} provides an enlarged view of the range 7.0--9.0 $\rm{\mu m}$ and 0.99--1.015 $R_{\rm sat}$; however, NS molecules exhibit only very subtle features within this range.
Although sulfur compounds tend to have large cross sections at UV wavelengths of $<0.4~{\rm {\mu}m}$ \citep[see e.g.,][]{2023Natur.617..483T}, the lack of UV cross section prevents us from investigating how NS affects the transmission spectrum at UV wavelengths.

\subsection{Atmospheric calculations for different equilibrium temperature}
\label{subsec:AC_temp}

\begin{figure*}[t]
    \begin{tabular}{cc}
    \hspace{-0.5cm}
      \begin{minipage}[t]{0.45\linewidth}
        \centering
        \includegraphics[width=\linewidth]{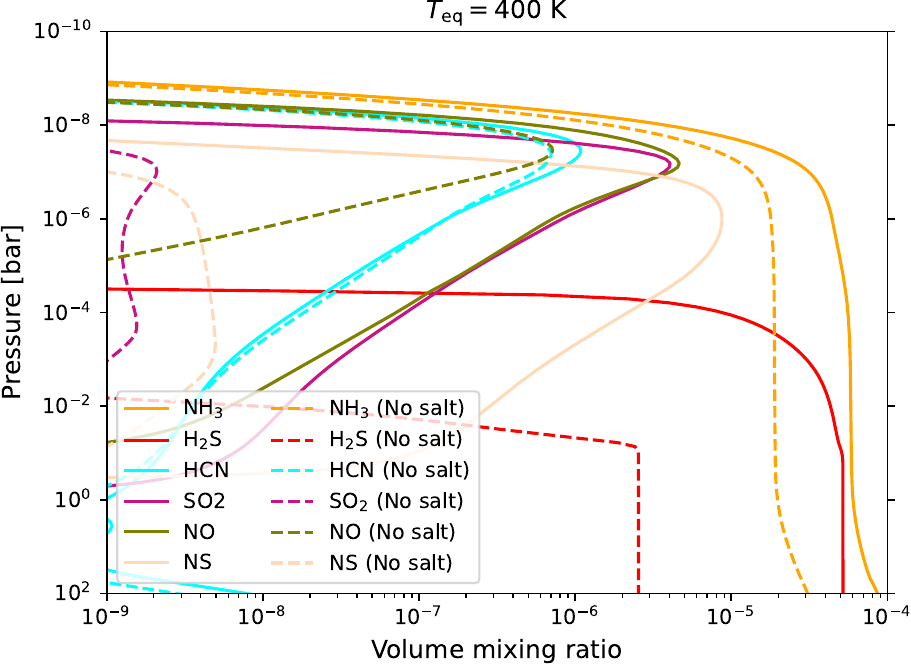}
      \end{minipage} &
        \hspace{0.5cm}
      \begin{minipage}[t]{0.45\linewidth}
        \centering
        \includegraphics[width=\linewidth]{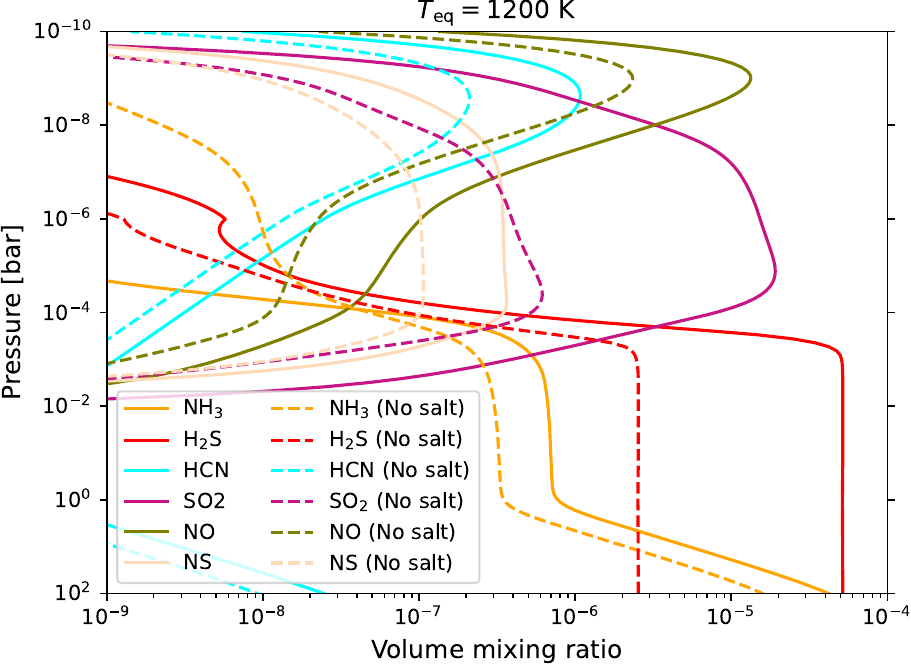}
      \end{minipage} \\
    \end{tabular}
    \caption{Same as Figure \ref{fig:AS_fid}, but equilibrium temperature of the planet is $T_{\rm eq} = 400$ K (left panel) and 1200 K (right panel)}
\label{fig:AS_temp}
\end{figure*}

Interactions with disk gas and/or planet–planet scattering can migrate planets across a wide range of orbits, producing diverse equilibrium temperatures. Even when planets accrete gas with the same composition, temperature differences greatly alter which molecules form most efficiently in their atmospheres and, therefore, determine observability. 
In this section, we show the atmospheric structures and transmission spectra for planets with equilibrium temperatures of 400 K and 1200 K.

For cooler planets of $T_{\rm eq}=400~{\rm K}$, we find that salt dissociation significantly affects the abundance of sulfur compounds.
The left panel of Figure \ref{fig:AS_temp} presents the atmospheric structure for $T_{\rm eq}=400$ K. At low planetary temperatures, NH$_3$ is the most abundant nitrogen-bearing species within the pressure range $10^{-10}$--10$^2$ bar, regardless of whether the dust grains contain salts. The dependence of NH$_3$ abundance on the planet's equilibrium temperature has been studied in detail by \citet{2020AJ....160..288F}, \citet{2023ApJ...946...18O}, and \citet{2025ApJ...985..209M}.
Our results can be attributed to the equilibrium reaction N$_2$ + 3H$_2$ $\rightleftharpoons$ 2NH$_3$, which is biased toward NH$_3$ production at low temperatures.

In the disk containing salts, cool gas planets exhibit the same stratification of sulfur species seen at $T_{\rm eq} = 800$ K: H$_2$S dominates the lower atmosphere, NS dominates in the middle layers, and SO$_2$ becomes dominant in the uppermost layers.
The maximum SO$_2$ mixing ratio reaches $5 \times 10^{-6}$, which is two to three times lower than in the 800 K case. Salts raise the H$_2$S mixing ratio by an order of magnitude and boost NS and SO$_2$ by up to three orders of magnitude. Compared to the 800 K case, the SO$_2$ mixing ratio declines more steeply with depth.
In cool planets, the scarcity of OH radicals limits SO$_2$ formation via SO + OH $\rightleftharpoons$ SO$_2$ + H \citep{2021ApJ...923..264T, 2023Natur.617..483T, 2024ApJ...970L..10B}. However, when salts are present, SO$_2$ still dominates the upper atmosphere at $P \sim 10^{-8}$ bar. 
This pattern may arise because, in the upper atmosphere, O radicals both destroy NS and promote subsequent SO$_2$ production. O radicals remove NS via NS + O $\rightleftharpoons$ NO + S, which supplies sulfur atoms. The supplied S then reacts with O to form SO through S + O $\rightleftharpoons$ SO, followed by SO + O + M $\rightleftharpoons$ SO$_2$ + M. The simultaneous enrichment of NO and SO$_2$ and depletion of NS at altitudes rich in SO$_2$ supports this production pathway.

Planetary atmospheres with higher equilibrium temperatures drive more vigorous photochemistry, which boosts the production of SO$_2$ and NO in the upper layers. The right panel of Figure \ref{fig:AS_temp} shows the atmospheric structure for $T_{\rm eq}=1200$ K. In hot gas planets, the SO$_2$ volume mixing ratio of with-salts case exceeds $10^{-5}$ between $P = 10^{-8}$ and $10^{-4}$ bar, so SO$_2$ dominates even across the pressure range where NS prevailed at $T_{\rm eq} = 800$ K. Salts raise the SO$_2$ mixing ratio by an order of magnitude or more, whereas they increase the NS mixing ratio by only about a factor of four. Therefore, the difference in NS abundance between models with and without salts is smaller than in the $T_{\rm eq} = 800$ K case. 
As in the $T_{\rm eq} = 800$ K case, salts raise the bulk nitrogen abundance in the atmosphere. In hot planets, however, the equilibrium N$_2$ + 3H$_2$ $\rightleftharpoons$ 2NH$_3$ shifts to the left, which lowers NH$_3$ and—because NS forms with NH$_3$ as a precursor—also reduces NS. Consequently, the difference in the mixing ratios of these species between models with and without salts remains small.

\begin{figure*}[t]
\centering
\includegraphics[width=0.9\hsize]{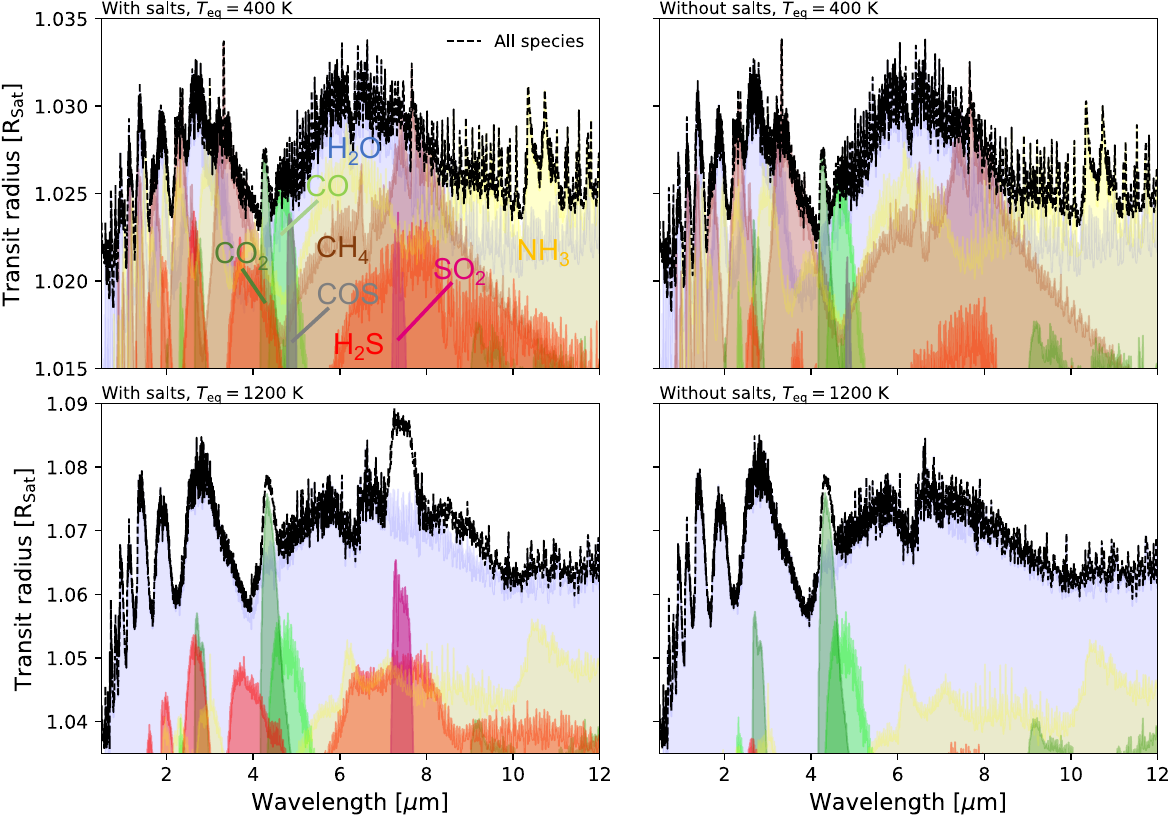}
\caption{Same as Figure \ref{fig:TS_fid_innerH2O}, but for different equilibrium temperature. The upper and lower panels shows the transmission spectra for $T_{\rm eq}=$ 400 and 1200K. The left and right columns correspond to the model with and without salts, respectively.
The corresponding atmospheric structures are shown in the Figure \ref{fig:AS_temp}.}
\label{fig:TS_temp}
\end{figure*}

Because the wavelength‑dependent change in transit radius, $\delta R_{\rm p} = Hf(\lambda)$, is proportional to the atmospheric scale height $H$ \citep{2024RvMG...90..411K}, hot planets with larger scale height display an inflated transmission spectrum whose baseline transit radius is higher than that of cooler gas giants.
Figure \ref{fig:TS_temp} present the spectra for planets with $T_{\rm eq} = 400$ K and 1200 K, respectively.
In cool planets, sulfur‑bearing molecules contribute only weakly to the transmission spectrum. Regardless of whether the model includes salts, SO$_2$ does not exhibit a prominent feature, and H$_2$S also has little influence on spectrum. Instead, CH$_4$ and NH$_3$ govern the opacity contributed by species other than H$_2$O. Salts modify the NH$_3$ absorption only slightly. In particular, the 1--2 $\rm \mu m$ band becomes marginally stronger when salts are present.
In hot planets, salts enhance the NH$_3$ and H$_2$S absorptions, yet these features remain hidden beneath the strong H$_2$O feature, so the 2--3 $\rm \mu m$ NH$_3$ and 4 $\rm \mu m$ H$_2$S bands seen at $T_{\rm eq} = 800$ K disappear. However, salts produce a clear SO$_2$ feature at 7--8 $\rm \mu m$ that stands out more sharply from the water bands than at any other equilibrium temperature.

\section{Discussion}
\label{sec:NSdiscussion}

\subsection{Implications for the origin of atmospheric sulfur}
\label{subsec:salt_effect_general}
Observations with JWST have begun to reveal sulfur-bearing species such as SO$_2$ and H$_2$S in exoplanet atmospheres. Most studies have attributed this sulfur to dissolution of refractory species (e.g., FeS) delivered by solid accretion. Recent comet and molecular-cloud observations, however, point to the possible presence of semi-volatile salts as additional elemental reservoirs, suggesting that dissolution of refractory material is not the only source.
In addition to being a candidate for the missing component that could account for tens of percent of the cosmic nitrogen budget, ammonium salt is a potential carrier of various volatile elements, including sulfur \citep[e.g.][]{2020Sci...367.7462P, 2020NatAs...4..533A, 2022MNRAS.516.3900A,2023NatAs.tmp...25M}. A recent study that examined the spectral features of NH$_4$SH in the laboratory and compared them with observational data suggests that NH$_4$SH may account for up to 20\% of the total sulfur budget \citep{2025A&A...693A.146S}.
As demonstrated in Section \ref{sec:NSresult}, if ammonium salts account for 20\% of the nitrogen and sulfur carriers, regions around the salt lines contain disk gas in which both nitrogen and sulfur exceed twice the solar value at the early stages of disk evolution (Figure \ref{fig:space-time_abundance}). These regions expand widely inside the H$_2$S snow line due to the diffusion of NH$_3$ and H$_2$S.
However, it is important to note that whether nitrogen and sulfur enrichment occurs on the salt line depends on the efficiency of dust drift and turbulence strength. For instance, low dust stickiness suppresses nitrogen and sulfur enrichment near the salt lines in the early stages of disk evolution because vapor diffusion outpaces its production via dust drift and sublimation. \citep{2025PASJ..tmp...29N}.
Moreover, the degree to which the atomic abundances of N and S increase at the salt line also depends on the assumed fraction of the elemental budgets locked in salts. Our fiducial choice, particularly for NH$_4$SH, corresponds to the upper limit inferred from comparisons between observations and laboratory spectra, and the actual salt content may be lower. If the salt budget were reduced by a factor of two, the enrichment at the salt dissociation lines would become comparable to that seen at the NH$_3$ and H$_2$S snow lines. Even in that case, the disk gas around the H$_2$O snow line could still be enriched in both N and S by up to a factor of $\sim$2 relative to the solar value.

A key feature of element transport by ammonium salts is their ability to enrich the disk gas in nitrogen and sulfur around the water snow line. This region has long been viewed as a favorable birthplace for gas-giant cores \citep[e.g., ][]{2015ApJ...806L...7Z, 2017A&A...608A..92D}. However, under typical disk conditions, temperatures there remain too low for refractory sulfur (FeS) to sublimate and too high for volatile nitrogen (N$_2$, NH$_3$) to remain trapped as ice. Our calculations suggest a new source of sulfur for exoplanetary atmospheres. Planets that accrete disk gas near the water snow line can intercept sulfur atoms released by ammonium salt dissociation, allowing sulfur-bearing molecules to form later without any additional solid input. 
For example, planets that acquire gas at $t = 0.5$ Myr and 1.5 au show mixing ratios of H$_2$S and SO$_2$ that rise by at least an order of magnitude at every equilibrium temperature examined in our parameter study. Planets with equilibrium temperatures above roughly 800 K display a clear SO$_2$ feature at 7--8 $\rm \mu m$ in transmission spectra; cooler planets do not, even when salts supply sulfur to the gas, because low equilibrium temperature suppresses the major SO$_2$ production path of SO + OH $\rightleftharpoons$ SO$_2$ + H \citep{2021ApJ...923..264T, 2023Natur.617..483T, 2024ApJ...970L..10B}. 

The dissociation of ammonium salts also increases the abundance of planetary nitrogen. In warm planets, salts enhances NH$_3$ mixing ratio in the lower atmosphere compared to the model without salts, while in the upper atmosphere, much of the additional nitrogen forms NS and NO. Consequently, NS and NO mixing ratios increase, while NH$_3$ falls above about 10$^{-4}$ bar. NS and NO have no prominent features in near- or mid-infrared bands, which limits direct confirmation of their enhancement. However, if a planetary atmosphere is strongly enriched in nitrogen as well as sulfur ($\rm{N/H} \gtrsim 5 \times \rm{solar}$) and the planet is not very hot ($T \lesssim 800$ K), NH$_3$ absorption near 1--2 $\rm{\mu m}$ and around 10 $\rm{\mu m}$ may appear together with the SO$_2$ band at 7--8 $\rm{\mu m}$ (Figures \ref{fig:TS_fid_innerH2O} and \ref{fig:TS_temp}). The warm Neptune WASP-107 b already shows simultaneous detections of NH$_3$ and SO$_2$. Although our results do not claim that salts are the sole source of these molecules, it demonstrates qualitatively that enrichment produced by salts offers a plausible common origin for both NH$_3$ and SO$_2$.

\subsection{Mechanisms and degeneracies in volatile elements enrichment}
\label{subsec:volatiles_enrichment}

\begin{figure*}[t]
\centering
\includegraphics[width=0.8\hsize]{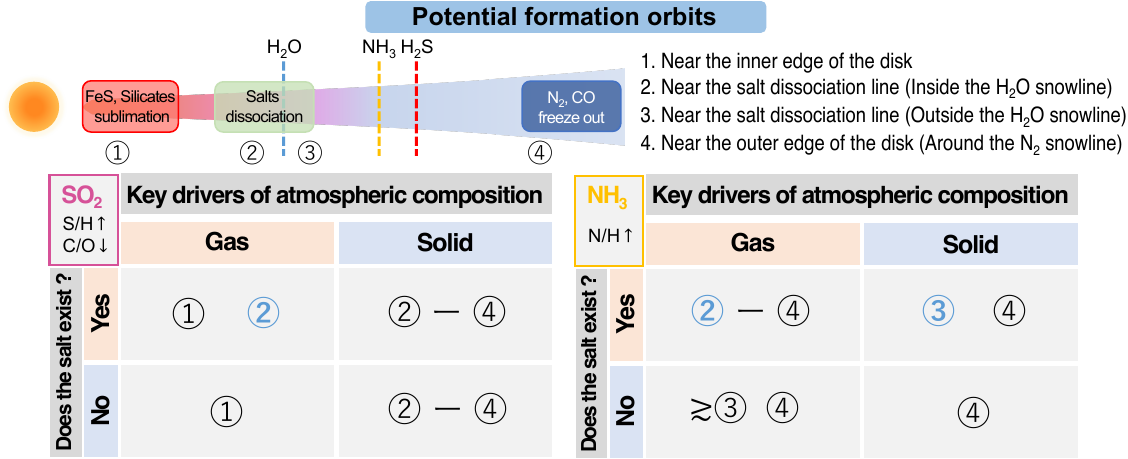}
\caption{Tables used to constrain candidate formation orbits when SO$_2$ and/or NH$_3$ are detected in an exoplanet's atmosphere. Four radial zones are considered: (1) near the inner edge of the disk, (2) just inside the salt line, (3) just outside it, and (4) the outskirts of the disk (see also Figure \ref{fig:elemental_ratios}). Formation paths branch according to two criteria: (i) whether the dominant contributors to the atmospheric composition are gas or solids, and (ii) whether a semi-volatile carrier capable of transporting S and N to the inner disk in solid form—then releasing them near the water snow line—is present. Orbits that appear only when such carriers are invoked are highlighted by blue numbers.}
\label{fig:potential_orbits}
\end{figure*}

\begin{figure*}[t]
\centering
\includegraphics[width=0.8\hsize]{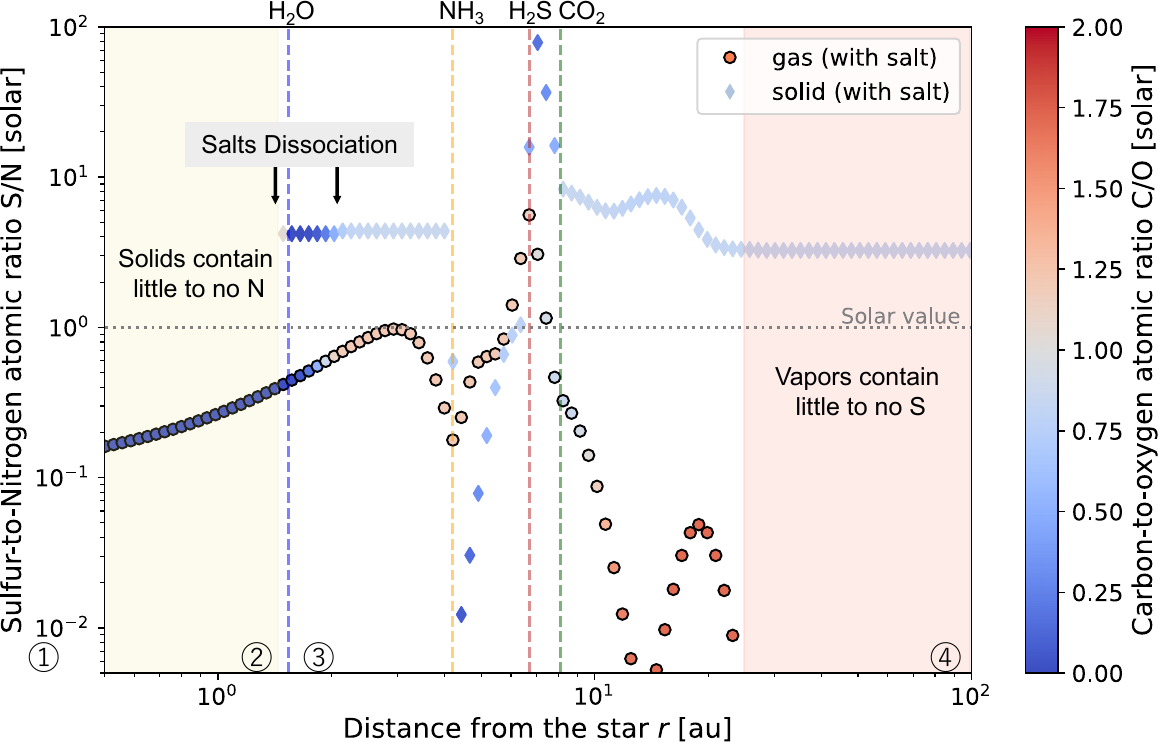}
\caption{S/N and C/O ratios of gas (circles) and solids (diamonds) derived from the spatial distributions of vapors, ices, and rocks at $t=0.5$ Myr.
The solid C/O ratio is a mass‑weighted average of the ice C/O ratio, which computed from the local abundances of CO, CO$_2$, and H$_2$O ices, and the C/O ratio of the rocky component.
The rocky component is assumed to have a uniform C/O ratio of 0.84 throughout the disk, obtained from the “others” fraction of C‑ and O‑bearing species listed in Table \ref{tb:elements_carrier}. Colored vertical lines mark the snow lines of each chemical species at $t=0.5$ Myr. The numbers along the bottom denote the four candidate formation orbits identified in Figure \ref{fig:potential_orbits}.}
\label{fig:elemental_ratios}
\end{figure*}

We have shown that salt dissociation can enrich nitrogen and sulfur in disk gas and open a new, gas‑driven pathway to SO$_2$ production in planets that form just inside the water snow line. This mechanism can also alter both the formation and detectability of nitrogen‑bearing species such as NH$_3$. However, there are several potential mechanisms that can deliver volatile elements to planetary atmospheres. These mechanisms create degeneracies in the formation orbits inferred from atmospheric composition. In this section, we discuss the pathways that can enrich exoplanet atmospheres in volatile elements. We evaluate each pathway with two criteria: (i) whether the dominant contributors to the atmospheric composition are gas or solids, and (ii) whether semi‑volatile carriers of sulfur and nitrogen are present. Figure \ref{fig:potential_orbits} summarizes the possible origins of SO$_2$ and NH$_3$ and the associated formation orbits. We then examine how to resolve the resulting degeneracies in Section \ref{subsec:degeneracy}.

\subsubsection{Volatile enrichment for SO$_2$ production}
Efficient SO$_2$ production needs to enrich the sulfur and oxygen abundance in planetary atmosphere.
If the atmosphere truly inherits the disk gas abundance, then an alternative route to S--O enrichment is the accretion of gas enriched by the sublimation of FeS and silicates in the innermost disk \citep{2021A&A...654A..72S,2025arXiv250616060O}. Because FeS and silicates would together supply more than half of the cosmic S and O budgets \citep{2005Icar..175....1P, 2011A&A...536A..91J, 2021PhR...893....1O}, their sublimation could, in principle, drive large enrichments, although it is unclear whether the required high temperatures can coexist with the formation of a gas-giant core and the onset of runaway gas accretion.

The strongest degeneracy between atmospheric composition and formation location arises when one considers the possible contributions of disk solid components to atmospheric compositions.
Planetesimal accretion and dissolution during or after formation \citep{2011MNRAS.416.1419H}, together with core erosion such as that inferred for Jupiter \citep{2018A&A...610L..14V, 2019ApJ...872..100D, 2022Icar..37814937H}, can inject elements in dust directly into the envelope. In the absence of carriers like NH$_4$SH, disk sulfur is dominated by refractory species; accordingly, the SO$_2$ detected in several exoplanet atmospheres can be attributed to sulfur delivered in solids \citep{2021ApJ...909...40T, 2022ApJ...937...36P}. In this scenario, sulfur enrichment can occur almost anywhere FeS or S$_n$ is incorporated into dust.
Left part of Figure \ref{fig:potential_orbits} summarizes the candidate formation orbits for SO$_2$ detection. Gas‑driven enrichment raises sulfur only in the innermost disk, whereas salts broaden the sulfur‑rich region to just inside the water snow line. If sulfur reaches the atmosphere through dissolution of accreted refractory solids,  sulfur enrichment may arise at any orbital distance.

\subsubsection{Nitrogen enrichment for NH$_3$ production}
Nitrogen is intrinsically harder to deliver to the inner disk because most of it is dominated in hyper-volatile carriers such as N$_2$. Refractory N-bearing organics exist, but they comprise only a small fraction of the total nitrogen reservoir \citep{2021PhR...893....1O} and are therefore unlikely to raise the gas-phase N/H ratio through dust drift and sublimation. Mechanisms capable of transporting nitrogen past its snow line include trapping in amorphous H$_2$O ice or clathration \citep{2020SSRv..216...77M}, but laboratory experiments indicate that N$_2$ is less readily trapped than other volatiles \citep{2007Icar..190..655B, 2023ApJ...955....5S}.
Thus, gas-driven processes are thought to enrich nitrogen only within a narrow region just interior to the N$_2$ snow line near the outer edge of the disk \citep{2019A&A...632L..11B, 2019AJ....158..194O}, unlike sulfur, which becomes enriched in the inner disk. If radial drift is efficient, transient nitrogen enrichment by a factor of a few above the solar value can occur only in the narrow region inside the NH$_3$ snow line, which contains $\sim$ 10\% of the nitrogen budget; however, this region still lies farther out than the water snow line. If semi-volatile ammonium salts constitute the primary nitrogen reservoir after N$_2$, however, the disk gas can become nitrogen‑rich interior to the water snow line.

The allowance of solid contributions also expands the range of orbits capable of producing nitrogen-rich atmospheres. If ammonium salts are present, N enrichment is possible beyond $\gtrsim$ 1.5 au; without salts it is confined to the orbit around or beyond the N$_2$ snow line in the far outer disk (\citealt{2019AJ....158..194O,2019A&A...632L..11B}, see right part of Figure \ref{fig:potential_orbits}), although the disk's own shadow potentially produces local cold regions where N enrichment is possible through N$_2$ ice accretion \citep{2021A&A...651L...2O,2022ApJ...936..188N}.

\subsection{Possible way to break the degeneracy}
\label{subsec:degeneracy}
In the previous section, we have shown that several orbits can produce the elemental enrichments required for efficient SO$_2$ and NH$_3$ production, which makes it difficult to infer a planet's formation orbit from its retrieved atmospheric composition. We now seek to resolve this degeneracy. The priority is to remove the degeneracy introduced by allowing both gas and solids as the contributor to the atmospheric composition. To achieve this, we have to identify compositional patterns characteristic of gas accretion and those characteristic of solid accretion.

A simple diagnostic is the enrichment of lithophilic metals such as Si, Fe, and Mg; substantial enrichment would imply the dissolution of the rocky material \citep{2021ApJ...914...12L, 2023ApJ...943..112C}. Phosphorus would serve as a similar tracer, since it resides predominantly in refractory phosphates within the disk \citep{2019AJ....158..194O, 2025MNRAS.tmp..849K}. 

Additionally, elemental ratios among the volatiles (C, O, N, and S) should also differ between the gas-dominated and solid-dominated scenarios. 
As an example, we examine the distribution of chemical species at $t = 0.5$ Myr in the model that includes salts.
Figure \ref{fig:elemental_ratios} plots the S/N and C/O ratios of disk solids and gas at $t = 0.5$ Myr as functions of the orbital radius. Inside the water snow line, H$_2$O sublimation drives the gas to a sub‑solar C/O ratio, whereas the solids attain a super‑solar C/O ratio. Nitrogen resides almost entirely in the gas, so solids contain little nitrogen. Outside the water snow line, the gas shows a super‑solar C/O ratio (by CO + CO$_2$) and the solids remain sub‑solar (by H$_2$O and rocks), except in narrow orbits near the H$_2$O and CO$_2$ snow lines where sublimation and recondensation cycle lower the C/O ratio. In addition, except near the NH$_3$ snow line, solids are S‑rich (by NH$_4$SH and FeS) and therefore exhibit higher S/N than the gas. These trends persist into the outer disk, where the gas ultimately becomes nearly sulfur‑free. Combinations of volatile‑element ratios help determine whether gas or solids contribute a planet's atmospheric composition. The idea that the ratios of volatile elements (O, C, N, and S) differ between the gas and solid components and vary with the planet’s formation location has also been explored in previous works \citep[e.g., ][]{2017MNRAS.469.3994B, 2019MNRAS.487.3998B, 2021A&A...654A..71S, 2021ApJ...909...40T, 2022ApJ...937...36P, 2025A&A...701A.194P, 2023ApJ...952L..18C, 2025arXiv250616060O}. A comparison between these works and our disk model is presented in Section \ref{subsec:comparison_with_prev}. An important feature of our model is that, when salts are present, the gas-phase S/N ratio inside the H$_2$O snow line (out to the FeS sublimation line), normalized by the stellar value that is typically taken to be $\sim 1$, becomes $<1$.

Once we establish whether gas or solids dominate a planet's atmospheric composition, we can infer its formation orbit by using two clues: (i) the presence or absence of diagnostic molecules such as SO$_2$ and NH$_3$ (Figure \ref{fig:potential_orbits}) and (ii) the volatile‑element ratios, both among C, O, N, and S and between each of these elements and hydrogen (Figures \ref{fig:evo_ices_and_vapors}, \ref{fig:space-time_abundance}, and \ref{fig:elemental_ratios}).

In this section, we examined the disk at $t=0.5$ Myr, and similar constraints should apply at later times as well.
In late stages, gas diffusion smooths the local enrichments near snow lines, yet solids and gas broadly keep their radial C/O and S/N patterns.
Although we simplify the disk chemistry, \citet{2019MNRAS.487.3998B} showed that, except for hydrocarbons, dust drift controls species distributions more strongly than chemical reactions. Consequently, adding minor species or introducing reaction networks should have little effect on the orbital dependence of the C/O and S/N ratios in our models.

Finally, coupling such compositional diagnostics with planet formation and orbital migration theory should yield even robust constraints on formation pathways. For example, type II migration allows a Jupiter-mass planet to move by only a few au \citep{2016ApJ...823...48T, 2020ApJ...891..143T}, potentially ruling out scenarios in which the S and N now observed in short-period planets were acquired far beyond the snow lines.

\paragraph{Caveats.}
In Sections \ref{subsec:volatiles_enrichment} and \ref{subsec:degeneracy}, we discussed the origin and diagnostic use of SO$_2$ and NH$_3$ detections in exoplanet atmospheres, but our model relies on several assumptions about planet formation. We assume a single-planet system in which gas accretion onto the planet is completed rapidly and in situ, without orbital migration during its growth. Consequently, our framework is most directly applicable to early-forming gas giants in disks in which pebble accretion occurs efficiently. For example, if such a planet forms in a salt-bearing disk inside the H$_2$O snow line, the accreted gas is enriched in S and N, with sub-solar S/N and C/O ratios and an atmospheric composition that favors efficient SO$_2$ and NH$_3$ production. By contrast, if the atmospheric composition of a planet forming at the same orbit is increasingly shaped by planetesimal accretion, the S/N and C/O ratios become higher and can reach super-solar values. We refer to our framework as “end-to-end” in the sense that it connects disk-composition simulations to the observability of atmospheric molecules, but this mapping is derived under the above assumptions about planetary growth.

\subsection{Exoplanet observations and the missing‑nitrogen problem}
\label{subsec:missingN}

Identifying the reservoirs of nitrogen and sulfur in the universe is crucial not only for linking planetary composition to formation pathways but also for understanding the synthesis of prebiotic molecules. Current inventories of N$_2$ and NH$_3$ do not close the cosmic nitrogen budget, a discrepancy known as the missing‑nitrogen problem \citep{2015ARA&A..53..541B}. Ammonium salts have been proposed as a promising hidden reservoir \citep{2020Sci...367.7462P, 2020NatAs...4..533A, 2022MNRAS.516.3900A, 2023NatAs.tmp...25M}, yet their prevalence in protoplanetary disks remains uncertain.

Observations of exoplanet atmospheres may provide clues for addressing this problem. So far, NH$_3$ has been detected in a single short‑period exoplanet with the inferred nitrogen abundance of $\sim 10 \times \rm{solar}$ \citep{2024Natur.630..836W}. 
The origin of such a high nitrogen abundance remains unclear, and low‑volatility nitrogen carriers such as ammonium salts may provide an explanation.
To test whether ammonium salts have served as this hidden reservoir, we need further laboratory experiments and theoretical studies to clarify the anion composition of salts that form in the interstellar medium. Quantifying the production rate of salts such as NH$_4$SH—capable of transporting elements with different volatility—and assessing their contribution to observable molecules like NH$_3$ and SO$_2$ will be essential for revealing the origin of nitrogen detected in short‑period exoplanets.

\subsection{Comparison with previous works}
\label{subsec:comparison_with_prev}
To date, coupled models of the compositional–dynamical evolution of protoplanetary disks have been developed in a variety of forms, differing in the diversity of chemical species included, whether chemical network calculations are implemented, whether dust and gas transport are decoupled by radial drift, and whether planet formation, orbital migration, and their feedback on disk evolution are taken into account \citep[e.g., ][]{2017MNRAS.469.3994B, 2019MNRAS.487.3998B, 2021A&A...654A..71S, 2021ApJ...909...40T, 2022MNRAS.517.2285C,2022ApJ...937...36P, 2025A&A...701A.194P, 2025arXiv250616060O}. In this section, we summarize the main differences between our model and previous works and review how the distributions of key volatile elements in disks have been predicted.

First, an important difference between our model and previous works is that, by extending the model of \citet{2025PASJ..tmp...29N}, we incorporate the transport of nitrogen and sulfur associated with the dissociation of semi-volatile ammonium salts. This allows us to show that, if salts account for 20\% of the nitrogen and sulfur budgets, the abundances of these elements in the disk gas around the H$_2$O snow line ($\sim 2$ au) can be enhanced to several up to $\sim 10$ times the solar values (Figure \ref{fig:space-time_abundance}). In conventional disk models, the main nitrogen and sulfur carriers are taken to be N$_2$, NH$_3$, and H$_2$S, FeS, respectively, and for typical disk temperature structures nitrogen enrichment occurs in the outer disk ($r \gtrsim 10$ au), whereas sulfur enrichment arises at $\sim 10$ au or $\sim 0.1$ au. The degree of enrichment can vary depending on what fraction of an element's budget is stored in a given carrier and on how efficient dust transport (i.e., the Stokes number) is relative to gas diffusion ($\alpha$). When dust drift is efficient, the abundances near snow lines can typically be enhanced to a few up to $\sim 10$ times the reference composition. Around the snow lines of carriers that contain most of a given element, such as N$_2$ for nitrogen and FeS for sulfur, the enrichment can be even more pronounced. For example, \citet{2025arXiv250616060O} predict sulfur enrichments of up to $\sim 100$ times the reference abundance at the FeS sublimation line when FeS is account for 90\% of the sulfur budget.

In this work we focus on the transport of N and S, while for C and O we adopt a minimal set of volatile ices that dominate in disks. By contrast, several previous works have focused on the C/O ratio in disks. Even prior to detailed chemo–dynamical modeling of disk dust and gas evolution, it had been predicted that the C/O ratio in the gas would be higher than the stellar value inside the H$_2$O snow line and lower outside it, with the opposite trend for solids \citep{2011ApJ...743L..16O}. Our model is consistent with this behavior. By including, in addition to volatile species, the sublimation of refractory carriers such as iron oxides and by modeling planetary growth and orbital migration, \citet{2021A&A...654A..71S} showed that planets migrating inward to within the water snow line acquire atmospheres with C/O $< 1$. Similarly, \citet{2022MNRAS.517.2285C}, who used a coupled model of disk gas and dust dynamics with a monodisperse grain-size distribution and an extensive gas--grain chemical network, showed that the gas composition inside $r<10$ au attains C/O $< 1$. In contrast to these predictions, JWST observations of the inner regions of disks around very low-mass stars have revealed cases with high C/O ratios \citep{2025ApJ...984L..62A, 2025A&A...702A.126G}. Using disk-composition models that account for both the sublimation of refractory carbon and the sequestration of volatiles into planetesimals, \citet{2025A&A...701A.194P} predicted that high C/O ratios can arise in the inner disk. While the combination of C/O and S/N ratios is a powerful diagnostic of planet formation pathways, it is important to keep in mind the diversity of model predictions discussed above. In the future, developing models that treat a broader range of physical processes and chemical species in a unified manner will be crucial for placing robust constraints on planet formation.

Another important aspect of our model is that we account for the atmospheric chemical evolution and observability of molecules in planetary atmospheres that inherit the disk gas composition. Gas incorporated into a planetary atmosphere undergoes chemical evolution under temperature–pressure conditions that differ from those in the disk, leading to the formation of species that are only minor constituents in the disk, such as SO$_2$ and NS. Furthermore, whether these species are detectable in transmission spectrum depends on their abundances and optical properties. Therefore, in addition to predicting compositions with detailed disk modeling, it is crucial to follow the subsequent chemical evolution in planetary atmospheres and assess the observability of the resulting species, in order to link the detected molecules in exoplanet atmospheres to planet formation pathways.
This is particularly true for the interpretation of SO$_2$, as SO$_2$ abundance does not reflect the bulk sulfur abundance, and photochemical modeling is necessary to assess whether SO$_2$ should be detectable for given atmospheric elemental abundances sculpted by planet formation.

\section{Conclusion}
\label{sec:NSconclusion}
The origin of sulfurs detected in close-in exoplanetary atmospheres through SO$_2$ remains under debate, and the recent simultaneous detection of SO$_2$ and NH$_3$ challenges the conventional view that relies on the accretion of solids such as planetesimals.
We propose a novel idea that ammonium salts such as NH$_4$SH---tentatively detected in comets and molecular clouds---act to yield sulfur- and nitrogen-rich gases in the inner regions of protoplanetary disks.
To test the hypothesis, we simulated the radial transport of dust containing volatile ices and ammonium salts, the dissociation of salts, and the sublimation and recondensation of volatiles. 
In addition, we perform photochemical simulations to investigate atmospheric chemical compositions on planets that formed around the ''salt line''.
We then discuss the observable signatures of those planets by calculating atmospheric transmission spectra. Our framework provides predictions for the observability of atmospheric molecules in planets with early core formation, and whose atmospheric elemental abundances follow the composition of the accreted disk gas.
Our key findings are summarized as follows.
\begin{enumerate}
    \item The vapor release from dust containing ammonium salts and volatile ices leads to sulfur and nitrogen abundances reaching up to 10 times the solar within 0.1 Myr, near the salt lines and volatile snow lines. After 1.0 Myr, these vapors diffuse both inward and outward within the disk, raising sulfur and nitrogen abundances to about 2–5 times the solar within 3 au of the disk (Figure \ref{fig:space-time_abundance}).
    The simultaneous enhancement of nitrogen and sulfur abundances to more than twice the solar extent occurs in regions near the NH$_4$SH salt line during the early stages of disk evolution and in regions inside the H$_2$S snow line during the later stages.

    \item The NH$_4$SH salt line is located near the water snow line. At 0.5 Myr, the O, N, and S abundances exceed twice the solar value at 1.5 au. Among sulfur‑ and nitrogen‑bearing species, NO and SO$_2$ dominate the upper atmosphere of short‑period exoplanets that inherit this composition; NS is most abundant in the middle atmosphere, and NH$_3$ together with H$_2$S prevails in the lower layers (Figure \ref{fig:AS_fid}).
    If the dust does not contain salt, the N and S abundances are not enriched in the same orbit, and the production of N- and S-bearing molecules, excluding NH$_3$, decreases by one to two orders of magnitude (Figure \ref{fig:AS_fid}).

    \item The nitrogen- and sulfur-bearing molecules that contribute to the transmission spectrum in the infrared band of 0.5–10 $\rm{\mu m}$ are NH$_3$, H$_2$S, and SO$_2$. Even when NO and NS are abundant in the atmosphere, they do not produce noticeable features in the spectrum (Figure \ref{fig:TS_fid_innerH2O}). Salts significantly highlight the SO$_2$ absorption feature at 7--8 $\rm{\mu m}$ and slightly boost H$_2$S and NH$_3$ features across 2--4 $\rm{\mu m}$ and 6--9 $\rm{\mu m}$.

    \item Relative to the $T_{\rm eq} =800$ K case, NH$_3$ becomes the dominant nitrogen‑bearing species in cool gas planets ($T_{\rm  eq}= 400$ K), whereas SO$_2$ becomes the dominant sulfur-bearing species in hot gas planets ($T_{\rm eq}= 1200$ K) across a broad pressure range. In cool planets, SO$_2$ contributes little to the transmission spectrum even when salts are present; NH$_3$ and CH$_4$ control the observable features. Hot planets display a larger transit radius, but strong H$_2$O bands bury most molecular features. Salts nevertheless sharpen the SO$_2$ absorption, making it more conspicuous at 7--8 $\rm{\mu m}$ than at any other equilibrium temperature.
    
    \item Transport of sulfur via NH$_4$SH salt provides a new, purely gas‑phase pathway for producing the SO$_2$ detected in some exoplanetary atmospheres: planets accrete disk gas that has been enriched near the water snow line. If the dissolution of accreted solids is permitted, the set of formation orbits capable of explaining the observed SO$_2$ and/or NH$_3$ becomes degenerate. Nevertheless, joint analysis of volatile-element ratios, such as S/N and C/O, in the planetary atmosphere would help to break the degeneracy (Figure \ref{fig:elemental_ratios}).

\end{enumerate}

Our results indicate that ammonium salts can serve as new carriers of nitrogen and sulfur in the disk, and their influence may appear as identifiable features in transit observations of planetary atmospheres.
By examining the effects of other potential carriers responsible for nitrogen and sulfur enrichment, and integrating our model with theories on the planetary migration, we can further elucidate the origins of planetary constituents.

\section*{Acknowledgements}
We appreciate the referee for providing constructive and useful reviews. We thank Satoshi Okuzumi, Hiroyuki Kurokawa, Shota Notsu, and Tetsuo Taki for helpful discussions and comments. This work was supported by JSPS KAKENHI Grant Numbers JP25KJ0093 and JP23K19072.
\bibliographystyle{aasjournal}
\bibliography{articles.bib} 

@ARTICLE{2024MNRAS.535..171P,
       author = {{Penzlin}, Anna B.~T. and {Booth}, Richard A. and {Kirk}, James and {Owen}, James E. and {Ahrer}, E. and {Christie}, Duncan A. and {Claringbold}, Alastair B. and {Esparza-Borges}, Emma and {L{\'o}pez-Morales}, M. and {Mayne}, N.~J. and {McCormack}, Mason and {Meech}, Annabella and {Panwar}, Vatsal and {Powell}, Diana and {Sergeev}, Denis E. and {Taylor}, Jake and {Wheatley}, Peter J. and {Zamyatina}, Maria},
        title = "{BOWIE-ALIGN: how formation and migration histories of giant planets impact atmospheric compositions}",
      journal = {\mnras},
         year = 2024,
        month = nov,
       volume = {535},
       number = {1},
        pages = {171-186},
          doi = {10.1093/mnras/stae2362},
archivePrefix = {arXiv},
       eprint = {2407.03199},
 primaryClass = {astro-ph.EP},
       adsurl = {https://ui.adsabs.harvard.edu/abs/2024MNRAS.535..171P}
}

@ARTICLE{2018A&A...613A..14E,
       author = {{Eistrup}, Christian and {Walsh}, Catherine and {van Dishoeck}, Ewine F.},
        title = "{Molecular abundances and C/O ratios in chemically evolving planet-forming disk midplanes}",
      journal = {\aap},
         year = 2018,
        month = may,
       volume = {613},
          eid = {A14},
        pages = {A14},
          doi = {10.1051/0004-6361/201731302},
archivePrefix = {arXiv},
       eprint = {1709.07863},
 primaryClass = {astro-ph.EP},
       adsurl = {https://ui.adsabs.harvard.edu/abs/2018A&A...613A..14E}
}

@ARTICLE{2022ApJ...936..188N,
       author = {{Notsu}, Shota and {Ohno}, Kazumasa and {Ueda}, Takahiro and {Walsh}, Catherine and {Eistrup}, Christian and {Nomura}, Hideko},
        title = "{The Molecular Composition of Shadowed Proto-solar Disk Midplanes Beyond the Water Snowline}",
      journal = {\apj},
         year = 2022,
        month = sep,
       volume = {936},
       number = {2},
          eid = {188},
        pages = {188},
          doi = {10.3847/1538-4357/ac87fa},
archivePrefix = {arXiv},
       eprint = {2208.06005},
 primaryClass = {astro-ph.EP},
       adsurl = {https://ui.adsabs.harvard.edu/abs/2022ApJ...936..188N}
}

@ARTICLE{2021A&A...651L...2O,
       author = {{Ohno}, Kazumasa and {Ueda}, Takahiro},
        title = "{Jupiter's ``cold'' formation in the protosolar disk shadow. An explanation for the planet's uniformly enriched atmosphere}",
      journal = {\aap},
         year = 2021,
        month = jul,
       volume = {651},
          eid = {L2},
        pages = {L2},
          doi = {10.1051/0004-6361/202141169},
archivePrefix = {arXiv},
       eprint = {2106.09084},
 primaryClass = {astro-ph.EP},
       adsurl = {https://ui.adsabs.harvard.edu/abs/2021A&A...651L...2O}
}

@ARTICLE{1969Icar...10..365L,
       author = {{Lewis}, John S.},
        title = "{The clouds of Jupiter and the NH $_{3}$H $_{2}$O and NH $_{3}$H $_{2}$S systems}",
      journal = {\icarus},
         year = 1969,
        month = may,
       volume = {10},
       number = {3},
        pages = {365-378},
          doi = {10.1016/0019-1035(69)90091-8},
       adsurl = {https://ui.adsabs.harvard.edu/abs/1969Icar...10..365L}
}

@ARTICLE{2021A&A...654A..71S,
       author = {{Schneider}, Aaron David and {Bitsch}, Bertram},
        title = "{How drifting and evaporating pebbles shape giant planets. I. Heavy element content and atmospheric C/O}",
      journal = {\aap},
         year = 2021,
        month = oct,
       volume = {654},
          eid = {A71},
        pages = {A71},
          doi = {10.1051/0004-6361/202039640},
archivePrefix = {arXiv},
       eprint = {2105.13267},
 primaryClass = {astro-ph.EP},
       adsurl = {https://ui.adsabs.harvard.edu/abs/2021A&A...654A..71S}
}

@ARTICLE{2017MNRAS.469.3994B,
       author = {{Booth}, Richard A. and {Clarke}, Cathie J. and {Madhusudhan}, Nikku and {Ilee}, John D.},
        title = "{Chemical enrichment of giant planets and discs due to pebble drift}",
      journal = {\mnras},
         year = 2017,
        month = aug,
       volume = {469},
       number = {4},
        pages = {3994-4011},
          doi = {10.1093/mnras/stx1103},
archivePrefix = {arXiv},
       eprint = {1705.03305},
 primaryClass = {astro-ph.EP},
       adsurl = {https://ui.adsabs.harvard.edu/abs/2017MNRAS.469.3994B}
}

@ARTICLE{2023ApJ...952L..18C,
       author = {{Crossfield}, Ian J.~M.},
        title = "{Volatile-to-sulfur Ratios Can Recover a Gas Giant's Accretion History}",
      journal = {\apjl},
         year = 2023,
        month = jul,
       volume = {952},
       number = {1},
          eid = {L18},
        pages = {L18},
          doi = {10.3847/2041-8213/ace35f},
archivePrefix = {arXiv},
       eprint = {2303.17622},
 primaryClass = {astro-ph.EP},
       adsurl = {https://ui.adsabs.harvard.edu/abs/2023ApJ...952L..18C}
}

@ARTICLE{2024Natur.630..831S,
       author = {{Sing}, David K. and {Rustamkulov}, Zafar and {Thorngren}, Daniel P. and {Barstow}, Joanna K. and {Tremblin}, Pascal and {Alves de Oliveira}, Catarina and {Beck}, Tracy L. and {Birkmann}, Stephan M. and {Challener}, Ryan C. and {Crouzet}, Nicolas and {Espinoza}, N{\'e}stor and {Ferruit}, Pierre and {Giardino}, Giovanna and {Gressier}, Am{\'e}lie and {Lee}, Elspeth K.~H. and {Lewis}, Nikole K. and {Maiolino}, Roberto and {Manjavacas}, Elena and {Rauscher}, Bernard J. and {Sirianni}, Marco and {Valenti}, Jeff A.},
        title = "{A warm Neptune's methane reveals core mass and vigorous atmospheric mixing}",
      journal = {\nat},
         year = 2024,
        month = jun,
       volume = {630},
       number = {8018},
        pages = {831-835},
          doi = {10.1038/s41586-024-07395-z},
archivePrefix = {arXiv},
       eprint = {2405.11027},
 primaryClass = {astro-ph.EP},
       adsurl = {https://ui.adsabs.harvard.edu/abs/2024Natur.630..831S}
}

@ARTICLE{2014ApJ...794L..12M,
       author = {{Madhusudhan}, Nikku and {Amin}, Mustafa A. and {Kennedy}, Grant M.},
        title = "{Toward Chemical Constraints on Hot Jupiter Migration}",
      journal = {\apjl},
         year = 2014,
        month = oct,
       volume = {794},
       number = {1},
          eid = {L12},
        pages = {L12},
          doi = {10.1088/2041-8205/794/1/L12},
archivePrefix = {arXiv},
       eprint = {1408.3668},
 primaryClass = {astro-ph.EP},
       adsurl = {https://ui.adsabs.harvard.edu/abs/2014ApJ...794L..12M}
}

@ARTICLE{2025arXiv250616060O,
       author = {{Ohno}, Kazumasa and {Ikoma}, Masahiro and {Okuzumi}, Satoshi and {Kimura}, Tadahiro},
        title = "{A Dichotomy of the Mass-Metallicity Relation of Exoplanetary Atmospheres Demarcated by their Birthplace}",
      journal = {arXiv e-prints},
         year = 2025,
        month = jun,
          eid = {arXiv:2506.16060},
        pages = {arXiv:2506.16060},
          doi = {10.48550/arXiv.2506.16060},
archivePrefix = {arXiv},
       eprint = {2506.16060},
 primaryClass = {astro-ph.EP},
       adsurl = {https://ui.adsabs.harvard.edu/abs/2025arXiv250616060O}
}

@ARTICLE{2021A&A...654A..72S,
       author = {{Schneider}, Aaron David and {Bitsch}, Bertram},
        title = "{How drifting and evaporating pebbles shape giant planets. II. Volatiles and refractories in atmospheres}",
      journal = {\aap},
         year = 2021,
        month = oct,
       volume = {654},
          eid = {A72},
        pages = {A72},
          doi = {10.1051/0004-6361/202141096},
archivePrefix = {arXiv},
       eprint = {2109.03589},
 primaryClass = {astro-ph.EP},
       adsurl = {https://ui.adsabs.harvard.edu/abs/2021A&A...654A..72S}
}

@ARTICLE{2024ApJ...970L..10B,
       author = {{Beatty}, Thomas G. and {Welbanks}, Luis and {Schlawin}, Everett and {Bell}, Taylor J. and {Line}, Michael R. and {Murphy}, Matthew and {Edelman}, Isaac and {Greene}, Thomas P. and {Fortney}, Jonathan J. and {Henry}, Gregory W. and {Mukherjee}, Sagnick and {Ohno}, Kazumasa and {Parmentier}, Vivien and {Rauscher}, Emily and {Wiser}, Lindsey S. and {Arnold}, Kenneth E.},
        title = "{Sulfur Dioxide and Other Molecular Species in the Atmosphere of the Sub-Neptune GJ 3470 b}",
      journal = {\apjl},
         year = 2024,
        month = jul,
       volume = {970},
       number = {1},
          eid = {L10},
        pages = {L10},
          doi = {10.3847/2041-8213/ad55e9},
archivePrefix = {arXiv},
       eprint = {2406.04450},
 primaryClass = {astro-ph.EP},
       adsurl = {https://ui.adsabs.harvard.edu/abs/2024ApJ...970L..10B}
}

@ARTICLE{2025ApJ...985..209M,
       author = {{Mukherjee}, Sagnick and {Fortney}, Jonathan J. and {Wogan}, Nicholas F. and {Sing}, David K. and {Ohno}, Kazumasa},
        title = "{Effects of Planetary Parameters on Disequilibrium Chemistry in Irradiated Planetary Atmospheres: From Gas Giants to Sub-Neptunes}",
      journal = {\apj},
         year = 2025,
        month = jun,
       volume = {985},
       number = {2},
          eid = {209},
        pages = {209},
          doi = {10.3847/1538-4357/adc7b3},
archivePrefix = {arXiv},
       eprint = {2410.17169},
 primaryClass = {astro-ph.EP},
       adsurl = {https://ui.adsabs.harvard.edu/abs/2025ApJ...985..209M}
}

@ARTICLE{2020AJ....160..288F,
       author = {{Fortney}, Jonathan J. and {Visscher}, Channon and {Marley}, Mark S. and {Hood}, Callie E. and {Line}, Michael R. and {Thorngren}, Daniel P. and {Freedman}, Richard S. and {Lupu}, Roxana},
        title = "{Beyond Equilibrium Temperature: How the Atmosphere/Interior Connection Affects the Onset of Methane, Ammonia, and Clouds in Warm Transiting Giant Planets}",
      journal = {\aj},
         year = 2020,
        month = dec,
       volume = {160},
       number = {6},
          eid = {288},
        pages = {288},
          doi = {10.3847/1538-3881/abc5bd},
archivePrefix = {arXiv},
       eprint = {2010.00146},
 primaryClass = {astro-ph.EP},
       adsurl = {https://ui.adsabs.harvard.edu/abs/2020AJ....160..288F}
}

@ARTICLE{2016ApJ...833..203P,
       author = {{Piso}, Ana-Maria A. and {Pegues}, Jamila and {{\"O}berg}, Karin I.},
        title = "{The Role of Ice Compositions for Snowlines and the C/N/O Ratios in Active Disks}",
      journal = {\apj},
         year = 2016,
        month = dec,
       volume = {833},
       number = {2},
          eid = {203},
        pages = {203},
          doi = {10.3847/1538-4357/833/2/203},
archivePrefix = {arXiv},
       eprint = {1611.00741},
 primaryClass = {astro-ph.EP},
       adsurl = {https://ui.adsabs.harvard.edu/abs/2016ApJ...833..203P}
}

@ARTICLE{2024Natur.632..752F,
       author = {{Fu}, Guangwei and {Welbanks}, Luis and {Deming}, Drake and {Inglis}, Julie and {Zhang}, Michael and {Lothringer}, Joshua and {Ih}, Jegug and {Moses}, Julianne I. and {Schlawin}, Everett and {Knutson}, Heather A. and {Henry}, Gregory and {Greene}, Thomas and {Sing}, David K. and {Savel}, Arjun B. and {Kempton}, Eliza M. -R. and {Louie}, Dana R. and {Line}, Michael and {Nixon}, Matt},
        title = "{Hydrogen sulfide and metal-enriched atmosphere for a Jupiter-mass exoplanet}",
      journal = {\nat},
         year = 2024,
        month = aug,
       volume = {632},
       number = {8026},
        pages = {752-756},
          doi = {10.1038/s41586-024-07760-y},
archivePrefix = {arXiv},
       eprint = {2407.06163},
 primaryClass = {astro-ph.EP},
       adsurl = {https://ui.adsabs.harvard.edu/abs/2024Natur.632..752F}
}

@ARTICLE{2024Natur.625...51D,
       author = {{Dyrek}, Achr{\`e}ne and {Min}, Michiel and {Decin}, Leen and {Bouwman}, Jeroen and {Crouzet}, Nicolas and {Molli{\`e}re}, Paul and {Lagage}, Pierre-Olivier and {Konings}, Thomas and {Tremblin}, Pascal and {G{\"u}del}, Manuel and {Pye}, John and {Waters}, Rens and {Henning}, Thomas and {Vandenbussche}, Bart and {Ardevol Martinez}, Francisco and {Argyriou}, Ioannis and {Ducrot}, Elsa and {Heinke}, Linus and {van Looveren}, Gwenael and {Absil}, Olivier and {Barrado}, David and {Baudoz}, Pierre and {Boccaletti}, Anthony and {Cossou}, Christophe and {Coulais}, Alain and {Edwards}, Billy and {Gastaud}, Ren{\'e} and {Glasse}, Alistair and {Glauser}, Adrian and {Greene}, Thomas P. and {Kendrew}, Sarah and {Krause}, Oliver and {Lahuis}, Fred and {Mueller}, Michael and {Olofsson}, Goran and {Patapis}, Polychronis and {Rouan}, Daniel and {Royer}, Pierre and {Scheithauer}, Silvia and {Waldmann}, Ingo and {Whiteford}, Niall and {Colina}, Luis and {van Dishoeck}, Ewine F. and {{\"O}stlin}, G{\"o}ran and {Ray}, Tom P. and {Wright}, Gillian},
        title = "{SO$_{2}$, silicate clouds, but no CH$_{4}$ detected in a warm Neptune}",
      journal = {\nat},
         year = 2024,
        month = jan,
       volume = {625},
       number = {7993},
        pages = {51-54},
          doi = {10.1038/s41586-023-06849-0},
archivePrefix = {arXiv},
       eprint = {2311.12515},
 primaryClass = {astro-ph.EP},
       adsurl = {https://ui.adsabs.harvard.edu/abs/2024Natur.625...51D}
}

@ARTICLE{2023Natur.614..659R,
       author = {{Rustamkulov}, Z. and {Sing}, D.~K. and {Mukherjee}, S. and {May}, E.~M. and {Kirk}, J. and {Schlawin}, E. and {Line}, M.~R. and {Piaulet}, C. and {Carter}, A.~L. and {Batalha}, N.~E. and {Goyal}, J.~M. and {L{\'o}pez-Morales}, M. and {Lothringer}, J.~D. and {MacDonald}, R.~J. and {Moran}, S.~E. and {Stevenson}, K.~B. and {Wakeford}, H.~R. and {Espinoza}, N. and {Bean}, J.~L. and {Batalha}, N.~M. and {Benneke}, B. and {Berta-Thompson}, Z.~K. and {Crossfield}, I.~J.~M. and {Gao}, P. and {Kreidberg}, L. and {Powell}, D.~K. and {Cubillos}, P.~E. and {Gibson}, N.~P. and {Leconte}, J. and {Molaverdikhani}, K. and {Nikolov}, N.~K. and {Parmentier}, V. and {Roy}, P. and {Taylor}, J. and {Turner}, J.~D. and {Wheatley}, P.~J. and {Aggarwal}, K. and {Ahrer}, E. and {Alam}, M.~K. and {Alderson}, L. and {Allen}, N.~H. and {Banerjee}, A. and {Barat}, S. and {Barrado}, D. and {Barstow}, J.~K. and {Bell}, T.~J. and {Blecic}, J. and {Brande}, J. and {Casewell}, S. and {Changeat}, Q. and {Chubb}, K.~L. and {Crouzet}, N. and {Daylan}, T. and {Decin}, L. and {D{\'e}sert}, J. and {Mikal-Evans}, T. and {Feinstein}, A.~D. and {Flagg}, L. and {Fortney}, J.~J. and {Harrington}, J. and {Heng}, K. and {Hong}, Y. and {Hu}, R. and {Iro}, N. and {Kataria}, T. and {Kempton}, E.~M. -R. and {Krick}, J. and {Lendl}, M. and {Lillo-Box}, J. and {Louca}, A. and {Lustig-Yaeger}, J. and {Mancini}, L. and {Mansfield}, M. and {Mayne}, N.~J. and {Miguel}, Y. and {Morello}, G. and {Ohno}, K. and {Palle}, E. and {Petit dit de la Roche}, D.~J.~M. and {Rackham}, B.~V. and {Radica}, M. and {Ramos-Rosado}, L. and {Redfield}, S. and {Rogers}, L.~K. and {Shkolnik}, E.~L. and {Southworth}, J. and {Teske}, J. and {Tremblin}, P. and {Tucker}, G.~S. and {Venot}, O. and {Waalkes}, W.~C. and {Welbanks}, L. and {Zhang}, X. and {Zieba}, S.},
        title = "{Early Release Science of the exoplanet WASP-39b with JWST NIRSpec PRISM}",
      journal = {\nat},
         year = 2023,
        month = feb,
       volume = {614},
       number = {7949},
        pages = {659-663},
          doi = {10.1038/s41586-022-05677-y},
archivePrefix = {arXiv},
       eprint = {2211.10487},
 primaryClass = {astro-ph.EP},
       adsurl = {https://ui.adsabs.harvard.edu/abs/2023Natur.614..659R}
}

@ARTICLE{2023Natur.614..664A,
       author = {{Alderson}, Lili and {Wakeford}, Hannah R. and {Alam}, Munazza K. and {Batalha}, Natasha E. and {Lothringer}, Joshua D. and {Adams Redai}, Jea and {Barat}, Saugata and {Brande}, Jonathan and {Damiano}, Mario and {Daylan}, Tansu and {Espinoza}, N{\'e}stor and {Flagg}, Laura and {Goyal}, Jayesh M. and {Grant}, David and {Hu}, Renyu and {Inglis}, Julie and {Lee}, Elspeth K.~H. and {Mikal-Evans}, Thomas and {Ramos-Rosado}, Lakeisha and {Roy}, Pierre-Alexis and {Wallack}, Nicole L. and {Batalha}, Natalie M. and {Bean}, Jacob L. and {Benneke}, Bj{\"o}rn and {Berta-Thompson}, Zachory K. and {Carter}, Aarynn L. and {Changeat}, Quentin and {Col{\'o}n}, Knicole D. and {Crossfield}, Ian J.~M. and {D{\'e}sert}, Jean-Michel and {Foreman-Mackey}, Daniel and {Gibson}, Neale P. and {Kreidberg}, Laura and {Line}, Michael R. and {L{\'o}pez-Morales}, Mercedes and {Molaverdikhani}, Karan and {Moran}, Sarah E. and {Morello}, Giuseppe and {Moses}, Julianne I. and {Mukherjee}, Sagnick and {Schlawin}, Everett and {Sing}, David K. and {Stevenson}, Kevin B. and {Taylor}, Jake and {Aggarwal}, Keshav and {Ahrer}, Eva-Maria and {Allen}, Natalie H. and {Barstow}, Joanna K. and {Bell}, Taylor J. and {Blecic}, Jasmina and {Casewell}, Sarah L. and {Chubb}, Katy L. and {Crouzet}, Nicolas and {Cubillos}, Patricio E. and {Decin}, Leen and {Feinstein}, Adina D. and {Fortney}, Joanthan J. and {Harrington}, Joseph and {Heng}, Kevin and {Iro}, Nicolas and {Kempton}, Eliza M. -R. and {Kirk}, James and {Knutson}, Heather A. and {Krick}, Jessica and {Leconte}, J{\'e}r{\'e}my and {Lendl}, Monika and {MacDonald}, Ryan J. and {Mancini}, Luigi and {Mansfield}, Megan and {May}, Erin M. and {Mayne}, Nathan J. and {Miguel}, Yamila and {Nikolov}, Nikolay K. and {Ohno}, Kazumasa and {Palle}, Enric and {Parmentier}, Vivien and {Petit dit de la Roche}, Dominique J.~M. and {Piaulet}, Caroline and {Powell}, Diana and {Rackham}, Benjamin V. and {Redfield}, Seth and {Rogers}, Laura K. and {Rustamkulov}, Zafar and {Tan}, Xianyu and {Tremblin}, P. and {Tsai}, Shang-Min and {Turner}, Jake D. and {de Val-Borro}, Miguel and {Venot}, Olivia and {Welbanks}, Luis and {Wheatley}, Peter J. and {Zhang}, Xi},
        title = "{Early Release Science of the exoplanet WASP-39b with JWST NIRSpec G395H}",
      journal = {\nat},
         year = 2023,
        month = feb,
       volume = {614},
       number = {7949},
        pages = {664-669},
          doi = {10.1038/s41586-022-05591-3},
archivePrefix = {arXiv},
       eprint = {2211.10488},
 primaryClass = {astro-ph.EP},
       adsurl = {https://ui.adsabs.harvard.edu/abs/2023Natur.614..664A}
}

@ARTICLE{2023Natur.617..483T,
       author = {{Tsai}, Shang-Min and {Lee}, Elspeth K.~H. and {Powell}, Diana and {Gao}, Peter and {Zhang}, Xi and {Moses}, Julianne and {H{\'e}brard}, Eric and {Venot}, Olivia and {Parmentier}, Vivien and {Jordan}, Sean and {Hu}, Renyu and {Alam}, Munazza K. and {Alderson}, Lili and {Batalha}, Natalie M. and {Bean}, Jacob L. and {Benneke}, Bj{\"o}rn and {Bierson}, Carver J. and {Brady}, Ryan P. and {Carone}, Ludmila and {Carter}, Aarynn L. and {Chubb}, Katy L. and {Inglis}, Julie and {Leconte}, J{\'e}r{\'e}my and {Line}, Michael and {L{\'o}pez-Morales}, Mercedes and {Miguel}, Yamila and {Molaverdikhani}, Karan and {Rustamkulov}, Zafar and {Sing}, David K. and {Stevenson}, Kevin B. and {Wakeford}, Hannah R. and {Yang}, Jeehyun and {Aggarwal}, Keshav and {Baeyens}, Robin and {Barat}, Saugata and {de Val-Borro}, Miguel and {Daylan}, Tansu and {Fortney}, Jonathan J. and {France}, Kevin and {Goyal}, Jayesh M. and {Grant}, David and {Kirk}, James and {Kreidberg}, Laura and {Louca}, Amy and {Moran}, Sarah E. and {Mukherjee}, Sagnick and {Nasedkin}, Evert and {Ohno}, Kazumasa and {Rackham}, Benjamin V. and {Redfield}, Seth and {Taylor}, Jake and {Tremblin}, Pascal and {Visscher}, Channon and {Wallack}, Nicole L. and {Welbanks}, Luis and {Youngblood}, Allison and {Ahrer}, Eva-Maria and {Batalha}, Natasha E. and {Behr}, Patrick and {Berta-Thompson}, Zachory K. and {Blecic}, Jasmina and {Casewell}, S.~L. and {Crossfield}, Ian J.~M. and {Crouzet}, Nicolas and {Cubillos}, Patricio E. and {Decin}, Leen and {D{\'e}sert}, Jean-Michel and {Feinstein}, Adina D. and {Gibson}, Neale P. and {Harrington}, Joseph and {Heng}, Kevin and {Henning}, Thomas and {Kempton}, Eliza M. -R. and {Krick}, Jessica and {Lagage}, Pierre-Olivier and {Lendl}, Monika and {Lothringer}, Joshua D. and {Mansfield}, Megan and {Mayne}, N.~J. and {Mikal-Evans}, Thomas and {Palle}, Enric and {Schlawin}, Everett and {Shorttle}, Oliver and {Wheatley}, Peter J. and {Yurchenko}, Sergei N.},
        title = "{Photochemically produced SO$_{2}$ in the atmosphere of WASP-39b}",
      journal = {\nat},
         year = 2023,
        month = may,
       volume = {617},
       number = {7961},
        pages = {483-487},
          doi = {10.1038/s41586-023-05902-2},
archivePrefix = {arXiv},
       eprint = {2211.10490},
 primaryClass = {astro-ph.EP},
       adsurl = {https://ui.adsabs.harvard.edu/abs/2023Natur.617..483T}
}

@ARTICLE{2021ApJ...923..264T,
       author = {{Tsai}, Shang-Min and {Malik}, Matej and {Kitzmann}, Daniel and {Lyons}, James R. and {Fateev}, Alexander and {Lee}, Elspeth and {Heng}, Kevin},
        title = "{A Comparative Study of Atmospheric Chemistry with VULCAN}",
      journal = {\apj},
         year = 2021,
        month = dec,
       volume = {923},
       number = {2},
          eid = {264},
        pages = {264},
          doi = {10.3847/1538-4357/ac29bc},
archivePrefix = {arXiv},
       eprint = {2108.01790},
 primaryClass = {astro-ph.EP},
       adsurl = {https://ui.adsabs.harvard.edu/abs/2021ApJ...923..264T}
}

@ARTICLE{2016ApJ...824..137Z,
       author = {{Zahnle}, K. and {Marley}, M.~S. and {Morley}, C.~V. and {Moses}, J.~I.},
        title = "{Photolytic Hazes in the Atmosphere of 51 Eri b}",
      journal = {\apj},
         year = 2016,
        month = jun,
       volume = {824},
       number = {2},
          eid = {137},
        pages = {137},
          doi = {10.3847/0004-637X/824/2/137},
archivePrefix = {arXiv},
       eprint = {1604.07388},
 primaryClass = {astro-ph.EP},
       adsurl = {https://ui.adsabs.harvard.edu/abs/2016ApJ...824..137Z}
}

@ARTICLE{2011ApJ...743L..16O,
       author = {{{\"O}berg}, Karin I. and {Murray-Clay}, Ruth and {Bergin}, Edwin A.},
        title = "{The Effects of Snowlines on C/O in Planetary Atmospheres}",
      journal = {\apjl},
         year = 2011,
        month = dec,
       volume = {743},
       number = {1},
          eid = {L16},
        pages = {L16},
          doi = {10.1088/2041-8205/743/1/L16},
archivePrefix = {arXiv},
       eprint = {1110.5567},
 primaryClass = {astro-ph.GA},
       adsurl = {https://ui.adsabs.harvard.edu/abs/2011ApJ...743L..16O}
}

@ARTICLE{2019MNRAS.487.3998B,
       author = {{Booth}, R.~A. and {Ilee}, J.~D.},
        title = "{Planet-forming material in a protoplanetary disc: the interplay between chemical evolution and pebble drift}",
      journal = {\mnras},
         year = 2019,
        month = aug,
       volume = {487},
       number = {3},
        pages = {3998-4011},
          doi = {10.1093/mnras/stz1488},
archivePrefix = {arXiv},
       eprint = {1905.12639},
 primaryClass = {astro-ph.EP},
       adsurl = {https://ui.adsabs.harvard.edu/abs/2019MNRAS.487.3998B}
}

@ARTICLE{2020Sci...367.7462P,
       author = {{Poch}, Olivier and {Istiqomah}, Istiqomah and {Quirico}, Eric and {Beck}, Pierre and {Schmitt}, Bernard and {Theul{\'e}}, Patrice and {Faure}, Alexandre and {Hily-Blant}, Pierre and {Bonal}, Lydie and {Raponi}, Andrea and {Ciarniello}, Mauro and {Rousseau}, Batiste and {Potin}, Sandra and {Brissaud}, Olivier and {Flandinet}, Laur{\`e}ne and {Filacchione}, Gianrico and {Pommerol}, Antoine and {Thomas}, Nicolas and {Kappel}, David and {Mennella}, Vito and {Moroz}, Lyuba and {Vinogradoff}, Vassilissa and {Arnold}, Gabriele and {Erard}, St{\'e}phane and {Bockel{\'e}e-Morvan}, Dominique and {Leyrat}, C{\'e}dric and {Capaccioni}, Fabrizio and {De Sanctis}, Maria Cristina and {Longobardo}, Andrea and {Mancarella}, Francesca and {Palomba}, Ernesto and {Tosi}, Federico},
        title = "{Ammonium salts are a reservoir of nitrogen on a cometary nucleus and possibly on some asteroids}",
      journal = {Science},
         year = 2020,
        month = mar,
       volume = {367},
       number = {6483},
          eid = {aaw7462},
        pages = {aaw7462},
          doi = {10.1126/science.aaw7462},
archivePrefix = {arXiv},
       eprint = {2003.06034},
 primaryClass = {astro-ph.EP},
       adsurl = {https://ui.adsabs.harvard.edu/abs/2020Sci...367.7462P}
}

@ARTICLE{2022MNRAS.516.3900A,
       author = {{Altwegg}, K. and {Combi}, M. and {Fuselier}, S.~A. and {H{\"a}nni}, N. and {De Keyser}, J. and {Mahjoub}, A. and {M{\"u}ller}, D.~R. and {Pestoni}, B. and {Rubin}, M. and {Wampfler}, S.~F.},
        title = "{Abundant ammonium hydrosulphide embedded in cometary dust grains}",
      journal = {\mnras},
         year = 2022,
        month = nov,
       volume = {516},
       number = {3},
        pages = {3900-3910},
          doi = {10.1093/mnras/stac2440},
archivePrefix = {arXiv},
       eprint = {2208.11396},
 primaryClass = {astro-ph.EP},
       adsurl = {https://ui.adsabs.harvard.edu/abs/2022MNRAS.516.3900A}
}

@ARTICLE{2023NatAs.tmp...25M,
       author = {{McClure}, M.~K. and {Rocha}, W.~R.~M. and {Pontoppidan}, K.~M. and {Crouzet}, N. and {Chu}, L.~E.~U. and {Dartois}, E. and {Lamberts}, T. and {Noble}, J.~A. and {Pendleton}, Y.~J. and {Perotti}, G. and {Qasim}, D. and {Rachid}, M.~G. and {Smith}, Z.~L. and {Sun}, Fengwu and {Beck}, Tracy L. and {Boogert}, A.~C.~A. and {Brown}, W.~A. and {Caselli}, P. and {Charnley}, S.~B. and {Cuppen}, Herma M. and {Dickinson}, H. and {Drozdovskaya}, M.~N. and {Egami}, E. and {Erkal}, J. and {Fraser}, H. and {Garrod}, R.~T. and {Harsono}, D. and {Ioppolo}, S. and {Jim{\'e}nez-Serra}, I. and {Jin}, M. and {J{\o}rgensen}, J.~K. and {Kristensen}, L.~E. and {Lis}, D.~C. and {McCoustra}, M.~R.~S. and {McGuire}, Brett A. and {Melnick}, G.~J. and {{\"O}berg}, Karin I. and {Palumbo}, M.~E. and {Shimonishi}, T. and {Sturm}, J.~A. and {van Dishoeck}, E.~F. and {Linnartz}, H.},
        title = "{An Ice Age JWST inventory of dense molecular cloud ices}",
      journal = {Nature Astronomy},
         year = 2023,
        month = jan,
          doi = {10.1038/s41550-022-01875-w},
archivePrefix = {arXiv},
       eprint = {2301.09140},
 primaryClass = {astro-ph.GA},
       adsurl = {https://ui.adsabs.harvard.edu/abs/2023NatAs.tmp...25M}
}

@ARTICLE{2002Icar..155..393L,
       author = {{Lodders}, Katharina and {Fegley}, Bruce},
        title = "{Atmospheric Chemistry in Giant Planets, Brown Dwarfs, and Low-Mass Dwarf Stars. I. Carbon, Nitrogen, and Oxygen}",
      journal = {\icarus},
         year = 2002,
        month = feb,
       volume = {155},
       number = {2},
        pages = {393-424},
          doi = {10.1006/icar.2001.6740},
       adsurl = {https://ui.adsabs.harvard.edu/abs/2002Icar..155..393L}
}

@ARTICLE{2011A&A...535A..47D,
       author = {{Danger}, G. and {Borget}, F. and {Chomat}, M. and {Duvernay}, F. and {Theul{\'e}}, P. and {Guillemin}, J. -C. and {Le Sergeant D'Hendecourt}, L. and {Chiavassa}, T.},
        title = "{Experimental investigation of aminoacetonitrile formation through the Strecker synthesis in astrophysical-like conditions: reactivity of methanimine (CH$_{2}$NH), ammonia (NH$_{3}$), and hydrogen cyanide (HCN)}",
      journal = {\aap},
         year = 2011,
        month = nov,
       volume = {535},
          eid = {A47},
        pages = {A47},
          doi = {10.1051/0004-6361/201117602},
       adsurl = {https://ui.adsabs.harvard.edu/abs/2011A&A...535A..47D}
}

@ARTICLE{2016ApJ...829...85B,
       author = {{Bergner}, Jennifer B. and {{\"O}berg}, Karin I. and {Rajappan}, Mahesh and {Fayolle}, Edith C.},
        title = "{Kinetics and Mechanisms of the Acid-base Reaction Between NH$_{3}$ and HCOOH in Interstellar Ice Analogs}",
      journal = {\apj},
         year = 2016,
        month = oct,
       volume = {829},
       number = {2},
          eid = {85},
        pages = {85},
          doi = {10.3847/0004-637X/829/2/85},
archivePrefix = {arXiv},
       eprint = {1608.00010},
 primaryClass = {astro-ph.IM},
       adsurl = {https://ui.adsabs.harvard.edu/abs/2016ApJ...829...85B}
}

@ARTICLE{2019ApJ...878L..20P,
       author = {{Potapov}, Alexey and {Theul{\'e}}, Patrice and {J{\"a}ger}, Cornelia and {Henning}, Thomas},
        title = "{Evidence of Surface Catalytic Effect on Cosmic Dust Grain Analogs: The Ammonia and Carbon Dioxide Surface Reaction}",
      journal = {\apjl},
         year = 2019,
        month = jun,
       volume = {878},
       number = {1},
          eid = {L20},
        pages = {L20},
          doi = {10.3847/2041-8213/ab2538},
archivePrefix = {arXiv},
       eprint = {1905.13471},
 primaryClass = {astro-ph.GA},
       adsurl = {https://ui.adsabs.harvard.edu/abs/2019ApJ...878L..20P}
}

@ARTICLE{1974MNRAS.168..603L,
       author = {{Lynden-Bell}, D. and {Pringle}, J.~E.},
        title = "{The evolution of viscous discs and the origin of the nebular variables.}",
      journal = {\mnras},
         year = 1974,
        month = sep,
       volume = {168},
        pages = {603-637},
          doi = {10.1093/mnras/168.3.603},
       adsurl = {https://ui.adsabs.harvard.edu/abs/1974MNRAS.168..603L}
}

@ARTICLE{1998ApJ...495..385H,
       author = {{Hartmann}, Lee and {Calvet}, Nuria and {Gullbring}, Erik and {D'Alessio}, Paola},
        title = "{Accretion and the Evolution of T Tauri Disks}",
      journal = {\apj},
         year = 1998,
        month = mar,
       volume = {495},
       number = {1},
        pages = {385-400},
          doi = {10.1086/305277},
       adsurl = {https://ui.adsabs.harvard.edu/abs/1998ApJ...495..385H}
}

@ARTICLE{1973A&A....24..337S,
       author = {{Shakura}, N.~I. and {Sunyaev}, R.~A.},
        title = "{Reprint of 1973A\&A....24..337S. Black holes in binary systems. Observational appearance.}",
      journal = {\aap},
         year = 1973,
        month = jun,
       volume = {500},
        pages = {33-51},
       adsurl = {https://ui.adsabs.harvard.edu/abs/1973A&A....24..337S}
}

@ARTICLE{2021ApJ...916...72M,
       author = {{Mori}, Shoji and {Okuzumi}, Satoshi and {Kunitomo}, Masanobu and {Bai}, Xue-Ning},
        title = "{Evolution of the Water Snow Line in Magnetically Accreting Protoplanetary Disks}",
      journal = {\apj},
         year = 2021,
        month = aug,
       volume = {916},
       number = {2},
          eid = {72},
        pages = {72},
          doi = {10.3847/1538-4357/ac06a9},
archivePrefix = {arXiv},
       eprint = {2105.13101},
 primaryClass = {astro-ph.EP},
       adsurl = {https://ui.adsabs.harvard.edu/abs/2021ApJ...916...72M}
}

@ARTICLE{1970PThPh..44.1580K,
       author = {{Kusaka}, T. and {Nakano}, T. and {Hayashi}, C.},
        title = "{Growth of Solid Particles in the Primordial Solar Nebula}",
      journal = {Progress of Theoretical Physics},
         year = 1970,
        month = dec,
       volume = {44},
       number = {6},
        pages = {1580-1595},
          doi = {10.1143/PTP.44.1580},
       adsurl = {https://ui.adsabs.harvard.edu/abs/1970PThPh..44.1580K}
}

@ARTICLE{1997ApJ...490..368C,
       author = {{Chiang}, E.~I. and {Goldreich}, P.},
        title = "{Spectral Energy Distributions of T Tauri Stars with Passive Circumstellar Disks}",
      journal = {\apj},
         year = 1997,
        month = nov,
       volume = {490},
       number = {1},
        pages = {368-376},
          doi = {10.1086/304869},
archivePrefix = {arXiv},
       eprint = {astro-ph/9706042},
 primaryClass = {astro-ph},
       adsurl = {https://ui.adsabs.harvard.edu/abs/1997ApJ...490..368C}
}

@ARTICLE{2016A&A...593A..99F,
       author = {{Feiden}, Gregory A.},
        title = "{Magnetic inhibition of convection and the fundamental properties of low-mass stars. III. A consistent 10 Myr age for the Upper Scorpius OB association}",
      journal = {\aap},
         year = 2016,
        month = sep,
       volume = {593},
          eid = {A99},
        pages = {A99},
          doi = {10.1051/0004-6361/201527613},
archivePrefix = {arXiv},
       eprint = {1604.08036},
 primaryClass = {astro-ph.SR},
       adsurl = {https://ui.adsabs.harvard.edu/abs/2016A&A...593A..99F}
}

@ARTICLE{2023ApJ...949..119K,
       author = {{Kondo}, Katsushi and {Okuzumi}, Satoshi and {Mori}, Shoji},
        title = "{The Roles of Dust Growth in the Temperature Evolution and Snow Line Migration in Magnetically Accreting Protoplanetary Disks}",
      journal = {\apj},
         year = 2023,
        month = jun,
       volume = {949},
       number = {2},
          eid = {119},
        pages = {119},
          doi = {10.3847/1538-4357/acc840},
archivePrefix = {arXiv},
       eprint = {2205.13511},
 primaryClass = {astro-ph.EP},
       adsurl = {https://ui.adsabs.harvard.edu/abs/2023ApJ...949..119K}
}

@ARTICLE{1985Icar...64..471P,
       author = {{Pollack}, J.~B. and {McKay}, C.~P. and {Christofferson}, B.~M.},
        title = "{A calculation of the Rosseland mean opacity of dust grains in primordial solar system nebulae}",
      journal = {\icarus},
         year = 1985,
        month = dec,
       volume = {64},
       number = {3},
        pages = {471-492},
          doi = {10.1016/0019-1035(85)90069-7},
       adsurl = {https://ui.adsabs.harvard.edu/abs/1985Icar...64..471P}
}

@ARTICLE{1994ApJ...421..640N,
       author = {{Nakamoto}, Taishi and {Nakagawa}, Yoshitsugo},
        title = "{Formation, Early Evolution, and Gravitational Stability of Protoplanetary Disks}",
      journal = {\apj},
         year = 1994,
        month = feb,
       volume = {421},
        pages = {640},
          doi = {10.1086/173678},
       adsurl = {https://ui.adsabs.harvard.edu/abs/1994ApJ...421..640N}
}

@ARTICLE{2016A&A...589A..15S,
       author = {{Sato}, Takao and {Okuzumi}, Satoshi and {Ida}, Shigeru},
        title = "{On the water delivery to terrestrial embryos by ice pebble accretion}",
      journal = {\aap},
         year = 2016,
        month = may,
       volume = {589},
          eid = {A15},
        pages = {A15},
          doi = {10.1051/0004-6361/201527069},
archivePrefix = {arXiv},
       eprint = {1512.02414},
 primaryClass = {astro-ph.EP},
       adsurl = {https://ui.adsabs.harvard.edu/abs/2016A&A...589A..15S}
}

@ARTICLE{2021A&A...646A..14H,
       author = {{Hyodo}, Ryuki and {Guillot}, Tristan and {Ida}, Shigeru and {Okuzumi}, Satoshi and {Youdin}, Andrew N.},
        title = "{Planetesimal formation around the snow line. II. Dust or pebbles?}",
      journal = {\aap},
         year = 2021,
        month = feb,
       volume = {646},
          eid = {A14},
        pages = {A14},
          doi = {10.1051/0004-6361/202039894},
archivePrefix = {arXiv},
       eprint = {2012.06700},
 primaryClass = {astro-ph.EP},
       adsurl = {https://ui.adsabs.harvard.edu/abs/2021A&A...646A..14H}
}

@ARTICLE{2021AandA...653A.141A,
       author = {{Asplund}, M. and {Amarsi}, A.~M. and {Grevesse}, N.},
        title = "{The chemical make-up of the Sun: A 2020 vision}",
      journal = {\aap},
         year = 2021,
        month = sep,
       volume = {653},
          eid = {A141},
        pages = {A141},
          doi = {10.1051/0004-6361/202140445},
archivePrefix = {arXiv},
       eprint = {2105.01661},
 primaryClass = {astro-ph.SR},
       adsurl = {https://ui.adsabs.harvard.edu/abs/2021A&A...653A.141A}
}

@ARTICLE{2017ApJS..228...20T,
       author = {{Tsai}, Shang-Min and {Lyons}, James R. and {Grosheintz}, Luc and {Rimmer}, Paul B. and {Kitzmann}, Daniel and {Heng}, Kevin},
        title = "{VULCAN: An Open-source, Validated Chemical Kinetics Python Code for Exoplanetary Atmospheres}",
      journal = {\apjs},
         year = 2017,
        month = feb,
       volume = {228},
       number = {2},
          eid = {20},
        pages = {20},
          doi = {10.3847/1538-4365/228/2/20},
archivePrefix = {arXiv},
       eprint = {1607.00409},
 primaryClass = {astro-ph.EP},
       adsurl = {https://ui.adsabs.harvard.edu/abs/2017ApJS..228...20T}
}

@ARTICLE{2022ExA....53..279M,
       author = {{Moses}, Julianne I. and {Tremblin}, Pascal and {Venot}, Olivia and {Miguel}, Yamila},
        title = "{Chemical variation with altitude and longitude on exo-Neptunes: Predictions for Ariel phase-curve observations}",
      journal = {Experimental Astronomy},
         year = 2022,
        month = apr,
       volume = {53},
       number = {2},
        pages = {279-322},
          doi = {10.1007/s10686-021-09749-1},
archivePrefix = {arXiv},
       eprint = {2103.07023},
 primaryClass = {astro-ph.EP},
       adsurl = {https://ui.adsabs.harvard.edu/abs/2022ExA....53..279M}
}

@ARTICLE{2023ApJ...946...18O,
       author = {{Ohno}, Kazumasa and {Fortney}, Jonathan J.},
        title = "{Nitrogen as a Tracer of Giant Planet Formation. I. A Universal Deep Adiabatic Profile and Semianalytical Predictions of Disequilibrium Ammonia Abundances in Warm Exoplanetary Atmospheres}",
      journal = {\apj},
         year = 2023,
        month = mar,
       volume = {946},
       number = {1},
          eid = {18},
        pages = {18},
          doi = {10.3847/1538-4357/acafed},
archivePrefix = {arXiv},
       eprint = {2211.16876},
 primaryClass = {astro-ph.EP},
       adsurl = {https://ui.adsabs.harvard.edu/abs/2023ApJ...946...18O}
}

@ARTICLE{2019A&A...627A..67M,
       author = {{Molli{\`e}re}, P. and {Wardenier}, J.~P. and {van Boekel}, R. and {Henning}, Th. and {Molaverdikhani}, K. and {Snellen}, I.~A.~G.},
        title = "{petitRADTRANS. A Python radiative transfer package for exoplanet characterization and retrieval}",
      journal = {\aap},
         year = 2019,
        month = jul,
       volume = {627},
          eid = {A67},
        pages = {A67},
          doi = {10.1051/0004-6361/201935470},
archivePrefix = {arXiv},
       eprint = {1904.11504},
 primaryClass = {astro-ph.EP},
       adsurl = {https://ui.adsabs.harvard.edu/abs/2019A&A...627A..67M}
}

@ARTICLE{2020SciPy-NMeth,
  author  = {Virtanen, Pauli and Gommers, Ralf and Oliphant, Travis E. and
            Haberland, Matt and Reddy, Tyler and Cournapeau, David and
            Burovski, Evgeni and Peterson, Pearu and Weckesser, Warren and
            Bright, Jonathan and {van der Walt}, St{\'e}fan J. and
            Brett, Matthew and Wilson, Joshua and Millman, K. Jarrod and
            Mayorov, Nikolay and Nelson, Andrew R. J. and Jones, Eric and
            Kern, Robert and Larson, Eric and Carey, C J and
            Polat, {\.I}lhan and Feng, Yu and Moore, Eric W. and
            {VanderPlas}, Jake and Laxalde, Denis and Perktold, Josef and
            Cimrman, Robert and Henriksen, Ian and Quintero, E. A. and
            Harris, Charles R. and Archibald, Anne M. and
            Ribeiro, Ant{\^o}nio H. and Pedregosa, Fabian and
            {van Mulbregt}, Paul and {SciPy 1.0 Contributors}},
  title   = {{{SciPy} 1.0: Fundamental Algorithms for Scientific
            Computing in Python}},
  journal = {Nature Methods},
  year    = {2020},
  volume  = {17},
  pages   = {261--272},
  adsurl  = {https://rdcu.be/b08Wh},
  doi     = {10.1038/s41592-019-0686-2},
}

@ARTICLE{2016JMoSp.327...73T,
       author = {{Tennyson}, Jonathan and {Yurchenko}, Sergei N. and {Al-Refaie}, Ahmed F. and {Barton}, Emma J. and {Chubb}, Katy L. and {Coles}, Phillip A. and {Diamantopoulou}, S. and {Gorman}, Maire N. and {Hill}, Christian and {Lam}, Aden Z. and {Lodi}, Lorenzo and {McKemmish}, Laura K. and {Na}, Yueqi and {Owens}, Alec and {Polyansky}, Oleg L. and {Rivlin}, Tom and {Sousa-Silva}, Clara and {Underwood}, Daniel S. and {Yachmenev}, Andrey and {Zak}, Emil},
        title = "{The ExoMol database: Molecular line lists for exoplanet and other hot atmospheres}",
      journal = {Journal of Molecular Spectroscopy},
         year = 2016,
        month = sep,
       volume = {327},
        pages = {73-94},
          doi = {10.1016/j.jms.2016.05.002},
archivePrefix = {arXiv},
       eprint = {1603.05890},
 primaryClass = {astro-ph.GA},
       adsurl = {https://ui.adsabs.harvard.edu/abs/2016JMoSp.327...73T}
}

@ARTICLE{2025A&A...693A.146S,
       author = {{Slavicinska}, K. and {Boogert}, A.~C.~A. and {Tychoniec}, {\L}. and {van Dishoeck}, E.~F. and {van Gelder}, M.~L. and {Navarro}, M.~G. and {Santos}, J.~C. and {Klaassen}, P.~D. and {Kavanagh}, P.~J. and {Chuang}, K.-J.},
        title = "{Ammonium hydrosulfide (NH$_{4}$SH) as a potentially significant sulfur sink in interstellar ices}",
      journal = {\aap},
         year = 2025,
        month = jan,
       volume = {693},
          eid = {A146},
        pages = {A146},
          doi = {10.1051/0004-6361/202451383},
archivePrefix = {arXiv},
       eprint = {2410.02860},
 primaryClass = {astro-ph.GA},
       adsurl = {https://ui.adsabs.harvard.edu/abs/2025A&A...693A.146S}
}

@ARTICLE{2025PASJ..tmp...29N,
       author = {{Nakazawa}, Kanon and {Okuzumi}, Satoshi},
        title = "{Nitrogen transport in protoplanetary disks by ammonium salts: A possible origin of Jupiter's nitrogen enrichment}",
      journal = {\pasj},
         year = 2025,
        month = apr,
          doi = {10.1093/pasj/psaf021},
archivePrefix = {arXiv},
       eprint = {2410.05743},
 primaryClass = {astro-ph.EP},
       adsurl = {https://ui.adsabs.harvard.edu/abs/2025PASJ..tmp...29N}
}

@ARTICLE{2017A&A...602A..21S,
       author = {{Schoonenberg}, Djoeke and {Ormel}, Chris W.},
        title = "{Planetesimal formation near the snowline: in or out?}",
      journal = {\aap},
         year = 2017,
        month = jun,
       volume = {602},
          eid = {A21},
        pages = {A21},
          doi = {10.1051/0004-6361/201630013},
archivePrefix = {arXiv},
       eprint = {1702.02151},
 primaryClass = {astro-ph.EP},
       adsurl = {https://ui.adsabs.harvard.edu/abs/2017A&A...602A..21S}
}

@ARTICLE{2011MNRAS.416.1419H,
       author = {{Hori}, Y. and {Ikoma}, M.},
        title = "{Gas giant formation with small cores triggered by envelope pollution by icy planetesimals}",
      journal = {\mnras},
         year = 2011,
        month = sep,
       volume = {416},
       number = {2},
        pages = {1419-1429},
          doi = {10.1111/j.1365-2966.2011.19140.x},
archivePrefix = {arXiv},
       eprint = {1106.2626},
 primaryClass = {astro-ph.EP},
       adsurl = {https://ui.adsabs.harvard.edu/abs/2011MNRAS.416.1419H}
}

@ARTICLE{2005Icar..175....1P,
       author = {{Pasek}, Matthew A. and {Milsom}, John A. and {Ciesla}, Fred J. and {Lauretta}, Dante S. and {Sharp}, Christopher M. and {Lunine}, Jonathan I.},
        title = "{Sulfur chemistry with time-varying oxygen abundance during Solar System formation}",
      journal = {\icarus},
         year = 2005,
        month = may,
       volume = {175},
       number = {1},
        pages = {1-14},
          doi = {10.1016/j.icarus.2004.10.012},
       adsurl = {https://ui.adsabs.harvard.edu/abs/2005Icar..175....1P}
}

@ARTICLE{2011A&A...536A..91J,
       author = {{Jim{\'e}nez-Escobar}, A. and {Mu{\~n}oz Caro}, G.~M.},
        title = "{Sulfur depletion in dense clouds and circumstellar regions. I. H$_{2}$S ice abundance and UV-photochemical reactions in the H$_{2}$O-matrix}",
      journal = {\aap},
         year = 2011,
        month = dec,
       volume = {536},
          eid = {A91},
        pages = {A91},
          doi = {10.1051/0004-6361/201014821},
archivePrefix = {arXiv},
       eprint = {1112.3240},
 primaryClass = {astro-ph.EP},
       adsurl = {https://ui.adsabs.harvard.edu/abs/2011A&A...536A..91J}
}

@ARTICLE{2019ApJ...885..114K,
       author = {{Kama}, Mihkel and {Shorttle}, Oliver and {Jermyn}, Adam S. and {Folsom}, Colin P. and {Furuya}, Kenji and {Bergin}, Edwin A. and {Walsh}, Catherine and {Keller}, Lindsay},
        title = "{Abundant Refractory Sulfur in Protoplanetary Disks}",
      journal = {\apj},
         year = 2019,
        month = nov,
       volume = {885},
       number = {2},
          eid = {114},
        pages = {114},
          doi = {10.3847/1538-4357/ab45f8},
archivePrefix = {arXiv},
       eprint = {1908.05169},
 primaryClass = {astro-ph.EP},
       adsurl = {https://ui.adsabs.harvard.edu/abs/2019ApJ...885..114K}
}

@ARTICLE{2021ApJS..257...12L,
       author = {{Le Gal}, Romane and {{\"O}berg}, Karin I. and {Teague}, Richard and {Loomis}, Ryan A. and {Law}, Charles J. and {Walsh}, Catherine and {Bergin}, Edwin A. and {M{\'e}nard}, Fran{\c{c}}ois and {Wilner}, David J. and {Andrews}, Sean M. and {Aikawa}, Yuri and {Booth}, Alice S. and {Cataldi}, Gianni and {Bergner}, Jennifer B. and {Bosman}, Arthur D. and {Cleeves}, L. Ilse and {Czekala}, Ian and {Furuya}, Kenji and {Guzm{\'a}n}, Viviana V. and {Huang}, Jane and {Ilee}, John D. and {Nomura}, Hideko and {Qi}, Chunhua and {Schwarz}, Kamber R. and {Tsukagoshi}, Takashi and {Yamato}, Yoshihide and {Zhang}, Ke},
        title = "{Molecules with ALMA at Planet-forming Scales (MAPS). XII. Inferring the C/O and S/H Ratios in Protoplanetary Disks with Sulfur Molecules}",
      journal = {\apjs},
         year = 2021,
        month = nov,
       volume = {257},
       number = {1},
          eid = {12},
        pages = {12},
          doi = {10.3847/1538-4365/ac2583},
archivePrefix = {arXiv},
       eprint = {2109.06286},
 primaryClass = {astro-ph.GA},
       adsurl = {https://ui.adsabs.harvard.edu/abs/2021ApJS..257...12L}
}

@ARTICLE{2021ApJ...909...40T,
       author = {{Turrini}, D. and {Schisano}, E. and {Fonte}, S. and {Molinari}, S. and {Politi}, R. and {Fedele}, D. and {Pani{\'c}}, O. and {Kama}, M. and {Changeat}, Q. and {Tinetti}, G.},
        title = "{Tracing the Formation History of Giant Planets in Protoplanetary Disks with Carbon, Oxygen, Nitrogen, and Sulfur}",
      journal = {\apj},
         year = 2021,
        month = mar,
       volume = {909},
       number = {1},
          eid = {40},
        pages = {40},
          doi = {10.3847/1538-4357/abd6e5},
archivePrefix = {arXiv},
       eprint = {2012.14315},
 primaryClass = {astro-ph.EP},
       adsurl = {https://ui.adsabs.harvard.edu/abs/2021ApJ...909...40T}
}

@ARTICLE{2021PhR...893....1O,
       author = {{{\"O}berg}, Karin I. and {Bergin}, Edwin A.},
        title = "{Astrochemistry and compositions of planetary systems}",
      journal = {\physrep},
         year = 2021,
        month = jan,
       volume = {893},
        pages = {1-48},
          doi = {10.1016/j.physrep.2020.09.004},
archivePrefix = {arXiv},
       eprint = {2010.03529},
 primaryClass = {astro-ph.EP},
       adsurl = {https://ui.adsabs.harvard.edu/abs/2021PhR...893....1O}
}

@ARTICLE{2024Natur.630..836W,
       author = {{Welbanks}, Luis and {Bell}, Taylor J. and {Beatty}, Thomas G. and {Line}, Michael R. and {Ohno}, Kazumasa and {Fortney}, Jonathan J. and {Schlawin}, Everett and {Greene}, Thomas P. and {Rauscher}, Emily and {McGill}, Peter and {Murphy}, Matthew and {Parmentier}, Vivien and {Tang}, Yao and {Edelman}, Isaac and {Mukherjee}, Sagnick and {Wiser}, Lindsey S. and {Lagage}, Pierre-Olivier and {Dyrek}, Achr{\`e}ne and {Arnold}, Kenneth E.},
        title = "{A high internal heat flux and large core in a warm Neptune exoplanet}",
      journal = {\nat},
         year = 2024,
        month = jun,
       volume = {630},
       number = {8018},
        pages = {836-840},
          doi = {10.1038/s41586-024-07514-w},
archivePrefix = {arXiv},
       eprint = {2405.11018},
 primaryClass = {astro-ph.EP},
       adsurl = {https://ui.adsabs.harvard.edu/abs/2024Natur.630..836W}
}

@ARTICLE{1991Icar...90..319L,
       author = {{Lichtenegger}, H.~I.~M. and {Komle}, N.~I.},
        title = "{Heating and evaporation of Icy particles in the vicinity of comets}",
      journal = {\icarus},
         year = 1991,
        month = apr,
       volume = {90},
       number = {2},
        pages = {319-325},
          doi = {10.1016/0019-1035(91)90110-F},
       adsurl = {https://ui.adsabs.harvard.edu/abs/1991Icar...90..319L}
}

@ARTICLE{2009P&SS...57.2053F,
       author = {{Fray}, N. and {Schmitt}, B.},
        title = "{Sublimation of ices of astrophysical interest: A bibliographic review}",
      journal = {\planss},
         year = 2009,
        month = dec,
       volume = {57},
       number = {14-15},
        pages = {2053-2080},
          doi = {10.1016/j.pss.2009.09.011},
       adsurl = {https://ui.adsabs.harvard.edu/abs/2009P&SS...57.2053F}
}

@ARTICLE{2018A&A...614A...1W,
       author = {{Woitke}, P. and {Helling}, Ch. and {Hunter}, G.~H. and {Millard}, J.~D. and {Turner}, G.~E. and {Worters}, M. and {Blecic}, J. and {Stock}, J.~W.},
        title = "{Equilibrium chemistry down to 100 K. Impact of silicates and phyllosilicates on the carbon to oxygen ratio}",
      journal = {\aap},
         year = 2018,
        month = jun,
       volume = {614},
          eid = {A1},
        pages = {A1},
          doi = {10.1051/0004-6361/201732193},
archivePrefix = {arXiv},
       eprint = {1712.01010},
 primaryClass = {astro-ph.EP},
       adsurl = {https://ui.adsabs.harvard.edu/abs/2018A&A...614A...1W}
}

@ARTICLE{2020NatAs...4..533A,
       author = {{Altwegg}, Kathrin and {Balsiger}, Hans and {H{\"a}nni}, Nora and {Rubin}, Martin and {Schuhmann}, Markus and {Schroeder}, Isaac and {S{\'e}mon}, Thierry and {Wampfler}, Susanne and {Berthelier}, Jean-Jacques and {Briois}, Christelle and {Combi}, Mike and {Gombosi}, Tamas I. and {Cottin}, Herv{\'e} and {De Keyser}, Johan and {Dhooghe}, Frederik and {Fiethe}, Bj{\"o}rn and {Fuselier}, Steven A.},
        title = "{Evidence of ammonium salts in comet 67P as explanation for the nitrogen depletion in cometary comae}",
      journal = {Nature Astronomy},
         year = 2020,
        month = jan,
       volume = {4},
        pages = {533-540},
          doi = {10.1038/s41550-019-0991-9},
archivePrefix = {arXiv},
       eprint = {1911.13005},
 primaryClass = {astro-ph.EP},
       adsurl = {https://ui.adsabs.harvard.edu/abs/2020NatAs...4..533A}
}

@ARTICLE{2020SSRv..216...77M,
       author = {{Mousis}, O. and {Aguichine}, A. and {Atkinson}, D.~H. and {Atreya}, S.~K. and {Cavali{\'e}}, T. and {Lunine}, J.~I. and {Mandt}, K.~E. and {Ronnet}, T.},
        title = "{Key Atmospheric Signatures for Identifying the Source Reservoirs of Volatiles in Uranus and Neptune}",
      journal = {\ssr},
         year = 2020,
        month = jun,
       volume = {216},
       number = {5},
          eid = {77},
        pages = {77},
          doi = {10.1007/s11214-020-00681-y},
archivePrefix = {arXiv},
       eprint = {2004.11061},
 primaryClass = {astro-ph.EP},
       adsurl = {https://ui.adsabs.harvard.edu/abs/2020SSRv..216...77M}
}

@ARTICLE{2011ARAandA..49..471M,
       author = {{Mumma}, Michael J. and {Charnley}, Steven B.},
        title = "{The Chemical Composition of Comets{\textemdash}Emerging Taxonomies and Natal Heritage}",
      journal = {\araa},
         year = 2011,
        month = sep,
       volume = {49},
       number = {1},
        pages = {471-524},
          doi = {10.1146/annurev-astro-081309-130811},
       adsurl = {https://ui.adsabs.harvard.edu/abs/2011ARA&A..49..471M}
}

@ARTICLE{2007Icar..190..655B,
       author = {{Bar-Nun}, A. and {Notesco}, G. and {Owen}, T.},
        title = "{Trapping of N $_{2}$, CO and Ar in amorphous ice{\textemdash}Application to comets}",
      journal = {\icarus},
         year = 2007,
        month = oct,
       volume = {190},
       number = {2},
        pages = {655-659},
          doi = {10.1016/j.icarus.2007.03.021},
       adsurl = {https://ui.adsabs.harvard.edu/abs/2007Icar..190..655B}
}

@ARTICLE{2023ApJ...955....5S,
       author = {{Simon}, Alexia and {Rajappan}, Mahesh and {{\"O}berg}, Karin I.},
        title = "{Entrapment of Hypervolatiles in Interstellar and Cometary H$_{2}$O and CO$_{2}$ Ice Analogs}",
      journal = {\apj},
         year = 2023,
        month = sep,
       volume = {955},
       number = {1},
          eid = {5},
        pages = {5},
          doi = {10.3847/1538-4357/aceaf8},
       adsurl = {https://ui.adsabs.harvard.edu/abs/2023ApJ...955....5S}
}

@ARTICLE{2018A&A...610L..14V,
       author = {{Vazan}, Allona and {Helled}, Ravit and {Guillot}, Tristan},
        title = "{Jupiter's evolution with primordial composition gradients}",
      journal = {\aap},
         year = 2018,
        month = feb,
       volume = {610},
          eid = {L14},
        pages = {L14},
          doi = {10.1051/0004-6361/201732522},
archivePrefix = {arXiv},
       eprint = {1801.08149},
 primaryClass = {astro-ph.EP},
       adsurl = {https://ui.adsabs.harvard.edu/abs/2018A&A...610L..14V}
}

@ARTICLE{2019ApJ...872..100D,
       author = {{Debras}, Florian and {Chabrier}, Gilles},
        title = "{New Models of Jupiter in the Context of Juno and Galileo}",
      journal = {\apj},
         year = 2019,
        month = feb,
       volume = {872},
       number = {1},
          eid = {100},
        pages = {100},
          doi = {10.3847/1538-4357/aaff65},
archivePrefix = {arXiv},
       eprint = {1901.05697},
 primaryClass = {astro-ph.EP},
       adsurl = {https://ui.adsabs.harvard.edu/abs/2019ApJ...872..100D}
}

@ARTICLE{2022Icar..37814937H,
       author = {{Helled}, Ravit and {Stevenson}, David J. and {Lunine}, Jonathan I. and {Bolton}, Scott J. and {Nettelmann}, Nadine and {Atreya}, Sushil and {Guillot}, Tristan and {Militzer}, Burkhard and {Miguel}, Yamila and {Hubbard}, William B.},
        title = "{Revelations on Jupiter's formation, evolution and interior: Challenges from Juno results}",
      journal = {\icarus},
         year = 2022,
        month = may,
       volume = {378},
          eid = {114937},
        pages = {114937},
          doi = {10.1016/j.icarus.2022.114937},
archivePrefix = {arXiv},
       eprint = {2202.10041},
 primaryClass = {astro-ph.EP},
       adsurl = {https://ui.adsabs.harvard.edu/abs/2022Icar..37814937H}
}

@ARTICLE{2016ApJ...823...48T,
       author = {{Tanigawa}, Takayuki and {Tanaka}, Hidekazu},
        title = "{Final Masses of Giant Planets. II. Jupiter Formation in a Gas-depleted Disk}",
      journal = {\apj},
         year = 2016,
        month = may,
       volume = {823},
       number = {1},
          eid = {48},
        pages = {48},
          doi = {10.3847/0004-637X/823/1/48},
archivePrefix = {arXiv},
       eprint = {1510.06848},
 primaryClass = {astro-ph.EP},
       adsurl = {https://ui.adsabs.harvard.edu/abs/2016ApJ...823...48T}
}

@ARTICLE{2020ApJ...891..143T,
       author = {{Tanaka}, Hidekazu and {Murase}, Kiyoka and {Tanigawa}, Takayuki},
        title = "{Final Masses of Giant Planets. III. Effect of Photoevaporation and a New Planetary Migration Model}",
      journal = {\apj},
         year = 2020,
        month = mar,
       volume = {891},
       number = {2},
          eid = {143},
        pages = {143},
          doi = {10.3847/1538-4357/ab77af},
archivePrefix = {arXiv},
       eprint = {1907.02627},
 primaryClass = {astro-ph.EP},
       adsurl = {https://ui.adsabs.harvard.edu/abs/2020ApJ...891..143T}
}

@ARTICLE{2016ApJ...820...89F,
       author = {{France}, Kevin and {Loyd}, R.~O. Parke and {Youngblood}, Allison and {Brown}, Alexander and {Schneider}, P. Christian and {Hawley}, Suzanne L. and {Froning}, Cynthia S. and {Linsky}, Jeffrey L. and {Roberge}, Aki and {Buccino}, Andrea P. and {Davenport}, James R.~A. and {Fontenla}, Juan M. and {Kaltenegger}, Lisa and {Kowalski}, Adam F. and {Mauas}, Pablo J.~D. and {Miguel}, Yamila and {Redfield}, Seth and {Rugheimer}, Sarah and {Tian}, Feng and {Vieytes}, Mariela C. and {Walkowicz}, Lucianne M. and {Weisenburger}, Kolby L.},
        title = "{The MUSCLES Treasury Survey. I. Motivation and Overview}",
      journal = {\apj},
         year = 2016,
        month = apr,
       volume = {820},
       number = {2},
          eid = {89},
        pages = {89},
          doi = {10.3847/0004-637X/820/2/89},
archivePrefix = {arXiv},
       eprint = {1602.09142},
 primaryClass = {astro-ph.SR},
       adsurl = {https://ui.adsabs.harvard.edu/abs/2016ApJ...820...89F}
}

@ARTICLE{2016ApJ...824..101Y,
       author = {{Youngblood}, Allison and {France}, Kevin and {Loyd}, R.~O. Parke and {Linsky}, Jeffrey L. and {Redfield}, Seth and {Schneider}, P. Christian and {Wood}, Brian E. and {Brown}, Alexander and {Froning}, Cynthia and {Miguel}, Yamila and {Rugheimer}, Sarah and {Walkowicz}, Lucianne},
        title = "{The MUSCLES Treasury Survey. II. Intrinsic LY{\ensuremath{\alpha}} and Extreme Ultraviolet Spectra of K and M Dwarfs with Exoplanets*}",
      journal = {\apj},
         year = 2016,
        month = jun,
       volume = {824},
       number = {2},
          eid = {101},
        pages = {101},
          doi = {10.3847/0004-637X/824/2/101},
archivePrefix = {arXiv},
       eprint = {1604.01032},
 primaryClass = {astro-ph.SR},
       adsurl = {https://ui.adsabs.harvard.edu/abs/2016ApJ...824..101Y}
}

@ARTICLE{2016ApJ...824..102L,
       author = {{Loyd}, R.~O.~P. and {France}, Kevin and {Youngblood}, Allison and {Schneider}, Christian and {Brown}, Alexander and {Hu}, Renyu and {Linsky}, Jeffrey and {Froning}, Cynthia S. and {Redfield}, Seth and {Rugheimer}, Sarah and {Tian}, Feng},
        title = "{The MUSCLES Treasury Survey. III. X-Ray to Infrared Spectra of 11 M and K Stars Hosting Planets}",
      journal = {\apj},
         year = 2016,
        month = jun,
       volume = {824},
       number = {2},
          eid = {102},
        pages = {102},
          doi = {10.3847/0004-637X/824/2/102},
archivePrefix = {arXiv},
       eprint = {1604.04776},
 primaryClass = {astro-ph.SR},
       adsurl = {https://ui.adsabs.harvard.edu/abs/2016ApJ...824..102L}
}

@ARTICLE{2024RvMG...90..411K,
       author = {{Kempton}, Eliza M. -R. and {Knutson}, Heather A.},
        title = "{Transiting Exoplanet Atmospheres in the Era of JWST}",
      journal = {Reviews in Mineralogy and Geochemistry},
         year = 2024,
        month = jul,
       volume = {90},
       number = {1},
        pages = {411-464},
          doi = {10.2138/rmg.2024.90.12},
archivePrefix = {arXiv},
       eprint = {2404.15430},
 primaryClass = {astro-ph.EP},
       adsurl = {https://ui.adsabs.harvard.edu/abs/2024RvMG...90..411K}
}

@ARTICLE{2025arXiv250601800W,
       author = {{Wiser}, Lindsey S. and {Bell}, Taylor J. and {Line}, Michael R. and {Schlawin}, Everett and {Beatty}, Thomas G. and {Welbanks}, Luis and {Greene}, Thomas P. and {Parmentier}, Vivien and {Murphy}, Matthew M. and {Fortney}, Jonathan J. and {Arnold}, Kenny and {Mehta}, Nishil and {Ohno}, Kazumasa and {Mukherjee}, Sagnick},
        title = "{A Precise Metallicity and Carbon-to-Oxygen Ratio for a Warm Giant Exoplanet from its Panchromatic JWST Emission Spectrum}",
      journal = {arXiv e-prints},
         year = 2025,
        month = jun,
          eid = {arXiv:2506.01800},
        pages = {arXiv:2506.01800},
          doi = {10.48550/arXiv.2506.01800},
archivePrefix = {arXiv},
       eprint = {2506.01800},
 primaryClass = {astro-ph.EP},
       adsurl = {https://ui.adsabs.harvard.edu/abs/2025arXiv250601800W}
}

@ARTICLE{2015ApJ...806L...7Z,
       author = {{Zhang}, Ke and {Blake}, Geoffrey A. and {Bergin}, Edwin A.},
        title = "{Evidence of Fast Pebble Growth Near Condensation Fronts in the HL Tau Protoplanetary Disk}",
      journal = {\apjl},
         year = 2015,
        month = jun,
       volume = {806},
       number = {1},
          eid = {L7},
        pages = {L7},
          doi = {10.1088/2041-8205/806/1/L7},
archivePrefix = {arXiv},
       eprint = {1505.00882},
 primaryClass = {astro-ph.EP},
       adsurl = {https://ui.adsabs.harvard.edu/abs/2015ApJ...806L...7Z}
}

@ARTICLE{2017A&A...608A..92D,
       author = {{Dr{\k{a}}{\.z}kowska}, J. and {Alibert}, Y.},
        title = "{Planetesimal formation starts at the snow line}",
      journal = {\aap},
         year = 2017,
        month = dec,
       volume = {608},
          eid = {A92},
        pages = {A92},
          doi = {10.1051/0004-6361/201731491},
archivePrefix = {arXiv},
       eprint = {1710.00009},
 primaryClass = {astro-ph.EP},
       adsurl = {https://ui.adsabs.harvard.edu/abs/2017A&A...608A..92D}
}

@ARTICLE{2019A&A...632L..11B,
       author = {{Bosman}, A.~D. and {Cridland}, A.~J. and {Miguel}, Y.},
        title = "{Jupiter formed as a pebble pile around the N$_{2}$ ice line}",
      journal = {\aap},
         year = 2019,
        month = dec,
       volume = {632},
          eid = {L11},
        pages = {L11},
          doi = {10.1051/0004-6361/201936827},
archivePrefix = {arXiv},
       eprint = {1911.11154},
 primaryClass = {astro-ph.EP},
       adsurl = {https://ui.adsabs.harvard.edu/abs/2019A&A...632L..11B}
}

@ARTICLE{2019AJ....158..194O,
       author = {{{\"O}berg}, Karin I. and {Wordsworth}, Robin},
        title = "{Jupiter's Composition Suggests its Core Assembled Exterior to the N$_{2}$ Snowline}",
      journal = {\aj},
         year = 2019,
        month = nov,
       volume = {158},
       number = {5},
          eid = {194},
        pages = {194},
          doi = {10.3847/1538-3881/ab46a8},
archivePrefix = {arXiv},
       eprint = {1909.11246},
 primaryClass = {astro-ph.EP},
       adsurl = {https://ui.adsabs.harvard.edu/abs/2019AJ....158..194O}
}

@ARTICLE{2021ApJ...914...12L,
       author = {{Lothringer}, Joshua D. and {Rustamkulov}, Zafar and {Sing}, David K. and {Gibson}, Neale P. and {Wilson}, Jamie and {Schlaufman}, Kevin C.},
        title = "{A New Window into Planet Formation and Migration: Refractory-to-Volatile Elemental Ratios in Ultra-hot Jupiters}",
      journal = {\apj},
         year = 2021,
        month = jun,
       volume = {914},
       number = {1},
          eid = {12},
        pages = {12},
          doi = {10.3847/1538-4357/abf8a9},
archivePrefix = {arXiv},
       eprint = {2011.10626},
 primaryClass = {astro-ph.EP},
       adsurl = {https://ui.adsabs.harvard.edu/abs/2021ApJ...914...12L}
}

@ARTICLE{2023ApJ...943..112C,
       author = {{Chachan}, Yayaati and {Knutson}, Heather A. and {Lothringer}, Joshua and {Blake}, Geoffrey A.},
        title = "{Breaking Degeneracies in Formation Histories by Measuring Refractory Content in Gas Giants}",
      journal = {\apj},
         year = 2023,
        month = feb,
       volume = {943},
       number = {2},
          eid = {112},
        pages = {112},
          doi = {10.3847/1538-4357/aca614},
archivePrefix = {arXiv},
       eprint = {2211.09080},
 primaryClass = {astro-ph.EP},
       adsurl = {https://ui.adsabs.harvard.edu/abs/2023ApJ...943..112C}
}

@ARTICLE{2025MNRAS.tmp..849K,
       author = {{Kama}, Mihkel and {Shorttle}, Oliver and {Borthakur}, Sandipan P.~D. and {Keyte}, Luke and {Bergner}, Jennifer B. and {Fossati}, Luca and {Folsom}, Colin P. and {Ramler}, Heleri},
        title = "{Refractory phosphorus in the HD 100546 protoplanetary disk}",
      journal = {\mnras},
         year = 2025,
        month = may,
          doi = {10.1093/mnras/staf823},
archivePrefix = {arXiv},
       eprint = {2504.14228},
 primaryClass = {astro-ph.EP},
       adsurl = {https://ui.adsabs.harvard.edu/abs/2025MNRAS.tmp..849K}
}

@ARTICLE{2015ARA&A..53..541B,
       author = {{Boogert}, A.~C. Adwin and {Gerakines}, Perry A. and {Whittet}, Douglas C.~B.},
        title = "{Observations of the icy universe.}",
      journal = {\araa},
         year = 2015,
        month = aug,
       volume = {53},
        pages = {541-581},
          doi = {10.1146/annurev-astro-082214-122348},
archivePrefix = {arXiv},
       eprint = {1501.05317},
 primaryClass = {astro-ph.GA},
       adsurl = {https://ui.adsabs.harvard.edu/abs/2015ARA&A..53..541B}
}

@ARTICLE{1978ApJ...224..132B,
       author = {{Bohlin}, R.~C. and {Savage}, B.~D. and {Drake}, J.~F.},
        title = "{A survey of interstellar H I from Lalpha absorption measurements. II.}",
      journal = {\apj},
         year = 1978,
        month = aug,
       volume = {224},
        pages = {132-142},
          doi = {10.1086/156357},
       adsurl = {https://ui.adsabs.harvard.edu/abs/1978ApJ...224..132B}
}

@ARTICLE{2015ApJ...813...99F,
       author = {{Flaherty}, Kevin M. and {Hughes}, A. Meredith and {Rosenfeld}, Katherine A. and {Andrews}, Sean M. and {Chiang}, Eugene and {Simon}, Jacob B. and {Kerzner}, Skylar and {Wilner}, David J.},
        title = "{Weak Turbulence in the HD 163296 Protoplanetary Disk Revealed by ALMA CO Observations}",
      journal = {\apj},
         year = 2015,
        month = nov,
       volume = {813},
       number = {2},
          eid = {99},
        pages = {99},
          doi = {10.1088/0004-637X/813/2/99},
archivePrefix = {arXiv},
       eprint = {1510.01375},
 primaryClass = {astro-ph.SR},
       adsurl = {https://ui.adsabs.harvard.edu/abs/2015ApJ...813...99F}
}

@ARTICLE{2017ApJ...843..150F,
       author = {{Flaherty}, Kevin M. and {Hughes}, A. Meredith and {Rose}, Sanaea C. and {Simon}, Jacob B. and {Qi}, Chunhua and {Andrews}, Sean M. and {K{\'o}sp{\'a}l}, {\'A}gnes and {Wilner}, David J. and {Chiang}, Eugene and {Armitage}, Philip J. and {Bai}, Xue-ning},
        title = "{A Three-dimensional View of Turbulence: Constraints on Turbulent Motions in the HD 163296 Protoplanetary Disk Using DCO$^{+}$}",
      journal = {\apj},
         year = 2017,
        month = jul,
       volume = {843},
       number = {2},
          eid = {150},
        pages = {150},
          doi = {10.3847/1538-4357/aa79f9},
archivePrefix = {arXiv},
       eprint = {1706.04504},
 primaryClass = {astro-ph.EP},
       adsurl = {https://ui.adsabs.harvard.edu/abs/2017ApJ...843..150F}
}

@ARTICLE{2016ApJ...816...25P,
       author = {{Pinte}, C. and {Dent}, W.~R.~F. and {M{\'e}nard}, F. and {Hales}, A. and {Hill}, T. and {Cortes}, P. and {de Gregorio-Monsalvo}, I.},
        title = "{Dust and Gas in the Disk of HL Tauri: Surface Density, Dust Settling, and Dust-to-gas Ratio}",
      journal = {\apj},
         year = 2016,
        month = jan,
       volume = {816},
       number = {1},
          eid = {25},
        pages = {25},
          doi = {10.3847/0004-637X/816/1/25},
archivePrefix = {arXiv},
       eprint = {1508.00584},
 primaryClass = {astro-ph.SR},
       adsurl = {https://ui.adsabs.harvard.edu/abs/2016ApJ...816...25P}
}

@ARTICLE{2021ApJ...912..164D,
       author = {{Doi}, Kiyoaki and {Kataoka}, Akimasa},
        title = "{Estimate on Dust Scale Height from the ALMA Dust Continuum Image of the HD 163296 Protoplanetary Disk}",
      journal = {\apj},
         year = 2021,
        month = may,
       volume = {912},
       number = {2},
          eid = {164},
        pages = {164},
          doi = {10.3847/1538-4357/abe5a6},
archivePrefix = {arXiv},
       eprint = {2102.06209},
 primaryClass = {astro-ph.EP},
       adsurl = {https://ui.adsabs.harvard.edu/abs/2021ApJ...912..164D}
}

@ARTICLE{2023MNRAS.524.3184P,
       author = {{Pizzati}, Elia and {Rosotti}, Giovanni P. and {Tabone}, Beno{\^\i}t},
        title = "{Constraining turbulence in protoplanetary discs using the gap contrast: an application to the DSHARP sample}",
      journal = {\mnras},
         year = 2023,
        month = sep,
       volume = {524},
       number = {2},
        pages = {3184-3200},
          doi = {10.1093/mnras/stad2057},
archivePrefix = {arXiv},
       eprint = {2307.11150},
 primaryClass = {astro-ph.EP},
       adsurl = {https://ui.adsabs.harvard.edu/abs/2023MNRAS.524.3184P}
}

@ARTICLE{2011ApJ...738..141O,
       author = {{Oka}, Akinori and {Nakamoto}, Taishi and {Ida}, Shigeru},
        title = "{Evolution of Snow Line in Optically Thick Protoplanetary Disks: Effects of Water Ice Opacity and Dust Grain Size}",
      journal = {\apj},
         year = 2011,
        month = sep,
       volume = {738},
       number = {2},
          eid = {141},
        pages = {141},
          doi = {10.1088/0004-637X/738/2/141},
archivePrefix = {arXiv},
       eprint = {1106.2682},
 primaryClass = {astro-ph.EP},
       adsurl = {https://ui.adsabs.harvard.edu/abs/2011ApJ...738..141O}
}

@ARTICLE{2025A&A...701A.194P,
       author = {{Pacetti}, E. and {Schisano}, E. and {Turrini}, D. and {Dullemond}, C.~P. and {Molinari}, S. and {Walsh}, C. and {Fonte}, S. and {Lebreuilly}, U. and {Klessen}, R.~S. and {Hennebelle}, P. and {Ivanovski}, S.~L. and {Politi}, R. and {Polychroni}, D. and {Simonetti}, P. and {Testi}, L.},
        title = "{Planet formation in chemically diverse and evolving discs: I. Composition of planetary building blocks}",
      journal = {\aap},
         year = 2025,
        month = sep,
       volume = {701},
          eid = {A194},
        pages = {A194},
          doi = {10.1051/0004-6361/202554012},
archivePrefix = {arXiv},
       eprint = {2506.17399},
 primaryClass = {astro-ph.EP},
       adsurl = {https://ui.adsabs.harvard.edu/abs/2025A&A...701A.194P}
}

@ARTICLE{2022ApJ...937...36P,
       author = {{Pacetti}, Elenia and {Turrini}, Diego and {Schisano}, Eugenio and {Molinari}, Sergio and {Fonte}, Sergio and {Politi}, Romolo and {Hennebelle}, Patrick and {Klessen}, Ralf and {Testi}, Leonardo and {Lebreuilly}, Ugo},
        title = "{Chemical Diversity in Protoplanetary Disks and Its Impact on the Formation History of Giant Planets}",
      journal = {\apj},
         year = 2022,
        month = sep,
       volume = {937},
       number = {1},
          eid = {36},
        pages = {36},
          doi = {10.3847/1538-4357/ac8b11},
archivePrefix = {arXiv},
       eprint = {2206.14685},
 primaryClass = {astro-ph.EP},
       adsurl = {https://ui.adsabs.harvard.edu/abs/2022ApJ...937...36P}
}

@ARTICLE{2022MNRAS.517.2285C,
       author = {{Cevallos Soto}, Arturo and {Tan}, Jonathan C. and {Hu}, Xiao and {Hsu}, Chia-Jung and {Walsh}, Catherine},
        title = "{Inside-out planet formation - VII. Astrochemical models of protoplanetary discs and implications for planetary compositions}",
      journal = {\mnras},
         year = 2022,
        month = dec,
       volume = {517},
       number = {2},
        pages = {2285-2308},
          doi = {10.1093/mnras/stac2650},
archivePrefix = {arXiv},
       eprint = {2202.02483},
 primaryClass = {astro-ph.EP},
       adsurl = {https://ui.adsabs.harvard.edu/abs/2022MNRAS.517.2285C}
}

@ARTICLE{2025A&A...702A.126G,
       author = {{Grant}, S.~L. and {Temmink}, M. and {van Dishoeck}, E.~F. and {Gasman}, D. and {Arabhavi}, A.~M. and {Tabone}, B. and {Henning}, T. and {Kamp}, I. and {Caratti o Garatti}, A. and {Christiaens}, V. and {Esteve}, P. and {G{\"u}del}, M. and {Jang}, H. and {Kaeufer}, T. and {Kurtovic}, N.~T. and {Morales-Calder{\'o}n}, M. and {Perotti}, G. and {Schwarz}, K. and {Sellek}, A.~D. and {Stapper}, L.~M. and {Vlasblom}, M. and {Waters}, L.~B.~F.~M.},
        title = "{MINDS: A transition from H$_{2}$O to C$_{2}$H$_{2}$ dominated disk spectra with decreasing stellar luminosity}",
      journal = {\aap},
         year = 2025,
        month = oct,
       volume = {702},
          eid = {A126},
        pages = {A126},
          doi = {10.1051/0004-6361/202555862},
archivePrefix = {arXiv},
       eprint = {2508.04692},
 primaryClass = {astro-ph.EP},
       adsurl = {https://ui.adsabs.harvard.edu/abs/2025A&A...702A.126G}
}

@ARTICLE{2025ApJ...984L..62A,
       author = {{Arabhavi}, Aditya M. and {Kamp}, Inga and {van Dishoeck}, Ewine F. and {Henning}, Thomas and {Jang}, Hyerin and {Christiaens}, Valentin and {Gasman}, Danny and {Pascucci}, Ilaria and {Perotti}, Giulia and {Grant}, Sierra L. and {Barrado}, David and {G{\"u}del}, Manuel and {Lagage}, Pierre-Olivier and {Caratti o Garatti}, Alessio and {Lahuis}, Fred and {Waters}, L.~B.~F.~M. and {Kaeufer}, Till and {Kanwar}, Jayatee and {Morales-Calder{\'o}n}, Maria and {Schwarz}, Kamber and {Sellek}, Andrew D. and {Tabone}, Beno{\^\i}t and {Temmink}, Milou and {Vlasblom}, Marissa},
        title = "{MINDS: The Very Low-mass Star and Brown Dwarf Sample Hidden Water in Carbon-dominated Protoplanetary Disks}",
      journal = {\apjl},
         year = 2025,
        month = may,
       volume = {984},
       number = {2},
          eid = {L62},
        pages = {L62},
          doi = {10.3847/2041-8213/adc692},
archivePrefix = {arXiv},
       eprint = {2504.11425},
 primaryClass = {astro-ph.EP},
       adsurl = {https://ui.adsabs.harvard.edu/abs/2025ApJ...984L..62A}
}

\appendix
    \section{Effect of C/O ratio}
\label{append:C/O}

\begin{figure}[t]
\centering
\includegraphics[width=0.5\hsize]{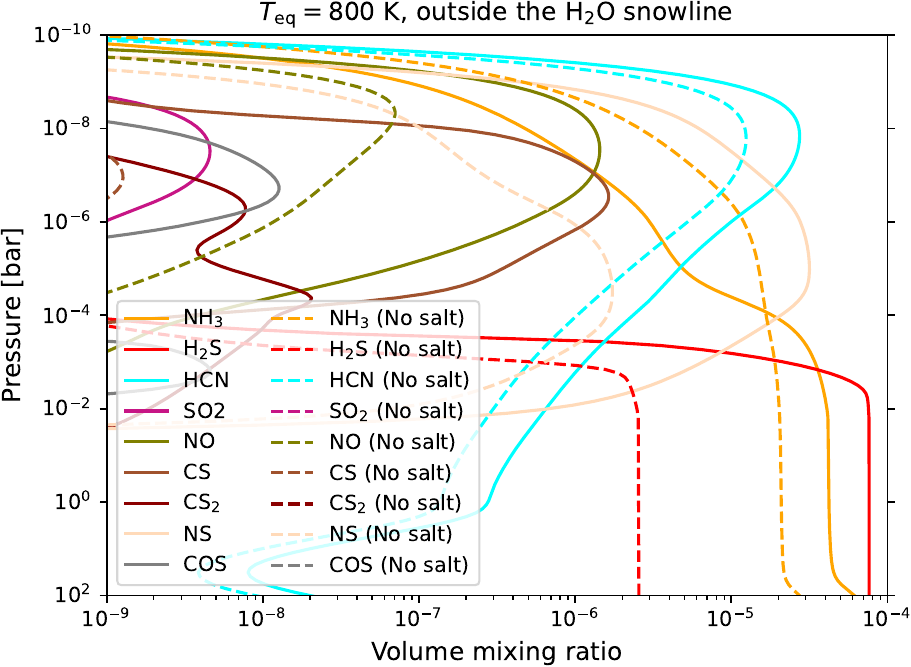}
\caption{Same as Figure \ref{fig:AS_temp}, but planet accreting disk gas at $r = 2.5$ au (outside the water snow line). The planet's equilibrium temperature is 800 K. Solid lines depict atmospheric structures for the planet that formed in the salt-bearing disk with elemental abundances (O/H, C/H, N/H, S/H)=(0.23, 0.28, 3.6, 3.0) $\times$ solar values, whereas dashed lines correspond to planets formed in the salt-free disk with (O/H, C/H, N/H, S/H)=(0.23, 0.27, 0.8, 0.1) $\times$ solar values.}
\label{fig:AS_outer}
\end{figure}

\begin{figure*}[t]
      \hspace{0.5cm}
      \begin{minipage}[c]{0.85\linewidth}
        \centering
        \includegraphics[width=0.5\linewidth]{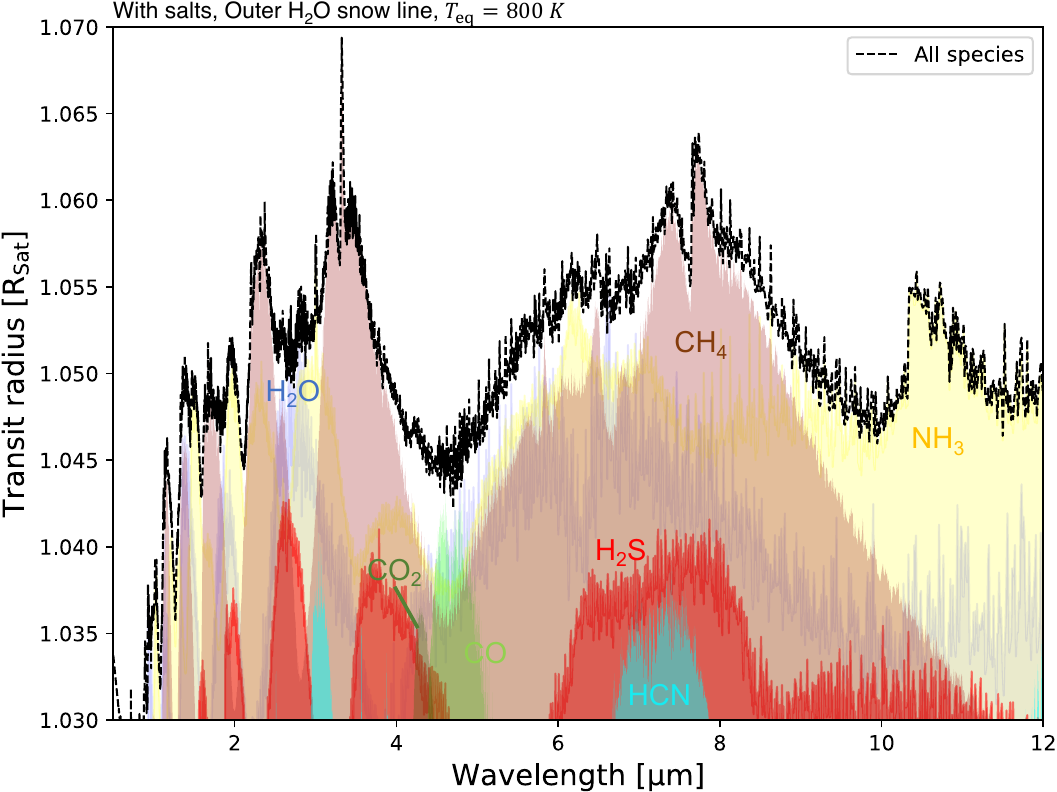}
      \end{minipage} \\
      \hspace{0.5cm}
      \begin{minipage}[c]{0.85\linewidth}
        \centering
        \includegraphics[width=0.5\linewidth]{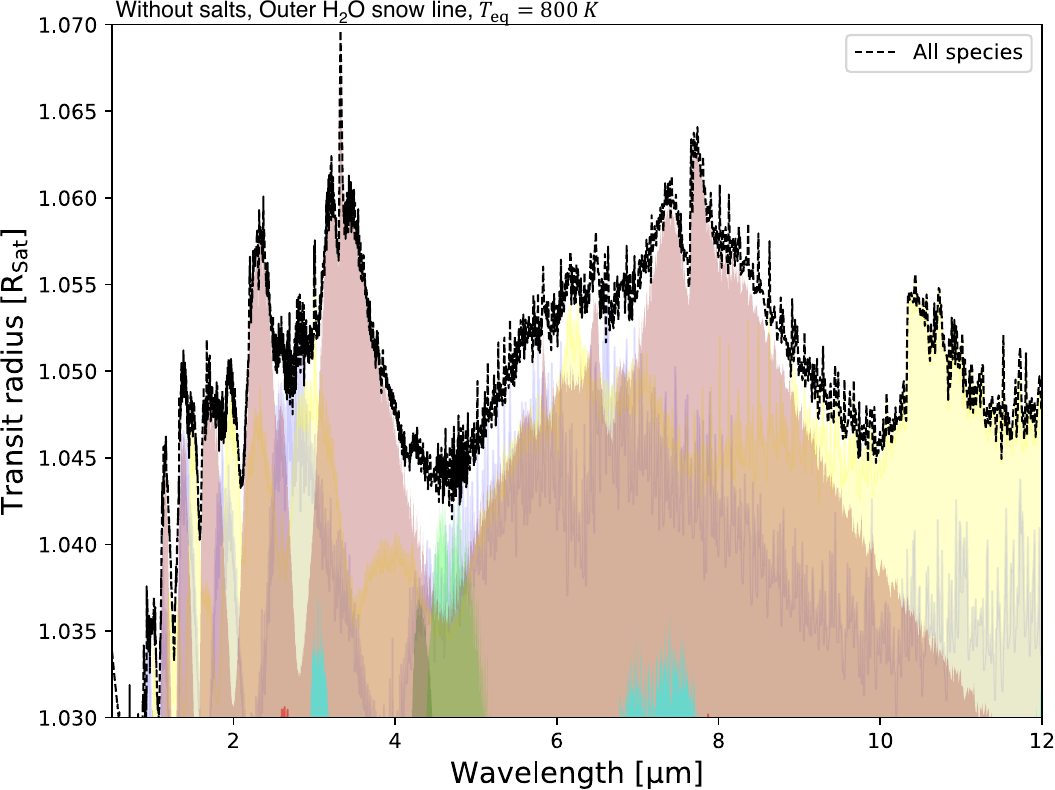}
      \end{minipage} \\
    \caption{The same configuration as Figure \ref{fig:TS_fid_innerH2O}, but transmission spectra of the planet formed outside the water snow line (2.5 au). The corresponding atmospheric structures are shown in the Figure \ref{fig:AS_outer}.}
\label{fig:TS_fid_outerH2O}
\end{figure*}

Radial drift of the major disk ices H$_2$O, CO, and CO$_2$ produces a broad range of C/O ratios, which record each planet's formation radius. Recent observations indicate that close-in gas giants generally show sub-stellar C/O values \citep{2024RvMG...90..411K, 2025arXiv250601800W}, consistent with accreting water-rich gas inside the water snow line. We therefore adopted a fiducial model in which planets form within water snow line. However, salt dissociation enriches sulfur and nitrogen on both sides of the water snow line, so we also calculate atmospheric structures and transmission spectra for planets that accrete salt-derived gas at 2.5 au, outside the water snow line, and examine how the C/O ratio modifies the results.

Disk gas just outside the water snow line has C/O $\sim 1$. Planets that accrete this gas hold HCN as the dominant nitrogen species and CS as the dominant sulfur species in their upper atmospheres, regardless of whether salts supply extra nitrogen and sulfur (Figure \ref{fig:AS_outer}). As in planets formed inside the snow line when salts are present, the mixing ratios of H$_2$S and NS rise by more than an order of magnitude relative to models without salts, and HCN increases by a factor of two. Figure \ref{fig:TS_fid_outerH2O} presents the transmission spectra derived from these atmospheric structures. Outside the snow line, the weak H$_2$O signature lets NH$_3$, HCN, and H$_2$S absorption stand out, and salts further strengthen these bands. 

SO$_2$ is absent even when salt-driven sulfur enrichment occurs.
SO$_2$ is produced through the following photochemical chain reactions \citep{2016ApJ...824..137Z, 2021ApJ...923..264T}:
\begin{equation}
\label{SO2_production}
    \begin{split}
        \rm{H_2O} &\overset{h \nu}{\longrightarrow} \rm{OH + H} \\
        \rm{S + OH} &\longrightarrow \rm{SO + H} \\
        \rm{SO + OH} &\longrightarrow \rm{SO_2 + H}.
    \end{split}
\end{equation}
The higher the O/H ratio (i.e., the more abundant the water molecules), the more efficient the production of OH radicals. 
The atomic sulfur in this reaction is supplied through the destruction of H$_2$S via hydrogen atom abstraction \citep{2021ApJ...923..264T}:
\begin{equation}
\label{S_production}
    \begin{split}
        \rm{H_2S + H} &\longrightarrow \rm{HS + H_2} \\
        \rm{HS + H} &\longrightarrow \rm{S + H_2}.
    \end{split}
\end{equation}
As a result, SO$_2$ is abundant in the upper layers of the atmosphere, where photochemical reactions are efficient, while H$_2$S is abundant in the lower layers, where such reactions are less efficient (Figure \ref{fig:AS_fid}).
In contrast, the following reactions inhibit the formation of the intermediate product SO:
\begin{equation}
\label{CS,OCS_production}
    \begin{split}
        \rm{S + C} &\longrightarrow \rm{CS} \\
        \rm{S + CO} &\longrightarrow \rm{COS}.
    \end{split}
\end{equation}
Thus, SO$_2$ production is more efficient when the S/H and O/H ratios are high and the C/O ratio is low.
The availability of salts corresponds to the first condition, while whether the planet formed inside or outside the water snow line relates to the second and third condition, influencing both the promotion and inhibition of SO$_2$ production.
Thus, orbits inside the water snowline provide a preferential site where the formed planets contain abundant SO$_2$ in their atmospheres if they inherit the C/O ratio of the disk gas.
Planet formation near the salt line, where simultaneous enrichment of oxygen and sulfur can naturally occur, is favorable for explaining the detection of SO$_2$. However, oxygen and sulfur in short-period exoplanet may also originate from vapor released by the sublimation of silicates and FeS near the inner edge of the disk. Other factors contributing to super-solar elemental abundances in planetary atmospheres are discussed in Section \ref{subsec:volatiles_enrichment}.

\end{document}